\documentclass[twocolumn,aps,prl,reprint, superscriptaddress,longbibliography]{revtex4-1}

\usepackage{amsmath}
\usepackage{amssymb}
\usepackage{graphicx}
\usepackage{color}
\usepackage{hyperref}

\usepackage{enumitem}

\usepackage[colorinlistoftodos]{todonotes}
\usepackage{lipsum}
\usepackage{booktabs}
\usepackage{todonotes}
\usepackage{titlesec}

\newcommand{\ket}[1]{\left|#1\right\rangle}
\newcommand{\bra}[1]{\left\langle#1\right|}

\newcommand{\Hdisp}{H_\mathrm{disp}}
\newcommand{\GHz}{\text{\,GHz}}
\newcommand{\MHz}{\text{\,MHz}}
\newcommand{\kHz}{\text{\,kHz}}
\newcommand{\ns}{\text{\,ns}}

\newcommand{\Phiext}{\Phi_\mathrm{ext}}
\newcommand{\PhiDC}{\Phi_\mathrm{DC}}
\graphicspath{ {main_figures/} {supplement_figures/} }

\makeatletter
\@ifundefined{textcolor}{}
{%
 \definecolor{BLACK}{gray}{0}
 \definecolor{WHITE}{gray}{1}
 \definecolor{RED}{rgb}{1,0,0}
 \definecolor{GREEN}{rgb}{0,1,0}
 \definecolor{BLUE}{rgb}{0,0,1}
 \definecolor{CYAN}{cmyk}{1,0,0,0}
 \definecolor{MAGENTA}{cmyk}{0,1,0,0}
 \definecolor{YELLOW}{cmyk}{0,0,1,0}
 \definecolor{darkgreen}{HTML}{1E821D}
}
\usepackage{dcolumn}
\usepackage{bm}

\DeclareMathOperator{\Tr}{Tr}

\usepackage{marginnote}

\newif\ifcomments
\commentstrue 

\makeatother

\begin{document}
\titleformat{\section}{\bfseries\small\centering}{\thesection.}{1em}{\MakeUppercase}

\title{Time-Dependent Hamiltonian Reconstruction \\ using Continuous Weak Measurements}

\author{Karthik~Siva}
\email{karthik\_siva@berkeley.edu}
\affiliation{Department of Physics, University of California, Berkeley, Berkeley, CA 94720, USA}

\author{Gerwin Koolstra}
\email{gkoolstra@lbl.gov}
\affiliation{Department of Physics, University of California, Berkeley, Berkeley, CA 94720, USA}
\affiliation{Computational Research Division, Lawrence Berkeley National Laboratory, Berkeley, CA 94720, USA}

\author{John Steinmetz}
\affiliation{Department of Physics and Astronomy, University of Rochester, Rochester, New York 14627, USA}
\affiliation{Center for Coherence and Quantum Optics, University of Rochester, Rochester, New York 14627, USA}

\author{William P. Livingston}
\affiliation{Department of Physics, University of California, Berkeley, Berkeley, CA 94720, USA}

\author{Debmalya Das}
\affiliation{Department of Physics and Astronomy, University of Rochester, Rochester, New York 14627, USA}
\affiliation{Center for Coherence and Quantum Optics, University of Rochester, Rochester, New York 14627, USA}

\author{L.~Chen}
\affiliation{Department of Physics, University of California, Berkeley, Berkeley, CA 94720, USA}

\author{J.M.~Kreikebaum}
\affiliation{Department of Physics, University of California, Berkeley, Berkeley, CA 94720, USA}
\affiliation{Materials Sciences Division, Lawrence Berkeley National Laboratory, Berkeley, CA 94720, USA}

\author{N.~J.~Stevenson}
\affiliation{Department of Physics, University of California, Berkeley, Berkeley, CA 94720, USA}

\author{C.~J\"unger}
\affiliation{Department of Physics, University of California, Berkeley, Berkeley, CA 94720, USA}
\affiliation{Computational Research Division, Lawrence Berkeley National Laboratory, Berkeley, CA 94720, USA}

\author{D. I.~Santiago}
\affiliation{Department of Physics, University of California, Berkeley, Berkeley, CA 94720, USA}
\affiliation{Computational Research Division, Lawrence Berkeley National Laboratory, Berkeley, CA 94720, USA}

\author{I. Siddiqi}
\affiliation{Department of Physics, University of California, Berkeley, Berkeley, CA 94720, USA}
\affiliation{Computational Research Division, Lawrence Berkeley National Laboratory, Berkeley, CA 94720, USA}
\affiliation{Materials Sciences Division, Lawrence Berkeley National Laboratory, Berkeley, CA 94720, USA}

\author{A. N.~Jordan}
\affiliation{Department of Physics and Astronomy, University of Rochester, Rochester, New York 14627, USA}
\affiliation{Center for Coherence and Quantum Optics, University of Rochester, Rochester, New York 14627, USA}
\affiliation{Institute for Quantum Studies, Chapman University, Orange, California 92866, USA}

\date{\today}

\begin{abstract}
    Reconstructing the Hamiltonian of a quantum system is an essential task for characterizing and certifying quantum processors and simulators. Existing techniques either rely on projective measurements of the system before and after coherent time evolution and do not explicitly reconstruct the full time-dependent Hamiltonian or interrupt evolution for tomography. Here, we experimentally demonstrate that an \textit{a priori} unknown, time-dependent Hamiltonian can be reconstructed from continuous weak measurements concurrent with coherent time evolution in a system of two superconducting transmons coupled by a flux-tunable coupler. In contrast to previous work, our technique does not require interruptions, which would distort the recovered Hamiltonian. We introduce an algorithm which recovers the Hamiltonian and density matrix from an incomplete set of continuous measurements and demonstrate that it reliably extracts amplitudes of a variety of single qubit and entangling two qubit Hamiltonians. We further demonstrate how this technique reveals deviations from a theoretical control Hamiltonian which would otherwise be missed by conventional techniques. Our work opens up novel applications for continuous weak measurements, such as studying non-idealities in gates, certifying analog quantum simulators, and performing quantum metrology.
\end{abstract}

\maketitle

\section{Introduction}
A central challenge in building reliable quantum computers and simulators is the characterization and validation of the Hamiltonians which they generate. This problem is of interest in both the quantum gate characterization and quantum simulation communities. The former seeks to quantify the errors in the quantum gates used in various experimental platforms and understand what limits their performance. Techniques such as quantum process tomography \cite{Chuang1997, Merkel2013, McKay2019}, gate set tomography \cite{Blume-Kohout2013, Greenbaum2015, Blume-Kohout2017, Rudinger2021}, and randomized benchmarking \cite{Emerson2005,Knill2008, Dankert2009, Magesan2011} have been developed and employed experimentally to characterize the type and level of errors present in sets of quantum gates in superconducting qubit, trapped ion, spin qubit, and quantum dot platforms \cite{Kim2014, Madzik2022, Xue2022}. However, such techniques do not necessarily elucidate what causes a gate to deviate from its idealization at the level of the Hamiltonian. For example, in superconducting qubit platforms, gates are performed on qubits by applying RF pulses generated by room temperature electronics. However, distortions in the pulse shape \cite{Yan2016, Willsch2017, Klimov2018, Arute2019, Lao2022} due to cryogenic microwave components, cross-talk between qubits, and deviations from approximations in the Hamiltonian cause the actual Hamiltonian as seen by the qubit to differ from the theoretical Hamiltonian generated by those pulses. We can then ask the following question: if we apply a pulse of duration $t$ intended to generate a quantum gate $U = \mathcal{T}\exp(-(i/\hbar)\int_0^t H(t')dt')$ (where $\mathcal{T}$ is the time-ordering operator), how does the actual time-dependent Hamiltonian $H(t)$ produced deviate from the ideal or theoretical one? Current gate characterization techniques do not answer this question, as they do not directly compute the Hamiltonian $H(t)$. A complete description of $H(t)$, though, would shed light on the sources of non-ideality of a gate, such as driving of undesirable terms in the Hamiltonian, or could be used to improve optimal control schemes \cite{Khaneja2005, Caneva2011,Doria2011, Muller2011}.

This question also arises naturally in the setting of analog quantum simulators, where the goal is to validate the Hamiltonians used in simulating the evolution of many-body quantum states \cite{Greiner2002, Choi2016, Bluvstein2021, Semeghini2021, Scholl2021, Altman2021, Monroe2021}. Here a system of qubits evolves under an interacting many-body Hamiltonian, such as in the Fermi-Hubbard model. For general many-body systems, the number of terms in the Hamiltonian can scale exponentially in the number of degrees of freedom. However, for systems with local interactions, this scaling is only polynomial. This observation has been used to propose several techniques for reconstructing Hamiltonians, for example by constructing the so-called ``correlation matrix" \cite{Bairey2019, Qi2019, Li2020}. Still, such techniques rely on interrupting the evolution to perform projective measurements. As the system controls, qubit evolution, and measurement apparatus cannot change instantaneously, the reconstructed Hamiltonian differs from the uninterrupted Hamiltonian seen by the qubit.


In this work, we introduce an experimental technique and theoretical tool for reconstructing an \textit{a priori} unknown time-dependent Hamiltonian using continuous weak measurements \cite{Diosi88,Wiseman1993,Jacobs2006,Gambetta2008, Murch2013, Jacobs2014, Korotkov2014, Korotkov2016}. Instead of interrupting the Hamiltonian evolution for projective measurements, we \textit{concurrently} measure the system and evolve it under the Hamiltonian, and our reconstruction technique accounts for the effect of the continuous measurement tone, thereby avoiding introducing spurious dynamics into the reconstructed Hamiltonian.  We experimentally apply and validate this technique for a system of two superconducting qubits coupled by a parametric coupler, which can be modulated to produce a variety of two-qubit entangling Hamiltonians~\cite{McKay2016, Lu2017,Yan2018, Abrams2020, Hong2020, Foxen2020, Sung2021, Sete2021}. By continuously measuring the field amplitude while applying pulses to the qubits and coupler, we can recover the qubits' $z$ coordinates $\langle \sigma_z(t)\rangle$ during the evolution. Next, to recover the Hamiltonian generated by those pulses, we introduce a reconstruction algorithm which takes as input the qubits' $z$ coordinates as a function of time and outputs an estimate of the Hamiltonian expressed in the Pauli basis. 

We validate our technique by applying it to various single-qubit and two-qubit Hamiltonians. We compare the reconstructed Hamiltonians to estimates made from independent calibrations and find good agreement. We also achieve an average of 97.5\% (97.1\%) fidelity between the final state as predicted by the reconstructed single (two) qubit Hamiltonians and tomography of the system at the end of the time evolution.

Compared to gate characterization techniques, which use only initial and final states as input, our method provides a complete time-dependent description of the dynamics produced by a pulse. Compared to previous time-dependent Hamiltonian reconstruction techniques~\cite{Krastanov2019,Bairey2019, Qi2019, Li2020}, our method does not rely on interrupting the Hamiltonian nor on having carefully calibrated Hamiltonian parameters \cite{Jacobs2006, Barchielli2009, Vijay2011, Vijay2012, Hatridge2013, Jacobs2014, Roch2014, Sun2014,Chantasri2016,Hacohen-Gourgy2016,Campagne-Ibarcq2016, Vool2016,Weber2016,Ficheux2018, Steinmetz2022}. The theoretical and experimental techniques introduced in this work therefore present an alternative approach for analyzing the dynamics of quantum systems, and our results suggest that this technique may prove useful in analyzing quantum gates, especially in emerging quantum platforms, certifying analog quantum simulators, and performing quantum metrology.


The paper is organized as follows. In Sec.~\ref{sec:experimental_setup}, we explain the experimental setup and how the continuous weak measurements are performed. We then formally state the problem, define the algorithm for reconstructing Hamiltonians for one and two qubits, and state the requirements. In Sec.~\ref{sec:SQHR}, we apply the Hamiltonian reconstruction algorithm to Hamiltonians generated by various single qubit pulses. In Sec.~\ref{sec:coupler}, we demonstrate the operation of the flux-tunable coupler to generate entangling two-qubit Hamiltonians. In Sec.~\ref{sec:2QHR}, we then experimentally validate the algorithm on two-qubit Hamiltonians and, as a novel application of this technique, estimate a dynamical coherent fidelity. Finally, in Sec.~\ref{sec:conclusions} we comment on the limitations of the technique and how they could be overcome and propose how this technique could be applied in other settings.

\begin{figure*}
\centering
    \includegraphics[width=\textwidth]{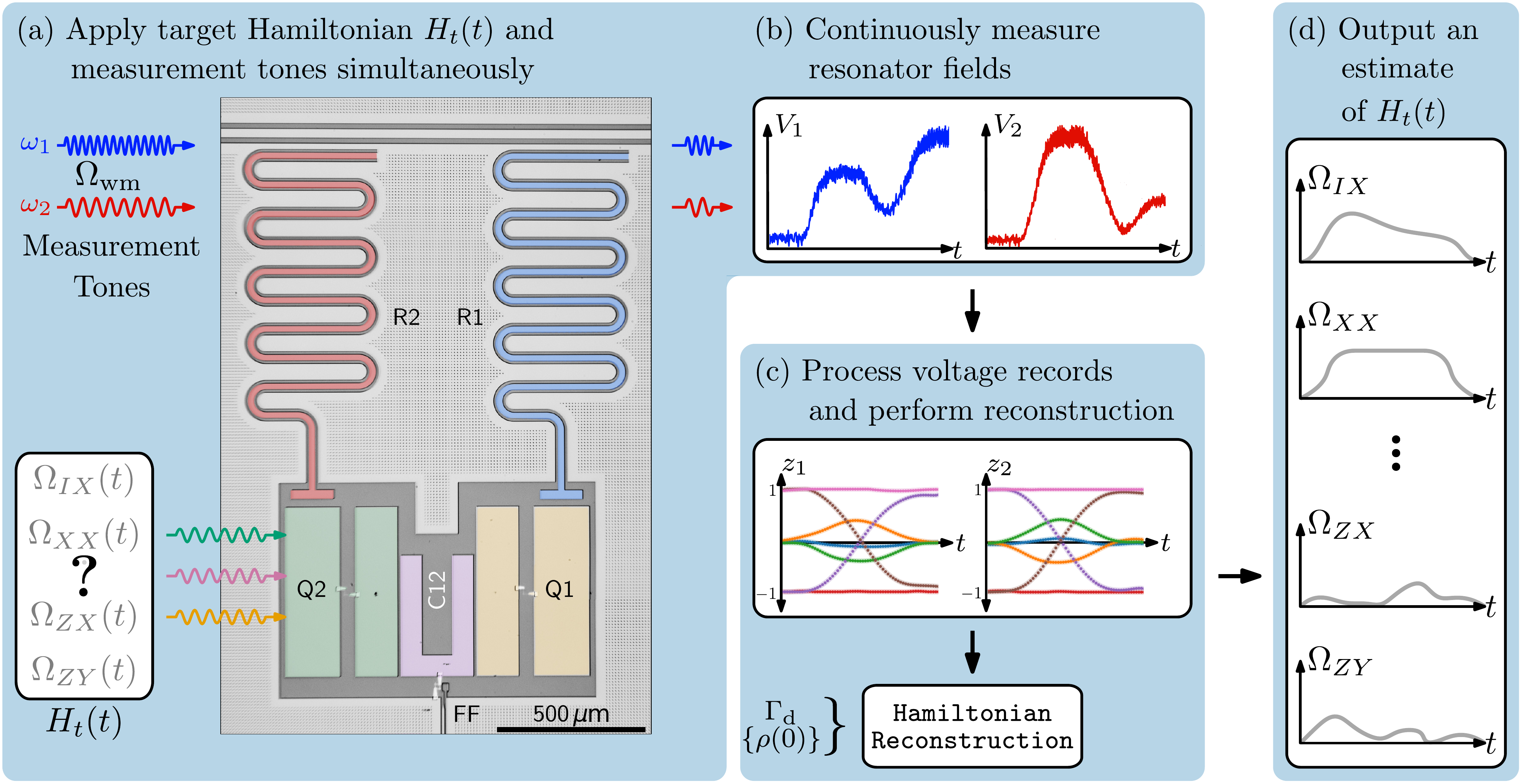}
    \caption{Hamiltonian reconstruction method. (a) To reconstruct an unknown target Hamiltonian $H_t (t)$ of a system composed of two fixed frequency transmons (Q1, Q2) coupled via flux-tunable coupler (C12), we continuously measure the qubits while applying pulses to generate $H_t(t)$. Single qubit target Hamiltonians are generated by pulses of duration of $t_p$ on the charge lines. Entangling two qubit target Hamiltonians are generated by applying RF pulses on the flux line (FF). In the adiabatic limit, the noisy resonator output field (schematically depicted in (b)) follows the evolution of the qubit $z$ coordinate during the pulse. (c) We average, filter and deconvolve voltage records $V_{1,2}$ and record the resulting qubits' $z$ evolution for $S\geq 2$ different initial states while keeping the qubit/coupler pulses fixed. The resulting ensemble averages $z_{1,2} (t)$ along with the calibrated measurement-induced dephasing rate $\Gamma_\mathrm{d}$ are inputted to the Hamiltonian reconstruction algorithm. (d) The output of the algorithm is an estimate of the time-dependent target Hamiltonian expanded in the Pauli operator basis.}
    \label{fig:fig_1}
\end{figure*}

\section{Experimental Procedure}\label{sec:experimental_setup}
In superconducting qubit experiments, Hamiltonians are applied to the device using RF pulses, which are generated by room temperature electronics and sent to the device via cryogenic wiring. However, the Hamiltonian as seen by the device may differ from an ideal or theoretical one if, for example, the pulse shape is distorted by the wiring. Cross-talk between qubits or drive lines and pulse miscalibration also drive undesirable terms in the Hamiltonian. Finally, the theoretical Hamiltonian describing the system generally only approximates the dynamics, and  high power pulses cause violations of the assumptions of those approximations. Our goal, then, is to reconstruct the true, time-dependent system Hamiltonian, which we refer to as the ``target" Hamiltonian $H_t(t)$. For example, a general single qubit Hamiltonian $H_t(t)$ can be expressed as

\begin{equation}
    H_t(t)/\hbar = \frac{1}{2}\vec{\Omega}(t)\cdot\vec{\sigma},
     \label{eq:H-general}
\end{equation}
where $\vec{\sigma} = (\sigma_x, \sigma_y, \sigma_z)$, a vector of the three Pauli operators, and we seek to reconstruct the drive amplitudes $\vec{\Omega}(t)$. We now introduce our procedure for using continuous weak measurements to reconstruct target Hamiltonians for systems of one and two qubits. We first detail how measurements on our device are performed and then explain Hamiltonian reconstruction algorithm.

\subsection{Experimental Setup}
Our experiment consists of two superconducting transmon qubits~\cite{Koch2007}, labeled Q1(Q2) in Fig.~\ref{fig:fig_1}(a), each dispersively coupled to a coplanar waveguide resonator, labeled R1(R2). The two qubits are coupled through a high frequency flux-tunable coupler consisting of an asymmetric SQUID and a large U-shaped paddle to achieve high capacitive coupling to the transmons~\cite{McKay2016,Lu2017, Yan2018, Abrams2020, Hong2020, Foxen2020, Sung2021, Sete2021}. The frequency of the coupler and qubits and the resulting static coupling between the two qubits can be tuned by applying a DC bias on the flux line, which changes the flux $\Phi_\mathrm{ext}$ through the SQUID. Throughout this work, we operate with $\Phi_\mathrm{ext}=0$. The frequencies of Q1(Q2) and R1(R2) at this point are $\omega_q/2\pi = 5.319\GHz(5.271\GHz)$ and $\omega_\mathrm{res}/2\pi = 6.631\GHz(6.529\GHz)$, where $\omega_q$ is the $\ket{0}\rightarrow\ket{1}$ transition frequency of the transmon and $\omega_\mathrm{res}$ is the resonator frequency. In Sec.~\ref{sec:coupler}, we demonstrate parametric modulation of the coupler, which generates a two-qubit entangling Hamiltonian, and then in Sec.~\ref{sec:2QHR}, we apply our reconstruction technique to both qubits evolving under such target Hamiltonians.

In the interaction picture, the static dispersive Hamiltonian $\Hdisp$ for the system is approximately
\begin{equation}
    \Hdisp/\hbar  = \sum_{i=1}^2 \frac{\chi_i}{2} a^\dagger_i a_i \sigma_z^{(i)},
    \label{eq:H_ss}
\end{equation}
where $\chi_i$ is the full dispersive shift of the $i$-th resonator frequency conditioned on the state of the $i$-th qubit, $a_i$ ($a_i^\dagger$) is the annihilation (creation) operator for the $i$-th resonator, and $\sigma_z^{(i)}$ is the Pauli $z$ operator on the $i$-th qubit. In our device, $\chi_1/2\pi$($\chi_2/2\pi$) is $0.64\MHz$($0.75\MHz$). The resonators are each coupled through a Purcell filter \cite{Jeffrey2014,Sete2015, Bronn2015} to a transmission line at a rate $\kappa/2\pi=11.78\MHz (14.47\MHz)$. A complete diagram of the experimental setup is provided in App.~\ref{supp:experimental_setup}, and a full table of device parameters is provided in App.~\ref{supp:device_parameters_and_stability}. 
We continuously measure the qubits as they evolve under $H_t(t)$ by simultaneously applying a weak measurement (WM) tone on the transmission line at the midpoint frequency between the two qubit state-dependent frequencies of the resonator (top of Fig.~\ref{fig:fig_1}(a)). The tone populates the resonator with mean photon number $\bar{n} = \langle a^\dagger a\rangle  \approx 0.94(0.82)$, and the reflected resonator field is then measured as follows \cite{Steinmetz2022}. We amplify both quadratures of the reflected field using a traveling wave parametric amplifier (TWPA) \cite{Macklin2015}, and after further amplification and subsequent demodulation,  digitize both quadratures at 1 gigasample per second (GSa/s) to generate voltage records. 

Each shot of the experiment produces a noisy voltage record (Fig.~\ref{fig:fig_1}(b)) carrying a small amount of information about the qubit $z$ coordinate. To estimate the ensemble average $z(t) = \Tr(\rho(t) \sigma_z)$ from the voltage records, we first average many records and filter and downsample them to $2\ns$ resolution (see App.~\ref{supp:data_processing} for more details). When $\chi/\kappa\ll 1$, the average resonator field is proportional to $z(t)$, retarded by a delay $\tau \approx 2/\kappa$:
\begin{equation}
    \label{eq:resonator_field}
    \text{Re}[a(t)] \approx -\frac{2\Omega_\mathrm{wm}}{\kappa}\frac{\chi}{\kappa}z(t-\tau),
\end{equation}
where $\Omega_\mathrm{wm}$ is the amplitude of the measurement tone. To extract $z(t)$ for Q1(Q2), we shift the averaged, filtered trace by $\tau = 27\ns$($22\ns$). When the qubit state $z(t)$ does not evolve too quickly compared to the resonator linewidth, the resonator field adiabatically follows the qubit~\cite{Koolstra2022}. For example, if the qubit is driven around the $x$ axis with amplitude $\Omega_X$, we require that $2 \Omega_X\ll \kappa$ (App.~\ref{supp:qubit_state_recovery}). We therefore strongly couple the resonators to the readout port.

To reconstruct $H_t(t)$, we perform this procedure for various initial states of the qubits, resulting in a collection of ensemble averages shown in Fig.~\ref{fig:fig_1}(c). To rescale the voltage traces, we additionally prepare the qubits in $\ket{0}$ and $\ket{1}$ and perform the continuous measurement described above but without any target Hamiltonian applied to the qubits. These correspond to the lower $z=1$ and upper $z=-1$ limits for the resonator field, which we use to rescale the voltage records in Fig.~\ref{fig:fig_1}(c). Finally, the extracted $z(t)$ for each qubit is inputted to the Hamiltonian reconstruction algorithm.


Besides allowing us to measure $z(t)$, the WM tone induces two additional effects on the qubits. First, it shifts the qubit frequency by $\chi\bar{n}\propto \Omega_\mathrm{wm}^2$ according to Eq.~\eqref{eq:H_ss}. Resonant single qubit pulses must therefore be applied at this shifted frequency. Second, the qubit dephases more rapidly, with the measurement-induced dephasing rate $\Gamma_\mathrm{d}\propto \bar{n}$. Using Ramsey measurements with the simultaneous WM tone, we measure $\Gamma_\mathrm{d}$ \cite{Gambetta2008} and supply it as an input to the algorithm, which takes this rate into account so that the extra dephasing does not impact the estimate of $H_t(t)$. This is explained in more detail in Sec.~\ref{sec:reconstruction_algo}.

Because the algorithm accounts for the additional dephasing, using higher measurement power appears optimal, as $\text{SNR}\propto\bar{n}$, and increasing SNR reduces the number of samples and acquisition time needed to estimate $z(t)$ accurately. However, because $\Gamma_\mathrm{d}\propto \bar{n}$, distinct initial states become indistinguishable more rapidly at higher measurement power. We will see in Sec.~\ref{sec:reconstruction_algo} how this poses a problem for the algorithm, but by considering the limit of projective measurement, this obstacle can be understood physically as a consequence of the quantum Zeno effect. In this limit, because the qubit state is pinned to the poles of the Bloch sphere, it no longer undergoes coherent dynamics. Therefore, we cannot recover those dynamics from the measurement record. Considering this tradeoff between acquisition time and qubit collapse, we calibrate our measurement strength to ensure $1/\Gamma_\mathrm{d} \gg t_p$ for all pulse durations $t_p$ that we study.

\subsection{Reconstruction Algorithm}\label{sec:reconstruction_algo}
We now explain the algorithm (Fig.~\ref{fig:fig_1}(c)), which takes as input the ensemble average $z(t)$ for various initial states, along with $\Gamma_\mathrm{d}$ and full tomography of the initial states $\{\rho(0)\}$, and estimates the time-dependent target Hamiltonian $H_t(t)$ (Fig.~\ref{fig:fig_1}(d)).

We first concentrate on the setup for a single qubit system, state the problem, and set up the notation. Formally, we consider an \textit{a priori} unknown time-dependent Hamiltonian $H_t(t)$, for a duration $t_p$.  The qubit, described by the ensemble-averaged density matrix $\rho$, evolves according to
\begin{align}
    \label{eq:Lindblad_master_eq}
    \dot{\rho} &= -\frac{i}{\hbar}[H_t(t), \rho] +  \frac{\Gamma_\mathrm{d}}{2}\mathcal{D}[\sigma_z]\rho, \\
    \mathcal{D}[L]\rho &= L\rho L^\dagger - \frac{1}{2}\{L^\dagger L, \rho\},
\end{align}
where we have ignored Lindblad operators other than measurement-induced dephasing for now. We express $H_t(t)$ in a complete operator basis, such as the Pauli operator basis in Eq.~\eqref{eq:H-general}, with time-dependent amplitudes $\{\Omega_i(t)\}$ to be determined. We measure $z(t)$ using continuous weak measurement simultaneous with $H_t(t)$ with time resolution $\Delta t$ and discretize time with index $n$ so that $t_{n+1} = t_n + \Delta t$. 

In the Heisenberg picture, the discretized equations of motion for a qubit at first order in $\Delta t$ are:
\begin{equation}
    \begin{split}
    \sigma_i(t_{n+1})  = \sigma_i(t_n) + \Delta t \big[\Omega_j(t_n)\sigma_k(t_n) \epsilon_{ijk}\\ - \Gamma_\mathrm{d} (1-\delta_{i,3})\sigma_i(t_n)\big],
    \label{eq:single_qubit_Heisenberg_EOM}        
    \end{split}
\end{equation}
where $\epsilon_{ijk}$ is the Levi-Civita symbol, repeated indices are summed, and the indices take values in $\{1,2,3\}$ which correspond to the three directions $\{x,y,z\}$. To recover the drive amplitudes $\Omega_j(t_n)$, it is convenient to express the Eq.~\eqref{eq:single_qubit_Heisenberg_EOM} more concisely in terms of matrices $M$, $D$, and $\Gamma$ as 

\begin{align}
    \label{eq:single_qubit_matrix_EOM}
    \vec{\sigma}(t_{n+1}) &= D(\Gamma_\mathrm{d})\vec{\sigma}(t_n) + \Delta t M(\vec{\sigma}(t_n)) \vec{\Omega}(t_n) \\
    [M(\vec{\sigma}(t_n))]_{ij} &= \epsilon_{ijk}\sigma_k(t_n)\\
    D(\Gamma_\mathrm{d}) &= (I - \Delta t\Gamma)\\
    [\Gamma]_{ij} &= \delta_{ij}(1-\delta_{i,3})\Gamma_\mathrm{d},
\end{align}

We now show that if the entire vector of operators $\vec{\sigma}(t_n)$ is known for all $t$ and at least $S\geq 2$ linearly independent $\vec{\sigma}(t_n)$ are used, then there exists a unique solution for $\vec{\Omega}(t_n)$ in Eq.~\eqref{eq:single_qubit_matrix_EOM}. Let $\vec{\sigma}_{[s]}(t_n)$ denote the $s$-th vector out of $S$ at time $t_n$, and let 
\begin{equation} \label{eq:stacked-vectors}
    \begin{split}
    \vec{\Sigma}_S(t_n) &= (\vec{\sigma}_{[1]}(t_n)^T, \dots, \vec{\sigma}_{[S]}(t_n)^T )^T \\
    D_S &= \bigoplus_{i=1}^S D(\Gamma_\mathrm{d}) \\
    M_S(t_n) &= (M(\vec{\sigma}_{[1]}(t_n))^T, \dots, M(\vec{\sigma}_{[S]}(t_n))^T)^T
    \end{split}
\end{equation} 
so that $\vec{\Sigma}(t_n)$ has $3S$ elements, $D_S$ is a block-diagonal matrix of dimensions $3S\times 3S$, and $M_S(t_n)$ has dimensions $3S\times3$. Then Eq.~\eqref{eq:single_qubit_matrix_EOM} becomes
\begin{equation}
    \label{eq:many_state_EOM}
   \vec{\Sigma}_S(t_{n+1})  = D_S\vec{\Sigma}_S(t_n)+ \Delta t M_S(t_n) \vec{\Omega}(t_n).
\end{equation}
This can be inverted to solve for $\vec{\Omega}(t_n)$ as long as $M_S(t_n)$ has a left inverse $(M_S(t_n))^+$, which is true if and only if it has rank 3 \footnote{The Moore-Penrose left pseudoinverse $A^+$ of an $m\times n$ matrix $A$ with rank $n$ is given by $A^+ = (A^T A)^{-1} A^T$ such that $A^+ A = I$.}. This condition can be satisfied for $S\geq2$. For example, $M_2(t_n)$ has rank 3 if it is constructed from the two initial states $\ket{0}$ and $\ket{+} = (\ket{0} + \ket{1})/\sqrt{2}$, but this condition cannot be satisfied for $S=1$. For $S \geq 2$, then:
\begin{equation}
    \label{eq:inversion_soln}
    \vec{\Omega}(t_n) = \frac{1}{\Delta t} (M_S(t_n))^+ \left(\vec{\Sigma}_S(t_{n+1}) - D_S\vec{\Sigma}_S(t_n) \right).
\end{equation}
Viewed in the Schrodinger picture, this yields a procedure by which the drive amplitudes $\vec{\Omega}(t_n)$ are recovered from a time series of Bloch vectors $\vec{\sigma}(t_n)$ starting from $S$ initial states. Moreover, the dephasing due to the weak measurement tone is taken into account, and the solution in Eq.~\eqref{eq:inversion_soln} may be extended to account for other dissipative effects.

If all three components of the Bloch vector are measured continuously, Eq.~\ref{eq:inversion_soln} can be used to reconstruct the Hamiltonian. However, in our experiment, we measure only $\sigma_z(t_n)$ from the resonator field as in Eq.~\ref{eq:resonator_field}, so we have only one component of the Bloch vector. In that case, the preceding analysis simplifies as follows. Let 
\begin{equation}
    \begin{split}
    \vec{\sigma}'(t_n) &= (\sigma_z(t_n))\\
    \vec{\Omega}'(t_n) &= (\Omega_X(t_n), \Omega_Y(t_n))^T \\
    M'(\vec{\sigma}'(t_n)) &= (\sigma_y(t_n), -\sigma_x(t_n)),
    \end{split}
\end{equation} 
be reduced versions of the Pauli vector, drive amplitude vector, and matrix $M$. Similarly to \eqref{eq:stacked-vectors}, let
\begin{equation}
    \begin{split}
    \vec{\Sigma}'_S(t_n) &= (\vec{\sigma}'_{[1]}(t_n)^T, \dots \vec{\sigma}'_{[S]}(t_n)^T )^T \\
    M'_S(t_n) &= (M'(\vec{\sigma}'_{[1]}(t_n))^T, \dots M'(\vec{\sigma}'_{[S]}(t_n))^T)^T.
    \end{split}
\end{equation}
The matrix $D_S$ is not necessary in this case because the measurement-induced dephasing does not appear in the evolution of the measured coordinate $\sigma_z$. Then the first-order evolution of all $S$ states can be summarized as
\begin{equation}
    \label{eq:many_state_EOM_truncated}
    \vec{\Sigma}'_S(t_{n+1}) = \vec{\Sigma}'_S(t_n) + \Delta t M'_S(t_n) \vec{\Omega}'(t_n).
\end{equation}
Again, if $S\geq 2$, then $\vec{\Omega}'(t_n)$ can be solved:
\begin{equation}
    \label{eq:inversion_soln_truncated}
    \vec{\Omega}'(t_n)= \frac{1}{\Delta t}(M'_S(t_n))^{+} \left(\vec{\Sigma}'_S(t_{n+1})  - \vec{\Sigma}'_S(t_n)\right).
\end{equation}

Note that $M'(\vec{\sigma}'(t_n))$ depends on $\sigma_x(t_n)$ and $\sigma_y(t_n)$, which are not measured. However, full tomography of $\rho(0)$, the initial state, gives the entire Bloch vector $\vec{\sigma}(0)$, and $\vec{\sigma}(t_n)$ for $t > 0 $ is computed by first solving for the amplitudes $\vec{\Omega}'(t_{n-1})$ using Eq.~\eqref{eq:inversion_soln_truncated} and then integrating Eq.~\eqref{eq:Lindblad_master_eq} for one timestep. Alternating between solving Eq.~\eqref{eq:inversion_soln_truncated} to obtain $\vec{\Omega}'(t_n)$ and integrating Eq.~\eqref{eq:Lindblad_master_eq} to obtain $\vec{\Sigma}_S(t_n)$, we recover $\Omega_X(t)$ and $\Omega_Y(t)$. Although $D_S$ does not appear in Eq.~\eqref{eq:inversion_soln_truncated},  the measurement-induced dephasing still affects the reconstruction during the integration step. Dephasing also shrinks the Bloch vector and will tend to make $M_S'$ become singular, so it is essential to keep $1/\Gamma_\mathrm{d} \gg t_p$.

The inability to recover $\Omega_Z(t)$ is a limitation of applying a first-order update: the qubit $z$ coordinate is insensitive at first order in time to drives around the $z$ axis. This limitation can be overcome by going to a second order update scheme, which we detail further in App.~\ref{supp:reconstruction_app}. In this work, however, we restrict our attention to the first-order update methods due to experimental limitations that we discuss later, taking care to ensure that $\Omega_Z(t) \approx 0$. Alternatively, if $\Omega_Z(t)$ is nonzero but known \textit{a priori}, it is possible to account for it by including it in $H_t(t)$ when integrating Eq.~\eqref{eq:Lindblad_master_eq}. We refer to this process as ``preconditioning" $\Omega_Z(t)$.

The discussion above can be generalized to Hamiltonians acting on $Q$ qubits with total Hilbert space dimension $d=2^Q$, and the generalization of Eq.~\eqref{eq:inversion_soln_truncated} then requires $S\geq d(d-1)/Q$ initial states. The amplitudes associated to Pauli terms which are tensor products of the identity and $\sigma_z$ cannot be recovered, of which there will be $d-1$ out of $d^2-1$ total. Just as for $\Omega_Z(t)$ in the single qubit case, if these amplitudes are known, they can still be preconditioned.

We summarize the requirements to perform Hamiltonian reconstruction and refer the reader to App.~\ref{supp:reconstruction_app} for implementation details. The algorithm takes as input the time series $z$ coordinate data and the initial Bloch vector for various initial states, calibrated values for dissipative rates, and any preconditioned drive amplitudes. It is not necessary that the initial states be prepared perfectly along any particular axis, but that  $M'_S$ is full rank. The algorithm returns an estimate of the drive amplitudes $\vec{\Omega}(t)$, except those for which it cannot solve using first-order updates, and an estimate of Bloch vector at all times for each initial state.

\begin{figure*}[t]
    \centering
    \includegraphics[width=0.9\linewidth]{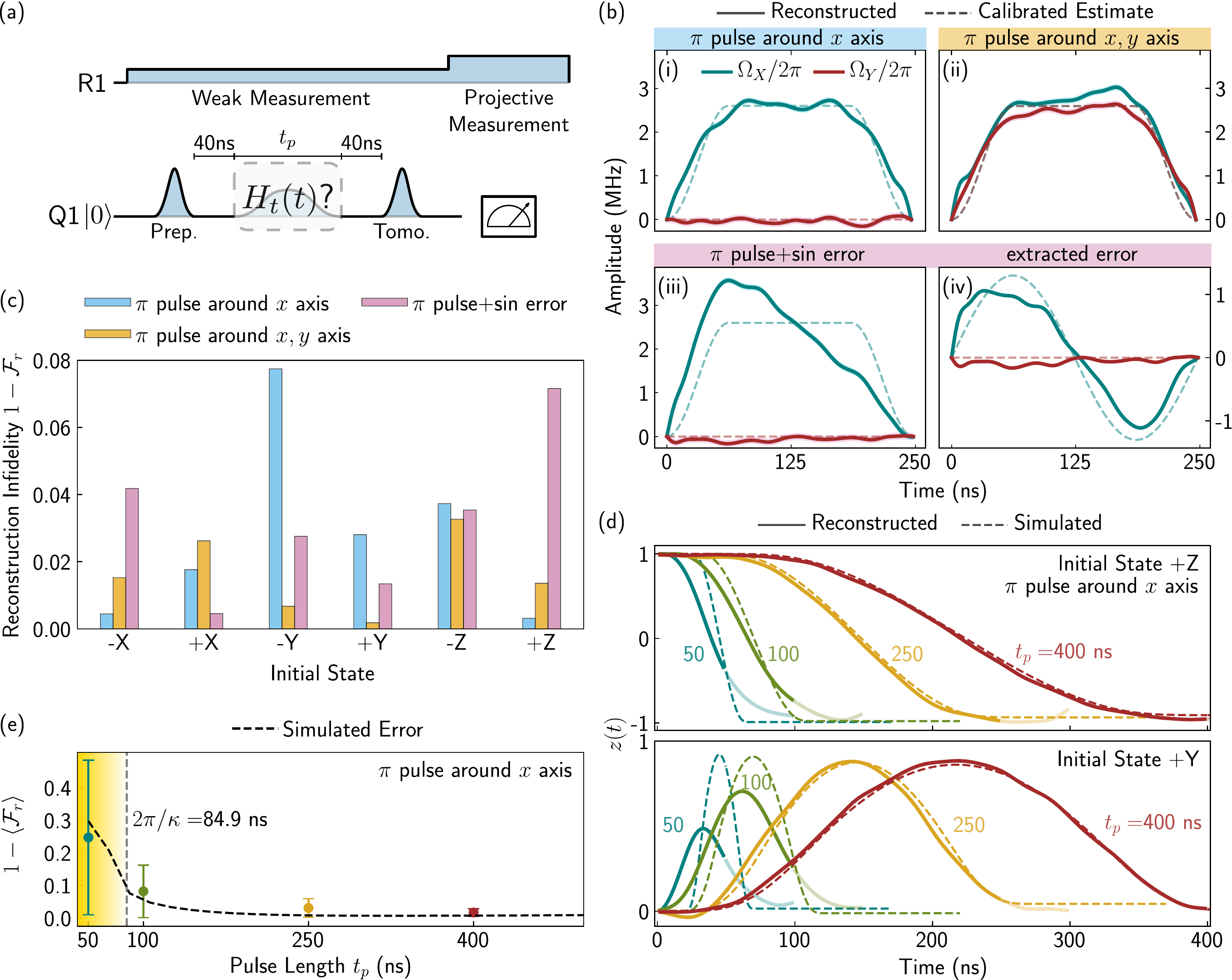}
    \caption{Single qubit Hamiltonian reconstruction. (a) Pulse sequence. We weakly measure Q1 while applying a qubit control pulse (duration $t_p$). After the pulse we use state tomography as independent validation of the reconstructed Hamiltonian $H_t (t)$. A 40\,ns buffer is added between consecutive pulses to ensure the resonator has reached a steady state. 
(b) Reconstructed Hamiltonian ($\Delta t = 2\ns$) for three different pulses: (i) Calibrated $\pi$-pulse with flat top and cosine ramp. (ii) Same as in (i) plus additional out of phase waveform (iii) Same as in (i) plus a sinusoidal amplitude ``error" pulse. (iv) Extracted $\pi$-pulse error estimated from the reconstructed Hamiltonian subtracted from calibrated estimate. (c) Post pulse reconstruction infidelity quantifying overlap between predicted final state and state tomography for the three pulses in (b). (d) Decoupling of resonator from qubit. For fast pulses, the qubit coordinate $z(t)$ obtained from the resonator evolution (solid lines, transparency indicates evolution after the end of the $\pi$ pulse) deviates from the QuTiP simulation (dashed lines). (e) Reconstruction infidelity averaged over all initial states for different gate durations. Error bars represent spread of $1-\mathcal{F}_r$ over different initial states used. The qubit simulation (dashed line) which includes the readout resonator captures the measured trend and shows that experimental value saturates the minimum for all pulse lengths. Low reconstruction fidelity for fast pulses is explained by the deviation observed in (d).}
    \label{fig:fig_2}
\end{figure*}

\section{Single Qubit Hamiltonian Reconstruction}\label{sec:SQHR}

We first apply and validate the algorithm on a single qubit (Q1) by applying three different target Hamiltonians. The pulse sequence for these experiments is shown schematically in Fig.~\ref{fig:fig_2}(a). The initial preparation pulse prepares the qubit in one of six initial states $\{\ket{\pm X}, \ket{\pm Y}, \ket{\pm Z}\}$, where, e.g. $\ket{+X}$ corresponds to the $+1$ eigenstate of the Pauli operator $\sigma_x$. The same target Hamiltonian is then applied to each initial state for $t_p=250\ns$ and takes the general form
\begin{equation}
    H_t(t)/\hbar = \frac{1}{2}\Omega_X(t)\sigma_x + \frac{1}{2}\Omega_Y(t)\sigma_y.
    \label{eq:H_sq}
\end{equation}
The sequence concludes with full state tomography, which  we use not as an input for the algorithm, but as validation of the algorithm's output. Full tomography of each initial state is also performed, as the complete density matrix at $t=0$ \textit{is} an input to the algorithm. 

For each target Hamiltonian, we vary the initial state and average, filter, and deconvolve $\sim 5.4\text{M}$ trajectories with $2\ns$ time resolution to obtain a smooth estimate of $z(t)$ (see App.~\ref{supp:data_processing}), which are then inputted to the Hamiltonian reconstruction algorithm.

We validate the output of the algorithm in three ways. First, we compare the reconstructed pulse shape to the applied waveform and confirm that the drive axis agrees with the phase(s) of the applied pulse(s). Next, we compare the reconstructed amplitude to an estimate obtained by an amplitude calibration, i.e. by fitting the rate of Rabi oscillations vs. drive amplitude. Lastly, the algorithm reconstructs the full density matrix at each timestep, so we compare the final state predicted by the algorithm to the experimental tomography of the final state. We call the agreement between these two states the ``reconstruction fidelity" $\mathcal{F}_r$, defined as follows:

\begin{equation}
    \mathcal{F}_r \equiv \Tr(\rho_\mathrm{rec}^{1/2}\rho_\mathrm{tomo} \rho_\mathrm{rec}^{1/2})^2,
    \label{eq:reconstruction_fidelity}
\end{equation}
where $\rho_\mathrm{tomo}$ is the density matrix as measured by full tomography of the final state and $\rho_\mathrm{rec}$ is the final density matrix as predicted by the algorithm, computed using the tomography of the initial states $\{\rho(0)\}$, the reconstructed Hamiltonian amplitudes $\Omega_X(t)$ and $\Omega_Y(t)$, and the measurement-induced dephasing $\Gamma_\mathrm{d}$.

We present the results for three target Hamiltonians in Fig.~\ref{fig:fig_2}(bc). The first target Hamiltonian (Fig.~\ref{fig:fig_2}(b)(i)) is generated by a $t_p=250\ns$ $\pi$ pulse around the $x$ axis with waveform consisting of a flat-top and cosine ramp edges. The reconstructed pulse shape $\Omega_X(t)$ qualitatively matches with that of the applied waveform and quantitatively agrees with the amplitude of the estimate. The algorithm also concludes that there is zero drive along the $y$ axis, as expected. Finally, tomographic validation of the reconstructed final state in Fig.~\ref{fig:fig_2}(c) shows that the algorithm achieves an average reconstruction fidelity of 97.2\%. 

Fig.~\ref{fig:fig_2}(c) also shows that the reconstructed Hamiltonian produces different fidelities for different initial states. For example, the $\ket{\pm X}$ states are insensitive to the shape of $\Omega_X(t)$ in pulse (i) and achieves the highest fidelity of the three preparation axes. To assess the performance of the algorithm when the target Hamiltonian acts non-trivially on all six initial states, we next apply it to a target Hamiltonian generated by two simultaneous out-of-phase $\pi$ pulses with the same waveform as in (i). The reconstructed Hamiltonian (shown in Fig.~\ref{fig:fig_2}(b)(ii)) again shows good agreement with the estimates of $\Omega_X(t)$ and $\Omega_Y(t)$ and achieves 98.4\% reconstruction fidelity. This indicates that the high fidelity achieved in (i) is not an artifact of the choice of initial states.

Finally, we demonstrate that the reconstructed Hamiltonian gives information beyond what is provided by characterization techniques which use as input only the initial and final states. We apply our technique to a target Hamiltonian generated by a sinusoidal ``error" pulse with period $t_p=250\ns$ superimposed on the $\pi$ pulse from (i) (Fig.~\ref{fig:fig_2}(b)(iii)). We motivate this experiment by viewing the superimposed error as a deviation from a theoretical control Hamiltonian $H_c(t)$ \cite{Khaneja2005, Caneva2011, Muller2011}, which we take to be a flat-top pulse with cosine ramp edges. Importantly, this deviation is undetectable in the final state tomography, as its effect integrates to zero. For visual clarity of the error pulse we use the estimate for $\Omega_X(t)$ from (i) as $H_c(t)$ and plot it as the dashed curves in  Fig.~\ref{fig:fig_2}(b)(iii). The difference between the reconstructed amplitude and this estimate, plotted in Fig.~\ref{fig:fig_2}(b)(iv), shows that the algorithm correctly reveals the sinusoidal deviation from $H_c(t)$. Thus although $H_t(t)$ and $H_c(t)$ produce the same unitary at $t=t_p$, our technique reveals the difference between them.

Our reconstruction technique can, in principle, be applied to target Hamiltonians generated by faster pulses. However, when the rate of qubit evolution exceeds the resonator linewidth $(\Omega \gg \kappa/2)$, the system no longer remains in the adiabatic regime, and the qubit $z$ coordinate can no longer be extracted easily from the reflected resonator field~\cite{Koolstra2022}. In Fig.~\ref{fig:fig_2}(d), we show that the reconstructed qubit evolution for $\pi$ pulses of varying length deviates from $z(t)$ obtained in QuTiP simulations of the qubit and resonator~\cite{Johansson2012, Johansson2013} when the adiabatic condition is violated. This results from the well-studied decoupling of the resonator from a rapidly driven qubit~\cite{Szombati2020,Koolstra2022}. 
Poor estimation of the qubit $z$ coordinate, in turn, leads to an overall lower reconstruction fidelity, shown averaged over the initial states in Fig.~\ref{fig:fig_2}(e). The dashed black line indicates the error in the reconstruction algorithm when the simulated resonator field is used to estimate $z(t)$ and then inputted to the algorithm. It represents the minimum reconstruction error one can expect from our protocol for each pulse duration. Fig.~\ref{fig:fig_2}(e) shows that for fast pulses, the primary source of error in reconstruction is violation of the adiabatic condition. The remaining difference between the experimental and simulated values for $\langle \mathcal{F}_r\rangle$ may be explained by additional transience and dephasing of the qubit state during the tomography and projective measurement pulses. Our technique therefore performs reliably when the amplitudes $\Omega \ll \kappa/2 = 5.89\MHz$.

Having validated the algorithm for a single qubit and established the regime in which our technique performs well, we next validate it on a system of two qubits. We apply the algorithm to two-qubit entangling target Hamiltonians generated by parametrically modulating the coupler, which we discuss next.

\begin{figure*}
    \centering
    \centering
    \includegraphics[width=0.9\linewidth]{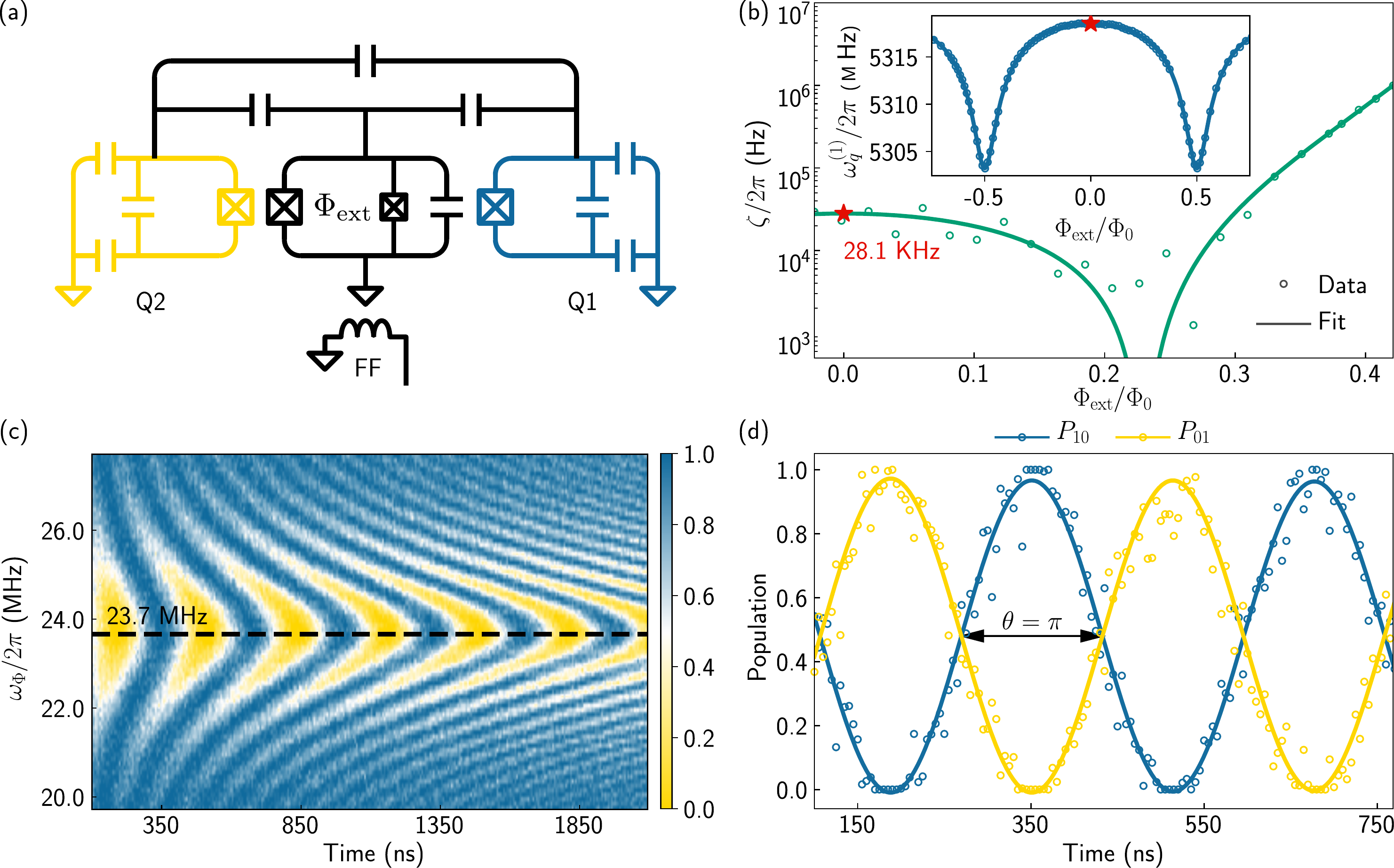}
    \caption{Two qubit gate operation. (a) Two fixed-frequency transmon qubits (Q1, Q2) are coupled to a tunable coupler (SQUID). The flux line FF is used to DC bias and modulate the coupler flux $\Phi_\mathrm{ext}$. (b)
    By DC biasing the coupler, the qubit and bus frequencies and couplings can be tuned. The coupler is biased at $\Phi_\mathrm{ext}=0$ (marked by red star) where the qubit frequency (fit to Eq.~\eqref{eq:tunable_qubit_freq}) is first-order insensitive to the flux (inset). The static $ZZ$ rate $\zeta$ is measured using a JAZZ measurement~\cite{Takita2017}, and the data are fit using models described in App.~\ref{supp:coupler}. (c) Population oscillations of the $\ket{10}$ state due to driving the flux line (FF) with  an AC flux drive at frequency $\omega_{\Phi}$ near $\frac{1}{2}(\omega_q^{(1)} - \omega_q^{(2)})$. (d) Populations of the $\ket{10}$ and $\ket{01}$ states after preparing the qubits in $\ket{10}$ and driving the coupler at $\omega_\Phi / 2\pi = 23.7\MHz$ for various pulse durations. The populations oscillate out of phase, indicating Q1 and Q2 are swapping an excitation. A full swap of an excitation occurs when $\theta=\pi$ (as in Eq.~\eqref{eq:XY_parameterization}), taking 163\,ns.}
    \label{fig:fig_3}
\end{figure*}

\section{XY Interaction}\label{sec:coupler}
The first-order update rule described in Sec.~\ref{sec:reconstruction_algo} does not allow for the reconstruction of $\Omega_{IZ}(t)$, $\Omega_{ZI}(t)$, and $\Omega_{ZZ}(t)$. The presence of such terms in the system can limit the performance of the reconstruction algorithm, so it is important to minimize and to characterize them. The first two unwanted terms correspond to off-resonant driving and can be controlled by tuning the drive frequency. The third term can arise due to $ZZ$ interactions between qubits. Our device uses a tunable coupler architecture (Fig.~\ref{fig:fig_3}(a)) in which the static $ZZ$ coupling between the qubits can be tuned by biasing the coupler SQUID flux $\Phi_\mathrm{ext}$ ~\cite{McKay2016, Lu2017,Yan2018, Abrams2020, Hong2020, Foxen2020, Sung2021, Sete2021}. Further, by modulating $\Phi_\mathrm{ext}$ using an additional RF pulse on the flux line FF, it is possible to activate a variety of entangling two-qubit gates such as iSWAP and CZ. Here, we review the physics of the coupler and demonstrate operation and measurements of this system.


We first examine the static properties. The two qubits Q1 and Q2 are coupled directly through capacitance between transmon pads as well as indirectly through the tunable coupler C12. The full two-level Hamiltonian $H_\mathrm{full}$ of this system is 
\begin{align}
    H_\mathrm{full} &= H_\mathrm{q} + H_\mathrm{c} + H_\mathrm{qq} + H_\mathrm{qc} \\
    H_\mathrm{q}/\hbar &= -\frac{1}{2}\sum_{i=1}^2\omega_q^{(i)}\sigma_z^{(i)} \\
    H_\mathrm{c}/\hbar &= -\frac{1}{2}\omega_c\sigma_z^{(c)} \\
    H_\mathrm{qq}/\hbar &= g_{12}(\sigma_+^{(1)}\sigma_-^{(2)} + \sigma_-^{(1)}\sigma_+^{(2)}) \\
    H_\mathrm{qc}/\hbar &=\sum_{i=1}^2 g_{ic} (\sigma_+^{(i)}\sigma_-^{(c)} + \sigma_-^{(i)}\sigma_+^{(c)}),
\end{align}
where $\sigma_{\pm}^{(i)}$ are the raising/lowering operators for the qubits, $\sigma_z^{(c)}$ is the Pauli $z$ operator for the coupler mode,  $\sigma_{\pm}^{(c)}$ are the raising/lowering operators for the coupler mode, $\omega_c$ is the coupler frequency, $g_{12}$ is the direct coupling between the qubits, and $g_{ic}$ is the coupling between qubit $i$ and the coupler. The coupler frequency $\omega_c$ tunes with the flux as
\begin{equation}
    \omega_c(\Phi_\mathrm{ext}) =
    \omega_{c,0}\sqrt[4]{\cos^2(\pi \tilde{\Phi}_\mathrm{ext}) + d^2\sin^2(\pi \tilde{\Phi}_\mathrm{ext})},
\end{equation}
where $\tilde{\Phi}_\mathrm{ext} = \Phi_\mathrm{ext}/\Phi_0$, $\Phi_0 = h/(2e)$, $d$ is the SQUID asymmetry, and $\omega_{c,0}$ is the coupler frequency at zero flux. Denote the qubit-coupler detuning $\Delta_i(\Phi_\mathrm{ext}) \equiv \omega_q^{(i)} - \omega_c(\Phi_\mathrm{ext})$ and $1/\Delta(\Phi_\mathrm{ext}) \equiv \sum_{i=1}^2 1/\Delta_i(\Phi_\mathrm{ext})$. By applying the Schrieffer-Wolff transformation, the effective, static two-qubit Hamiltonian $H_\mathrm{eff}$ to second order in $|g_{ic}/\Delta_i(\Phi_\mathrm{ext})|$ is approximately \cite{Yan2018}: 

\begin{align}
    H_\mathrm{eff} &= \tilde{H}_\mathrm{q} + \tilde{H}_\mathrm{qq}\\
    \tilde{H}_\mathrm{q}/\hbar &= -\frac{1}{2}\sum_{i=1}^{2} \tilde{\omega}_q^{(i)}(\Phi_\mathrm{ext})\sigma_z^{(i)} \\
    \tilde{H}_{\mathrm{qq}}/\hbar &= \left(J(\Phi_\mathrm{ext}) + g_{12}\right)(\sigma_+^{(1)}\sigma_-^{(2)} + \sigma_-^{(1)}\sigma_+^{(2)})\\
    J(\Phi_\mathrm{ext}) &= \frac{g_\mathrm{1c} g_\mathrm{2c}}{2\Delta(\Phi_\mathrm{ext})},
    \label{eq:tunable_2q_Hamiltonian}
\end{align}
where $g_{12}$ is the direct coupling between the two qubits, $g_{ic}$ is the bare coupling between qubit $i$ and the coupler, $J(\Phi_\mathrm{ext})$ is the indirect coupling between the qubits, and $\tilde{\omega}_q^{(i)}(\Phi_\mathrm{ext})$ is the Lamb-shifted qubit frequency which also tunes with $\Phi_\mathrm{ext}$:
\begin{align}
    \tilde{\omega}_q^{(i)}(\Phi_\mathrm{ext}) &= \omega_q^{(i)} + \frac{g_{ic}^2}{\Delta_i(\Phi_\mathrm{ext})}.
    \label{eq:tunable_qubit_freq}
\end{align}
Because the coupler frequency is much higher than the qubit frequencies in this device ($\Delta(\Phi_\mathrm{ext})<0$), the two sources of coupling can have opposite sign and the static $ZZ$ interaction can therefore be eliminated by biasing $\Phi_\mathrm{ext}$. We use Ramsey measurements and JAZZ measurement \cite{Takita2017} to measure the qubit frequencies and static $ZZ$ rate $\zeta$ along the tuning curve, shown in Fig.~\ref{fig:fig_3}(b). Biasing $\Phi_\mathrm{ext}$ to $0.23\Phi_0$ achieves $\zeta= 0$, but exposes the qubits to greater flux noise than at $\Phi_\mathrm{ext}=0$, as $|\partial \omega_q^{(i)}/\partial \Phi_\mathrm{ext}|_{\Phi_\mathrm{ext}=0.23\Phi_0} > 0$. We therefore operate at $\Phi_\mathrm{ext}=0$ and find $\zeta/2\pi = 28.1\kHz$. However, since $\zeta$ is much smaller than the typical drive amplitudes for our $t_p=250\ns$ pulses (Fig.~\ref{fig:fig_2}(b)), we expect that the corrections will be negligible.

We next turn to the dynamical properties of the coupler (see App.~\ref{supp:coupler} for details). Modulating the coupler flux $\Phi_\mathrm{ext}$ with frequency $\omega_\Phi$ near $(\omega_q^{(1)} - \omega_q^{(2)})/2$ and amplitude $\varepsilon$ generates the following effective two-qubit Hamiltonian \cite{McKay2016}: 

\begin{equation}
    H_\mathrm{ex}/\hbar =  \frac{\varepsilon^2}{8}\left(\frac{\partial^2 J}{\partial \Phi_\mathrm{ext}^2}\bigg|_{\Phi_\mathrm{ext}=0} \right)(\sigma_+^{(1)}\sigma_-^{(2)} + \sigma_-^{(1)}\sigma_+^{(2)}).
    \label{eq:exchange_interaction}
\end{equation}

To measure the exchange interaction generated by Eq.~\eqref{eq:exchange_interaction} we prepare the two qubits in the state $\ket{10}$ ($\ket{Q_1 Q_2}$) and measure the populations for varying pulse lengths and modulation frequencies around $(\omega_q^{(1)} - \omega_q^{(2)})/2$. The resulting population oscillations between the two states are shown in Fig.~\ref{fig:fig_3}(c), and a linecut shows the populations of the $\ket{01}$ and $\ket{10}$ states oscillating out of phase with each other, taking 82\,ns to share the excitation equally between the two qubits, generating a maximally entangled state.

By calibrating the phase and frequency of the coupler drive, we use Eq.~\eqref{eq:exchange_interaction} to generate an approximate entangling Hamiltonian $H$ in the Pauli basis as
\begin{equation}
    H(t)/\hbar = \frac{1}{4}\sum_{\alpha,\beta \in \{X,Y\}}\Omega_{\alpha\beta}(t)\sigma_\alpha^{(1)} \sigma_\beta^{(2)},
    \label{eq:two_qubit_Hamiltonian}
\end{equation}
where $\Omega_{XX}(t) = \Omega_{YY}(t)$ and $\Omega_{XY}(t) = - \Omega_{YX}(t)$. By further calibrating the amplitude and pulse duration, we then generate the so-called $XY$ gates \cite{Abrams2020} parameterized by two angles $\beta$ and $\theta$: 
\begin{equation}
    XY(\beta, \theta) = \begin{pmatrix}
    1 & 0 & 0 & 0 \\
    0 & \cos\left(\frac{\theta}{2}\right) & ie^{-i\beta}\sin\left(\frac{\theta}{2}\right) & 0 \\
    0 & ie^{i\beta}\sin\left(\frac{\theta}{2}\right) & \cos\left(\frac{\theta}{2}\right) & 0 \\
    0 & 0 & 0 & 1
    \end{pmatrix}
    \label{eq:XY_parameterization}
\end{equation}
for $\beta \in [0,\pi)$ and $\theta \in [0, 2\pi)$, which correspond to the phase and amplitude, respectively. 

In the next section, we apply the Hamiltonian reconstruction algorithm to the two qubits evolving under the exchange interaction generated by modulation of the coupler. For several choices of $(\beta, \theta)$, we evaluate the performance of the algorithm, connect the deviations in the recovered dynamics to the calibration, and evaluate the fidelity of the corresponding two-qubit gate.

\begin{figure*}[t]
    \centering
    \includegraphics[width=0.9\linewidth]{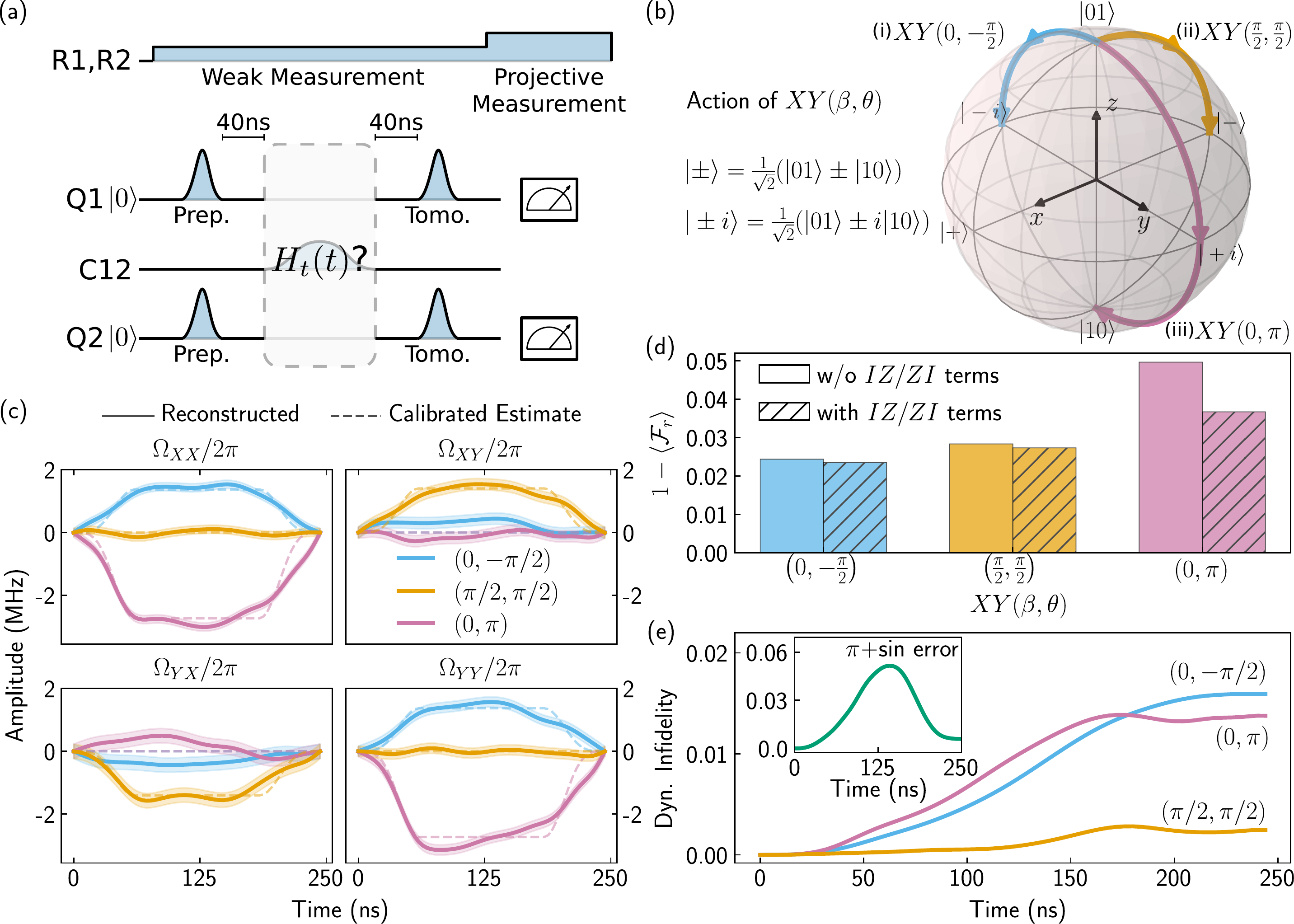}
    \caption{Two qubit Hamiltonian reconstruction. (a) Pulse sequence. The coupler RF pulse generates an entangling two-qubit Hamiltonian. (b) We study three different pulses $XY(\beta,\theta)$ whose ideal action in the $\{|01\rangle,|10\rangle\}$ subspace is schematically depicted on the Bloch sphere. (c) Reconstructed Hamiltonians for $XY(0,-\frac{\pi}{2})$ (blue), $XY(\frac{\pi}{2},\frac{\pi}{2})$ (orange), and $XY(0,\pi)$ (pink) ($\Delta t = 4\ns$). The shaded band around each Pauli term represents the confidence interval, obtained from the average Pauli amplitude when no coupler pulse was applied. Single qubit and additional two qubit Pauli terms have no appreciable signal (see App.~\ref{supp:extended_2_qubit_data}) (d) The reconstruction infidelity (averaged over initial states) computed with respect to tomography after the pulse is small for all three pulses. Preconditioning the Hamiltonian reconstruction algorithm with an $IZ$ and $ZI$ term (magnitude derived from QPT) reduces this error further (right, shaded bars),  particularly for $XY(0,\pi)$. (e) Dynamical gate infidelity during the three pulses showing the accuracy of the reconstructed dynamics with respect to the assumed pulse. Inset shows dynamical gate fidelity for pulse (iii) in Fig.~\ref{fig:fig_2}(b) and reveals large deviation from an ``ideal" control Hamiltonian despite high final gate fidelity.}
    \label{fig:fig_4}
\end{figure*}

\section{Two Qubit Hamiltonian Reconstruction}\label{sec:2QHR}
To validate this technique on a system of two qubits, we apply target Hamiltonians to both Q1 and Q2 and apply separate weak measurement tones on R1 and R2. We explore both non-interacting and interacting Hamiltonians but reserve the results of the former for App.~\ref{supp:extended_2_qubit_data}. The pulse sequence for reconstructing interacting Hamiltonians is shown in Fig.~\ref{fig:fig_4}(a). For each Hamiltonian, we prepare 16 initial states $\{\ket{+X},\ket{+Y},\ket{\pm Z}\}^{\otimes 2}$. Note that each initial state is a product state, which can be prepared quickly and with high fidelity using single qubit pulses. Target Hamiltonians $H_t(t)$ are generated by modulating the coupler for a duration of $t_p = 250\ns$. Full tomography of the final state is performed for comparison with the output of the reconstruction algorithm. For each target Hamiltonian and initial state, we average $\sim 2.7\text{M}$ trajectories to generate the input for the algorithm.

Modulating the coupler as described in Sec.~\ref{sec:coupler} ideally produces nontrivial action only in the single-excitation subspace as in Eq.~\eqref{eq:two_qubit_Hamiltonian} and can be summarized by the Bloch sphere with poles corresponding to $\{\ket{01},\ket{10}\}$ shown in Fig.~\ref{fig:fig_4}(b). The three ideal paths shown correspond to pulses as follows:

\begin{enumerate}[label=(\roman*)]
    \item a $\pi/2$ pulse around the $x$ axis with flat top and cosine ramp edges, generating $XY(0,-\pi/2 )$,
     \item a -$\pi/2$ pulse around the $y$ axis, generating $XY(\pi/2, \pi/2)$,  
    \item a -$\pi$ pulse around the $x$ axis, generating $XY(0, \pi)$.
\end{enumerate}

A general two-qubit Hamiltonian $H_t(t)$ takes the form
\begin{equation}
    H_t(t)/\hbar = \frac{1}{4}\sum_{\alpha,\beta \in \{I,X,Y,Z\}}\Omega_{\alpha\beta}(t)\sigma_\alpha^{(1)} \sigma_\beta^{(2)},
    \label{eq:H_2q}
\end{equation}
and contains 15 nontrivial amplitudes of which we reconstruct 12. The four most relevant reconstructed amplitudes for the three pulses are shown in Fig.~\ref{fig:fig_4}(c), as we find negligible signal in the others. For the full set of 12 terms, see App.~\ref{supp:extended_2_qubit_data}. 

We now discuss the results for each Hamiltonian in turn. Hamiltonian (i) (shown in blue in Fig.~\ref{fig:fig_4}(b)) is generated by the term $\sigma_x^{(1)} \sigma_x^{(2)} + \sigma_y^{(1)}\sigma_y^{(2)}$. We therefore expect that $\Omega_{XX}(t) = \Omega_{YY}(t)$ and that $\Omega_{XY}(t) = \Omega_{YX}(t) = 0$. We estimate an expected amplitude profile by measuring the rate of $\ket{01}\leftrightarrow\ket{10}$ oscillations at the drive power used, similar to Fig.~\ref{fig:fig_3}(d). We note that Eqs.~\eqref{eq:exchange_interaction} and~\eqref{eq:two_qubit_Hamiltonian} suggest that $\Omega_{XX}(t)\propto \varepsilon(t)^2$, so the amplitudes will be approximately the square of the waveform. Fig.~\ref{fig:fig_4}(c) confirms that the reconstructed Hamiltonian agrees well with this expected amplitude profile. We suspect that the small reconstructed amplitudes $\Omega_{XY}(t)$ and $\Omega_{YX}(t)$ are due to a miscalibration in the phase of the coupler pulse. The reconstruction infidelity in Fig.~\ref{fig:fig_4}(d) also confirms that the final state as predicted by the algorithm agrees with the experimental tomography with a fidelity of $97.6\%$. 

We next investigate how this technique performs on $XY$ and $YX$ terms by driving the coupler to generate the Hamiltonian (ii), which is proportional to $\sigma_x^{(1)} \sigma_y^{(2)} - \sigma_y^{(1)}\sigma_x^{(2)}$ (shown in orange in Fig.~\ref{fig:fig_4}(b)). Again, Fig.~\ref{fig:fig_4}(c) shows that the amplitudes of the $XY$ and $YX$ terms are in good qualitative agreement with the expected amplitude shapes and that the $XX$ and $YY$ terms are not driven. This suggests that, compared to the pulse used for Hamiltonian (i),  the phase of the coupler pulse was more accurately calibrated. We find a reconstruction fidelity of 97.2\%, providing a second validation of the algorithm.

Finally, we investigate the performance of this technique when $\Omega_{IZ}$ and $\Omega_{ZI}$ are non-negligible. We drive the coupler to try to generate Hamiltonian (iii) (shown in purple in Fig.~\ref{fig:fig_4}(b)) and find that although the qualitative agreement in Fig.~\ref{fig:fig_4}(c) is good, the quantitative agreement in Fig.~\ref{fig:fig_4}(d) is only 95\%, noticeably worse than for the slower pulses. To determine the source of the error, we precondition the algorithm with nonzero values for $\Omega_{IZ}$ and $\Omega_{ZI}$. By optimizing the reconstruction fidelity over these values, we find that choosing $\Omega_{IZ}/2\pi = 763\kHz$ and $\Omega_{ZI}/2\pi = 392\kHz $ improves the reconstruction fidelity by 1.3\%. On the other hand, performing this optimization for Hamiltonians (i) and (ii) improves the reconstruction fidelity by only 0.1\% and estimates  $\Omega_{IZ}/2\pi = 191\kHz$ $\Omega_{ZI}/2\pi =112\kHz$ for case (i) and $\Omega_{IZ}/2\pi = 120\kHz$ $\Omega_{ZI}/2\pi =60\kHz$ for case (ii). We can understand the source of these single qubit Pauli terms as follows. Modulating the coupler with amplitude $\varepsilon$ induces dynamical single qubit frequency shifts ($\sigma^{(1)}_Z$, $\sigma^{(2)}_Z$) terms proportional to $\varepsilon^2 (\partial^2\tilde{\omega}_q^{(i)}/\partial \Phi_\mathrm{ext}^2)$ (evaluated at $\Phi_\mathrm{ext}=0$)~\cite{McKay2016}. When the coupler is driven resonantly, these shifts vanish in an appropriate frame, yielding Eq.~\eqref{eq:exchange_interaction}, but when it is driven off-resonantly, their difference cannot be eliminated completely. The improved reconstruction fidelity for pulse (iii) suggests that the modulation frequency of the pulse was off-resonant by $((\Omega_{IZ}-\Omega_{ZI})/4)/2\pi=93\kHz$, resulting in an imperfect $XY(0,\pi)$ pulse. While the relative values of $\Omega_{IZ}$ and $\Omega_{ZI}$ stem from imperfect calibration of the coupler modulation frequency, their absolute value is due to imperfect calibration of virtual $Z$ phases \cite{McKay2017} following the pulse and the optimization of these amplitudes with respect to post-pulse tomography. Using the Hamiltonian reconstruction, we can thus isolate sources of coherent error at the Hamiltonian level and connect them to errors in the controls, such as phase and frequency miscalibration. This experiment also demonstrates how significant single qubit frequency shifts in (iii) lead to model error due to using a first-order algorithm, which can be remedied by preconditioning.


The reconstructed time-dependent Hamiltonian $H_r(t)$ also allows us to compare the coherent dynamics of the qubits to those generated by a theoretical control Hamiltonian $H_c(t)$, such as one found by an optimal control algorithm \cite{Khaneja2005, Caneva2011, Muller2011}. To quantify how different the dynamics produced by $H_r(t)$ and $H_c(t)$ are, we introduce the dynamical coherent fidelity $\mathcal{F}(H_r(t), H_c(t))$ as follows:
\begin{align}
    \mathcal{F}(H_r(t), H_c(t)) &\equiv \int d\psi |\bra{\psi}U_c^\dagger(t)U_r(t)\ket{\psi}|^2,\\
    U_{r,c}(t) &= \mathcal{T}\exp\left(-\frac{i}{\hbar}\int_0^t H_{r,c}(t')dt'\right),
    \label{eq:dynamical_gate_fidelity}
\end{align}
where  $U_r(t)$ is the unitary generated by the reconstructed Hamiltonian, $U_c(t)$ is the unitary generated by the theoretical control Hamiltonian, and $\mathcal{T}$ is the time-ordering operator. Operationally, it measures on average how different the actions of the two unitaries are when acting on a pure state, during the course of evolution.  Note that $\mathcal{F}(H_r(t), H_c(t))$ need not decrease monotonically with $t$, as multiple Hamiltonians can ultimately generate  the same final unitary $U(t_p)$.  We plot the dynamical coherent infidelity $1-\mathcal{F}(H_r(t), H_c(t))$ in Fig.~\ref{fig:fig_4}(e) for theoretical control Hamiltonians which give the three ideal trajectories on the single excitation Bloch sphere in Fig.~\ref{fig:fig_4}(b). We observe that pulses (i) and (ii) steadily diverge from their ideal Hamiltonians due to imperfect phase and frequency calibrations and the dynamical single qubit frequency shifts to achieve 98.4\% and 98.7\% final coherent gate fidelity, respectively. However, in these cases, the final coherent gate fidelities do not differ appreciably from their lowest values during the pulses. Instead, we demonstrate in the inset of Fig.~\ref{fig:fig_4}(d) how $\mathcal{F}(H_r(t), H_c(t))$ reveals a significant divergence from $H_c(t)$ which the final gate fidelity cannot detect by evaluating it for the single qubit flat-top $\pi$ pulse with a superimposed sinusoidal pulse (Fig.~\ref{fig:fig_2}(b)(iii)). Although the final coherent gate fidelity is 99.4\%, the dynamical coherent fidelity falls as far as 94.8\% at $t=142\ns$ due to the superimposed sine pulse. This ``worst case" value quantifies how far the true Hamiltonian strayed from the theoretical control Hamiltonian and indicates suboptimal control despite ultimately achieving high fidelity. Therefore, the full dynamical coherent fidelity calculated from the reconstructed Hamiltonian reveals more information about how closely the experimental Hamiltonian matches the theoretical control Hamiltonian than can be obtained from only initial and final measurements and enables us to quantify the discrepancy.
\section{Conclusions}\label{sec:conclusions}
We have introduced a technique for reconstructing the time-dependent Hamiltonian of an interacting system using continuous weak measurements and have experimentally applied and validated it for systems of one and two superconducting transmon qubits. We have shown that we are able to reconstruct nontrivial two-qubit Hamiltonians with a time resolution of 4\,ns and achieve qualitative agreement with expected values and quantitative agreement between the predicted final state and experimental tomography of the final state. 

We now comment on the limitations of this technique and how they may be overcome. First, performance of the algorithm suffers if terms in the Hamiltonian which are not calculated by the algorithm are significant. In our experiments, we are most likely to suffer from this when there are significant dynamics in the $IZ, ZI,\text{ and }ZZ$ terms. We demonstrated in Sec.~\ref{sec:2QHR} that when the coupler was driven with higher power to perform $XY(0,\pi)$, imperfect calibration of the modulation frequency produced significant $IZ$ and $ZI$ dynamics, and by 
preconditioning $\Omega_{IZ}(t)$ and $\Omega_{ZI}(t)$ to be nonzero, we could improve the reconstruction fidelity. This limitation stems from the fact that calculating amplitudes for these three terms in the algorithm requires an update equation which is second order in time, which amplifies shot noise. Second-order update may therefore be achievable with more repetitions or with faster pulses and stronger measurement, provided that the adiabatic condition still holds. Naively increasing the number of repetitions requires the experiment to remain stable for longer periods of time. However, recent advances in active qubit reset ~\cite{Magnard2018} can be used to increase the repetition rate, thus reducing the total time requirement. Finally, although we reconstruct the time-dependent Hamiltonian from ensemble averages in this work, developing a reconstruction technique which uses a small set of noisy trajectories is an interesting future direction, as it may prove more efficient in terms of required repetitions.

We also showed that when the adiabatic condition $2\Omega \ll \kappa$ is violated, the resonator field does not follow the qubit as it evolves under the target Hamiltonian. This results in a poor estimate of the $z$ coordinates, which leads to poor performance of the reconstruction algorithm. To reconstruct faster dynamics, it may be possible to increase the coupling $\kappa$, but the resonator and Purcell filter must be designed carefully to preserve qubit lifetimes. Alternatively, the technique developed in \cite{Koolstra2022} could be used to reconstruct the qubit state even when the adiabatic condition is violated. The reconstructed states could then be used for Hamiltonian reconstruction.


The most immediate application of Hamiltonian reconstruction is to study novel quantum gates, particularly in nascent quantum platforms which do not yet have standard gate sets. For example, classical optimization can be used to estimate an optimal control Hamiltonian for a gate, and our technique can be used to validate it experimentally. By extending this technique to qutrits, one could further investigate leakage errors during gates. Second, our technique may be used to validate analog quantum simulators with locally tunable interactions by isolating subsystems and performing Hamiltonian reconstruction. Finally, our technique is well-suited for quantum metrology in which the qubit behaves as a sensor and drive amplitudes come from the environment rather than user control.

\section*{Acknowledgements}
We are grateful for illuminating conversations with Justin Dressel and comments on the manuscript figures from Long Nguyen which greatly improved their clarity. This work was supported by the U.S. Army Research Laboratory
and the U.S. Army Research Office under contract/grant number W911NF-17-S-0008.

\newpage
\appendix
\titleformat{\section}{\bfseries\normalsize\centering}{\thesection.}{1em}{\MakeUppercase}
\renewcommand{\thefigure}{\arabic{figure}}
\renewcommand\theequation{\Alph{section}\arabic{equation}}
\renewcommand{\thesection}{\Alph{section}}
\renewcommand{\thesubsection}{\arabic{subsection}}
\setcounter{equation}{0}


\section{Experimental setup}
\label{supp:experimental_setup}
\begin{figure*}[t]
    \centering
    \includegraphics[width=\linewidth]{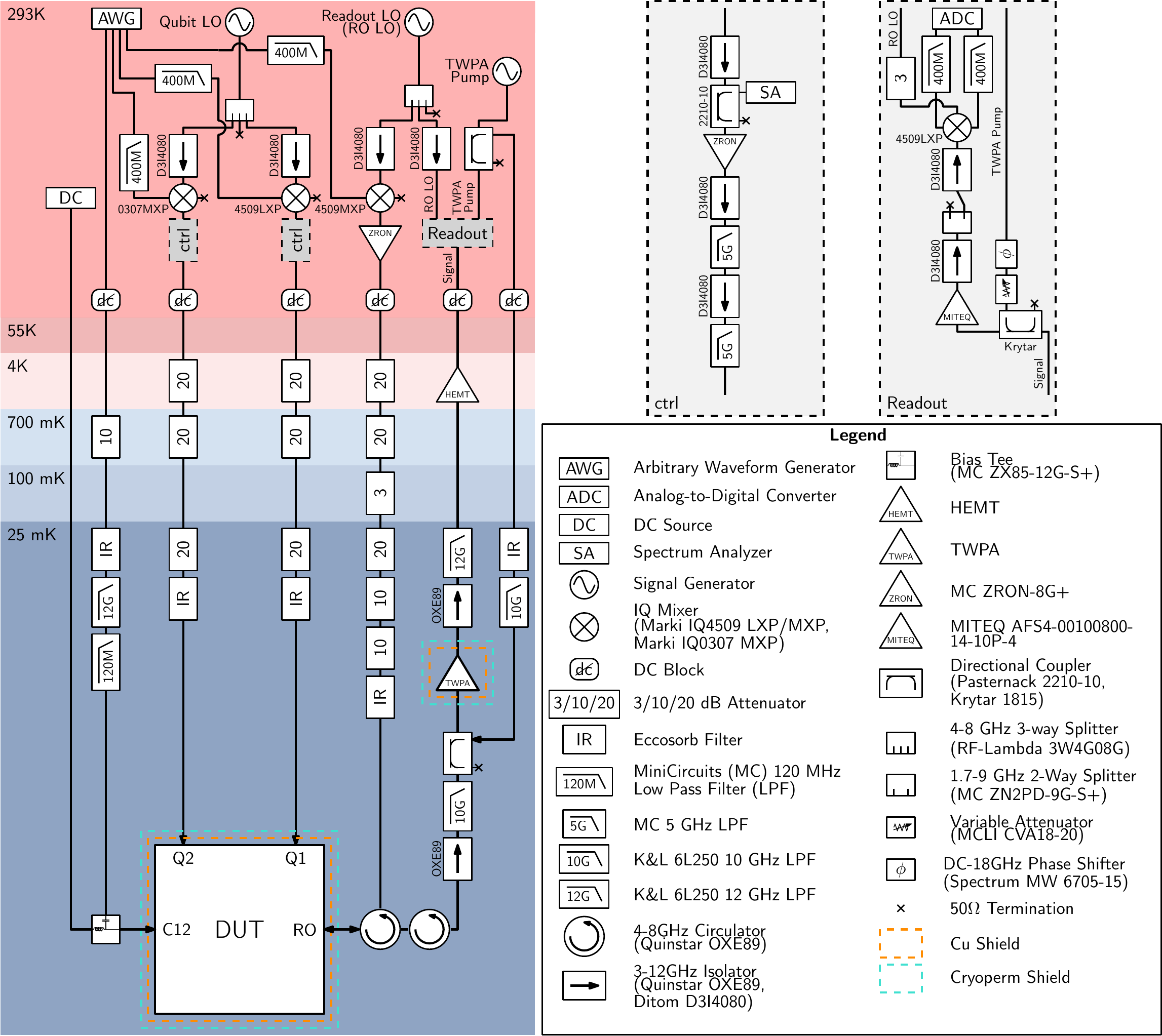}
    \caption{Experimental setup. Details and specification of larger electronics are given in Appendix~\ref{supp:experimental_setup}.}
    \label{fig:experimental_setup}
\end{figure*}

The experiment presented in this work was performed with a chip of three superconducting transmon qubits, one of which was not used. The chip was cooled to 25 mK in an Oxford Instruments dilution refrigerator. The cryogenic wiring and room temperature electronics used in the setup are shown in Fig.~\ref{fig:experimental_setup}. Single qubit pulses are generated by upconversion of IF pulses generated using a Tektronix 5014c Arbitrary Waveform Generator (AWG) at 1 GSa/s through IQ mixing with a 5.5 GHz local oscillator (LO) tone sourced by a Keysight MXG N5183B Signal Generator (SG). Pulses are then amplified and low-pass filtered at room temperature. Readout pulses are also generated by the same AWG and similarly upconverted by mixing the signal from the AWG with a 6.8 GHz LO tone sourced by an Agilent E8257D SG and then amplified at room temperature. Finally, low-frequency tunable coupler modulation pulses are generated directly by the AWG. All pulses pass through a DC block, are then attenuated at each temperature stage, and finally are low-pass filtered at the base stage before being passed to the chip.

DC bias for the flux-tunable coupler is sourced by a Yokogawa GS200. The experiment is performed at a bias of -0.0752 mA at which point the qubit frequencies are first-order insensitive to the the coupler flux and we observe the longest coherence times.

The reflected measurement signals are redirected through cryogenic circulators, amplified by a traveling wave parametric amplifier (TWPA) at 25 mK, a high electron mobility transistor (HEMT) at 4 K, and a low-noise room-temperature  amplifier. The room temperature signal is then downconverted to $I$ and $Q$ components at the IF by mixing with the same 6.8 GHz LO tone used to generate the readout pulses. The IF signals are then filtered and digitized at 1 GSa/s by an Alazar ATS 9373 Analog-to-Digital Converter (ADC). Finally, the digitized signals are demodulated in software.

The TWPA is pumped by a 7.98 GHz tone generated by a Hewlett-Packard 83732B SG. The pump tone is low-pass filtered at the base stage before entering the TWPA. The pump amplitude and frequency are chosen to maximize gain in the reflected readout signals at the two frequencies used to measure the qubits. Most of the power of the pump tone is used as a nulling tone to cancel the pump tone in the reflected readout signal at room temperature. This is achieved by phase-shifting and attenuating the nulling tone and then coupling to the amplified, reflected readout signal before downconversion to IF.

\section{Device parameters}
\label{supp:device_parameters_and_stability}
\begin{table*}[t!]
\begin{tabular}{@{}llcc@{}}
\toprule
\hline \hline
\multicolumn{1}{c}{Parameter}     & Description & Q1 & Q2  \\ \midrule
\hline \hline
$\omega_\mathrm{res}/2\pi$       & Bare resonator frequency & 6631 MHz & 6529 MHz           \\
$\kappa/2\pi$              & Resonator linewidth & 11.78 MHz & 14.47 MHz     \\
$\omega_q/2\pi$       & Transmon $\ket{0}\rightarrow \ket{1}$ frequency  & 5319 MHz      & 5271 MHz    \\
$\alpha_q/2\pi$              & Transmon anharmonicity & -187.17 MHz & -185.03 MHz          \\
$T_1$                     & Transmon relaxation time &  $61\pm 8$ $\mu$s &  $41\pm 10$ $\mu$s\\
$T_{2, \mathrm{Ramsey}}$  & Ramsey dephasing time (empty cavity) & $60\pm 5$ $\mu$s & $50\pm 14$ $\mu$s     \\
$\chi/ 2\pi$               & Full dispersive cavity shift & 0.64 MHz  &  0.75 MHz   \\
$\eta$ & Measurement efficiency & 0.41 & 0.39  \\
\midrule
\end{tabular}
\caption{Calibrated parameters of the two transmon qubits and their respective resonators. All parameters are measured in the absence of any measurement tone. Uncertainties reflect standard deviation of parameters measured during 8 hours of data acquisition for each experiment.}
\label{tab:res_qubit_params}
\end{table*}

The sample consists of three fixed-frequency transmons of which the two used in this work are coupled directly as well as through a flux-tunable coupler. Each transmon is coupled to a $\lambda/4$ coplanar waveguide resonator (CPW) used for dispersive readout. The resonators are further coupled to a Purcell filter which is then coupled to the feedline used to measure the resonator fields. We define and report the most relevant parameters for the two qubits used in this work in Table~\ref{tab:res_qubit_params}. The resonator frequency $\omega_\mathrm{res}$ and linewidth $\kappa$ are measured by spectroscopy. $T_1$ is measured by repeatedly preparing a qubit in the excited state and measuring the qubit population at different times. $T_2^*$ and $\omega_q$ are measured by repeated Ramsey experiments on the $\ket{0}\rightarrow \ket{1}$ transition, and the anharmonicity $\alpha_q$ is measured by Ramsey  measurements on the $\ket{1}\rightarrow\ket{2}$ transition. 

Acquiring a dataset for a given pulse requires repeatedly preparing qubits in the chosen initial states and performing the pulse sequence presented in the main text. For our single (two) qubit datasets, we average over $\sim 5.4\text{M}$ ($\sim 2.7\text{M}$) repetitions, taking 8 hours. This method is limited by residual shot noise in the data, which in theory can be reduced by more measurements. Practically, however, stability may limit the number of repetitions that can be carried out. We find that for our system, $T_1$ and $T_2^*$ fluctuate by 13\% and 28\%, respectively, over the acquisition time but still remain much longer than the duration of each repetition $(\sim 500\ns)$.

\section{Recovering qubit coordinates from continuous weak measurements}
\label{supp:qubit_state_recovery}
In this section we derive analytically how the resonator response encodes information about the $z$ coordinate of a qubit. The starting point for our derivation is the driven, dispersive qubit-resonator Hamiltonian in the rotating frame of the qubit and resonator drives. For simplicity, we assume that both drives are on resonance:

\begin{align}
    H &= H_q + H_r + H_{rq} \\
    H_q/\hbar &= \frac{1}{2}\Omega_X(t) \sigma_x + \frac{1}{2}\Omega_Y(t) \sigma_y \\
    H_r/\hbar &= \Omega_\mathrm{wm}(a + a^\dagger) \\
    H_{rq}/\hbar &= -\frac{\chi}{2} \sigma_z a^\dagger a ,
    \label{eq:disp_H}
\end{align}
where $H_q$ is the single-qubit Hamiltonian, $\Omega_x(t)$ and $\Omega_y(t)$ are the drive envelopes, $H_r$ is the resonator Hamiltonian, $\Omega_\mathrm{wm}$ is the measurement strength, $H_{qr}$ is the dispersive coupling between the qubit and resonator, and $\chi$ is the full dispersive shift. 

The resulting Heisenberg equation of motion is
\begin{equation}
    \dot{a} = -\frac{\kappa}{2}\left(i\frac{\chi}{\kappa}\sigma_z + 1\right)a -i \Omega_\text{wm}.
    \label{eq:res_diff_eq}
\end{equation}
It can be checked that the following implicit equation is a solution to Eq.~\eqref{eq:res_diff_eq}:
\begin{widetext}
\begin{align}
        a(t) &= -i \bigg[\frac{2\Omega_\text{wm}}{\kappa}+ \frac{\chi}{\kappa}\int_0^t \sigma_z(t-\tau)a(t-\tau)d\mu(\tau)\bigg] \\
        \mu(\tau) &= 1-\frac{\kappa}{2}e^{-\kappa \tau/2}.
\end{align}
\end{widetext}
This recurrence relation has solution
\begin{widetext}
\begin{equation}
    a(t) = -i \frac{2\Omega_\text{wm}}{\kappa}\sum_{n=0}^\infty \mathcal{T}\bigg[\left(-i\frac{\chi}{\kappa}\int_0^t \sigma_z(t-\tau)\frac{\kappa}{2}e^{-\kappa \tau/2}d\tau\right)^n\bigg].
\end{equation}
\end{widetext}
We can understand this equation as follows: the resonator state is a series of filtered, time-ordered averages of the qubit $z$ coordinate. At lowest non-trivial order, this gives
\begin{widetext}
\begin{equation}
    a(t) =  -i \frac{2\Omega_\text{wm}}{\kappa} - \frac{2\Omega_\text{wm}}{\kappa}\frac{\chi}{\kappa}\left(\int_0^t \sigma_z(t-\tau)\frac{\kappa}{2}e^{-\kappa \tau/2}d\tau\right) + O\left(\frac{\chi^2}{\kappa^2}\right).
\end{equation}
\end{widetext}
Expanding the integrand, $\sigma_z(t-\tau) = \sigma_z(t)-\dot{\sigma}_z(t)\tau + \dots$:
\begin{widetext}
\begin{align}
\begin{split}
    a(t) &\approx   -\frac{2\Omega_\text{wm}}{\kappa}\left(i + \frac{\chi}{\kappa}\left(\sigma_z(t)-\dot{\sigma}_z(t)(2/\kappa)\right)\right) \\
    &\approx - \frac{2\Omega_\text{wm}}{\kappa}\left(i+ \frac{\chi}{\kappa}\sigma_z(t-\tau)  + \frac{\chi}{\kappa}\sum_{m=2}^\infty\left(\frac{d^{m}\sigma_z}{dt'^m}\right)\bigg|_{t'=t}\left(\frac{2}{\kappa}\right)^{m} \Gamma\left(m+1,\frac{\kappa}{2}t)\right)\right),
    \label{eq:resonator_explicit_approx}
\end{split}
\end{align}
\end{widetext}
where $\tau \equiv 2/\kappa$, $\Gamma(x,y)$ is the lower incomplete gamma function, and boundary terms proportional to $e^{-t/\tau}$ have been dropped. This result reveals the following intuition. Provided that the qubit state does not evolve too rapidly compared to  the linewidth of the resonator, $\tau$, the real part of the resonator state is proportional to the qubit state.  Finally, the qubit state which appears in Eq.~\eqref{eq:resonator_explicit_approx} is \textit{shifted} by $\tau$. In this experiment, the resonator linewidth ($\kappa/2\pi$) qubit 1(2) is 11.78\,MHz(14.47\,MHz) which corresponds to a shift of 27\,ns(22\,ns). When we use the resonator field to estimate the ensemble average $z$ component of the Bloch vector $z(t)$ as the input for the Hamiltonian reconstruction algorithm, we must rescale the voltage records by identifying the values that correspond to $z =\pm 1$ as well as shift the records by $\tau$.

The approximation in Eq.~\eqref{eq:resonator_explicit_approx} applies when $\dot{z}(t)< \tau = 2/\kappa$. The fastest source of qubit evolution arises from $H_q$, so $|\dot{z}(t)| \leq \Omega_\mathrm{max} = \max_{0\leq t\leq t_p}\max(|\Omega_X(t)|,|\Omega_Y(t))$.  This leads us to the ``adiabatic condition", $2\Omega_\mathrm{max} < \kappa$ which ensures that the resonator field follows the qubit evolution. We show in Fig.~\ref{fig:fig_2}(d-e) that when the qubit is driven too rapidly compared to the resonator linewidth, the resonator field ``averages"  the qubit evolution too much to be able to recover the qubit state from it, and the algorithm consequently produces a poor estimate of the Hamiltonian.

\section{Reconstruction Procedure}
\label{supp:reconstruction_app}
In this Appendix, we first give a more explicit description of the first-order update algorithm for a single qubit, which calculates the drive amplitudes $\Omega_X(t)$ and $\Omega_Y(t)$. We then outline all of the steps of the algorithm. Finally, we introduce a second-order algorithm which solves for all three drive amplitudes. 

\subsection{First-Order Update Algorithm}
Eq.~\eqref{eq:H-general} describes the coherent evolution of a qubit expressed in the Pauli operator basis. Eq.~\eqref{eq:inversion_soln} can be used to reconstruct the drive amplitudes when all three components of the Bloch vector are measured at all times, but Eq.~\eqref{eq:inversion_soln_truncated} is more useful in our context, as only the qubit $z$ coordinate is measured simultaneously with the target Hamiltonian. We therefore concentrate on detailing the derivation and algorithm for this case. Below we denote the time at discrete time step $n$ as $t_n$. The Heisenberg equation of motion for $\sigma_z$ is
\begin{equation}
    \label{eq:Heisenberg_z}
    \dot{\sigma}_z(t) = \Omega_X(t) \sigma_y(t) - \Omega_Y(t) \sigma_x(t) - \Gamma_1 \sigma_z(t) + \Gamma_\Delta,
\end{equation}
where $\Gamma_1 = \gamma_\downarrow + \gamma_\uparrow= 1/T_1$ is the sum of the rates of spontaneous relaxation and absorption, and $\Gamma_\Delta =  \gamma_\downarrow - \gamma_\uparrow$. 
Discretizing Eq.~\eqref{eq:Heisenberg_z} at lowest order in $\Delta t$, 
\begin{widetext}
\begin{equation}
    \label{eq:Heisenberg_z_discrete}
    \sigma_z(t_{n+1}) = \sigma_z(t_n)+\Delta t\left( \Omega_X(t_n) \sigma_y(t_n) - \Omega_Y(t_n) \sigma_x(t_n)  - \Gamma_1 \sigma_z(t_n) + \Gamma_\Delta\right).
\end{equation}
\end{widetext}
Eq.~\eqref{eq:Heisenberg_z_discrete} is linear in the drive amplitudes and can be solved simply if we always have the Bloch vector $(x(t_n), y(t_n), z(t_n))$ and $z(t_{n+1})$ for at least two states such that $(y(t_n), -x(t_n))$ are linearly independent. We achieve this by performing the continuous weak measurement with the target Hamiltonian on $S \geq 2$ initial states. Then, labeling $z_s(t_n)$ as the $z$ coordinate of the qubit evolved to time $t_n$ from the $s$-th initial state, we can rewrite Eq.~\eqref{eq:Heisenberg_z_discrete} in terms of the components of the Bloch vector for different initial states as
\begin{widetext}
\begin{equation}
    \label{eq:Heisenberg_z_discrete_matrix}
    \begin{pmatrix}
    z_1(t_{n+1}) \\
    z_2(t_{n+1}) \\
    \vdots \\
    z_S(t_{n+1})
    \end{pmatrix} = 
    \Delta t
    \begin{pmatrix}
    y_1 (t_n) & -x_1(t_n) \\
    y_2 (t_n) & -x_2(t_n) \\
    \vdots & \vdots\\
    y_S (t_n) & -x_S(t_n) 
    \end{pmatrix}
    \begin{pmatrix}
    \Omega_X(t_n) \\
    \Omega_Y(t_n)
    \end{pmatrix}
     -(\Gamma_1 \Delta t -1)\begin{pmatrix}
     z_1(t_n) \\
     z_2(t_n) \\
     \vdots \\
     z_S(t_n)
     \end{pmatrix} + \Delta t\Gamma_\Delta.
\end{equation}
\end{widetext}
The $S\times 2$ matrix in the first term of the right hand side (RHS) is the matrix $M_S'(t_n)$ in Eq.~\eqref{eq:many_state_EOM_truncated}, and has a left inverse when it has rank 2, so Eq.~\eqref{eq:Heisenberg_z_discrete_matrix} can be solved for the vector of amplitudes $(\Omega_X(t_n), \Omega_Y(t_n))^T$ by standard linear algebra and is equivalent to least-squares regression. We now outline the steps of the algorithm. We assume we are given $\{\rho_1(0), \rho_2(0),\dots \rho_S(0)\}$, the full density matrix of each initial state, the decoherence rates $\Gamma_1, \Gamma_\mathrm{d}$, and $z(t_n)$ for $1\leq n \leq N$ timesteps $(t_0 = 0)$. The algorithm then proceeds as follows, starting at $n=0$:
\begin{enumerate}
    \item Solve for the drive amplitudes $\Omega_X(t_n)$ and $\Omega_Y(t_n)$ by solving Eq.~\eqref{eq:Heisenberg_z_discrete_matrix}.
    \item Integrate the Lindblad master equation for the density matrix in Eq.~\eqref{eq:Lindblad_master_eq} to obtain the full density matrix at timestep $t_n$. This provides access to all of the components of the Bloch vector at $t_n$ for computing the drive amplitudes at the next time step.
    \item Increment $n$ and proceed to Step 1.
\end{enumerate}
The output of this algorithm is the drive amplitudes at $N$ timesteps $t_0,t_1,\dots t_{N-1}$ as well as the density matrix at each timestep. Note that although the measurement-induced dephasing rate $\Gamma_\mathrm{d}$ does not appear in Eq.~\eqref{eq:Heisenberg_z_discrete_matrix}, it affects the density matrix during Step 2, and $\Gamma_\mathrm{d}$ is absent from  Eq.~\eqref{eq:Heisenberg_z_discrete_matrix} only because it is a first derivative of $z(t)$. Similarly, if $\Omega_Z(t)$ is nonzero but known \textit{a priori}, it can be supplied to the algorithm, which then uses it in Step 2.

\subsection{Second-Order Update Algorithm}
The approach given thus far suffers from the absence of $\Omega_Z(t)$ in Eq.~\eqref{eq:Heisenberg_z_discrete_matrix}. By proceeding to an update equation which is second order in $\Delta t$, we can solve for $\Omega_Z(t)$. We use the Heisenberg equations for $\sigma_x(t)$ and $\sigma_y(t)$:
\begin{align}
    \dot{\sigma}_x(t) &= \Omega_Y(t) \sigma_z(t) - \Omega_Z(t) \sigma_y(t) - \Gamma_2 \sigma_x(t)\\
    \dot{\sigma}_y(t) &= \Omega_Z(t) \sigma_x(t) - \Omega_X(t) \sigma_z(t) - \Gamma_2 \sigma_y(t)
\end{align}
where $\Gamma_2 = \Gamma_\mathrm{d} + \Gamma_\mathrm{\phi} +  \Gamma_1/2$ and $\Gamma_\phi$ is the intrinsic dephasing rate. Eq.~\eqref{eq:Heisenberg_z_discrete} can be promoted to a second-order equation as follows:
\begin{widetext}
\begin{equation}
    \label{eq:Heisenberg_discrete_second_order}
    \begin{split}
        \sigma_z(t_{n+2}) = \sigma_z(t_{n+1}) + &\Delta t \bigg(\Omega_X(t_{n+1}) \left(\sigma_y(t_{n}) + \Delta t(\Omega_Z(t_{n})\sigma_x(t_{n}) -\Omega_X(t_{n})\sigma_z(t_{n}) - \Gamma_2 \sigma_y(t_{n}))\right)\\ &-\Omega_Y(t_{n+1})\left(\sigma_x(t_{n}) + \Delta t(\Omega_Y(t_{n})\sigma_z(t_{n}) -\Omega_Z(t_{n})\sigma_y(t_{n})- \Gamma_2 \sigma_x(t_{n}))\right) \bigg)    \\
        &-\Delta t \Gamma_1 \sigma_z(t_{n+1}) -\Delta t \Gamma_\Delta
    \end{split}
\end{equation}
\end{widetext}
We now treat $\Omega_X(t_n)$, $\Omega_Y(t_n)$, and $\rho(t_n)$ as ``known" and solve for the three unknown components $\Omega_X(t_{n+1}), \Omega_Y(t_{n+1}), \text{ and } \Omega_Z(t_n)$. To satisfy the base case, we use Eq.~\eqref{eq:Heisenberg_z_discrete_matrix} to solve for $\Omega_X(t_0)$, $\Omega_Y(t_0)$ and as before assume that we have tomography of the initial state $\rho(t_0)$. However, Eq.~\eqref{eq:Heisenberg_discrete_second_order} is now nonlinear in the unknowns, containing terms like $\Omega_X(t_{n+1})\Omega_Z(t_n)$ and $\Omega_Y(t_{n+1})\Omega_Z(t_n)$, and requires a nonlinear solver, such as Gauss-Newton method, gradient descent, or other methods of nonlinear regression. This method requires $S\geq 3$ equations to solve for the amplitudes at each time step.

One limitation of this method is that when coupling between $\sigma_z(t)$ and the other qubit coordinates vanishes, i.e. $\Omega_{X}(t)=\Omega_{Y}(t)=0$ for some $t$, $z(t)$ is totally unaffected by $\Omega_Z(t)$ and cannot be recovered from Eq.~\eqref{eq:Heisenberg_discrete_second_order}. However, if they are known to vanish for all $t$, then we can apply a known \emph{spectroscopy drive}, i.e. a drive around the $x$ or $y$ axes with known nonzero amplitude, to re-introduce coupling to $\Omega_Z(t)$. A second drawback is that this method is second order in $\Delta t$, as we are effectively computing the second derivative of $z(t)$. However, each derivative we take amplifies high frequency noise. In our experiment, we find that applying the second order update rule results in a poor estimate of the target Hamiltonian. However, this is not a fundamental limitation and can in theory be overcome by more averaging and longer acquisition, provided that the system remains stable.

We now comment on how this technique extends to a system of two qubits. At first order,  Eq.~\eqref{eq:Heisenberg_z_discrete} for each qubit now contains eight terms, with four in common ($\sigma_x^{(1)}\sigma_x^{(2)}$, $\sigma_x^{(1)}\sigma_y^{(2)}$, $\sigma_y^{(1)}\sigma_x^{(2)}$, and $\sigma_y^{(1)}\sigma_y^{(2)}$). The first order update requires a minimum of six states, as each yields two linearly independent equations. Terms which are tensor products of only identity and Pauli $z$ operators, i.e. $\sigma_z^{(1)}$, $\sigma_z^{(2)}$, and $\sigma_z^{(1)}\sigma_z^{(2)}$ do not appear at first order but do appear at second order, similar to Eq.~\eqref{eq:Heisenberg_discrete_second_order}. More generally, for a system of $Q$ qubits (total Hilbert space dimension $d = 2^Q$) each continuously measured along the $z$ axis, a minimum of $d(d-1)/Q$ initial states to apply the first order update to recover $d^2-d$ out of $d^2-1$ amplitudes. By going to second order we can recover all of the drives using $d^2-1$ initial states. As before, we can use a spectroscopy drive to introduce deliberate coupling if necessary. Moreover, this technique requires only single qubit drives to ensure coupling to all drive terms.

Finally, we note that the formalism introduced can be extended to the setting where multiple qubit coordinates are continuously measured. For example, if two coordinates of a qubit, such as $y(t)$ and $z(t)$, are known for all $t$, then a first-order update method suffices and all three amplitudes can be recovered, provided that $S \geq 2$ initial states are used. It can also be extended to higher dimensional systems by expanding the Hamiltonian in a complete basis, such as the generalized Gell-Mann matrices. 

\subsection{Efficient Reconstruction for Fast Evolution}
Eqs.~\eqref{eq:Heisenberg_z_discrete_matrix} and ~\eqref{eq:Heisenberg_discrete_second_order} provide an efficient means for estimating the amplitudes of the target Hamiltonian at each time step. However, their accuracy suffers at large drive amplitudes (relative to $\Delta t$). On the other hand, minimizing an objective function which calls a matrix exponentiation function is computationally intensive. We combine the two approaches by splitting the Hamiltonian into a ``fast" initial guess $H_g(t)$ and a ``slow" perturbation $\delta H(t)$, where the speeds refer to $\Omega/(1/\Delta t)$, and optimize the perturbation $\delta H(t)$. We then view the RHS of the update equations as estimating $\dot{z}(t)$ by $\dot{z}_\mathrm{est}(t; H_g(t), \delta H(t))$, a function of $H_g(t)$ and $\delta H(t)$, and minimize the squared sums (over initial states) of $\dot{z}_\mathrm{est}(t; H_g(t), \delta H(t)) - \dot{z}(t)$ over $\delta H(t)$. The update equations in the Heisenberg picture are now written as 
\begin{equation} \label{eq:fast_slow_update}
\begin{split}
\sigma_z(t_{n+1})&=U_g^\dagger(t_n)\sigma_z(t_n) U_g(t_n) \\
&+\Delta t\bigg(i[\delta H(t_n), \sigma_z(t_n)]\\
&+\sum_i\gamma_i\mathcal{D}^\dagger[\sigma_i]\sigma_z(t_n)\bigg)
\end{split}
\end{equation}
\begin{align}
    \mathcal{D}^\dagger[\sigma_i]\sigma_z(t_n) &= \sigma_i^\dagger \sigma_z(t_n) \sigma_i - \frac{1}{2} \{\sigma_i^\dagger \sigma_i, \sigma_z(t_n)\},    \\
    U_g(t_n)&=e^{-i\int_{t_n}^{t_{n+1}} H_g(t) dt},
\end{align}
and we optimize over $\delta H(t)$ expressed in the Pauli operator basis. Since the matrix exponential does not depend on $\delta H(t)$, it is not evaluated repeatedly by the optimizer. We find that this approach yields more accurate reconstruction, particularly when the dynamics are fast compared to the timestep $\Delta t$. In our experiment, we are limited by the resonator linewidth $\kappa$ due to the requirement that the resonator field adiabatically follow the qubit state, and it is not strictly necessary to use this approach. However, if the resonator linewidth far exceeds the time resolution of the data after filtering, then this technique is essential to model and reconstruct the dynamics accurately.

\section{Details of Data Processing and Filtering}
\label{supp:data_processing}
\begin{figure}
    \centering
    \includegraphics[width=0.9\columnwidth]{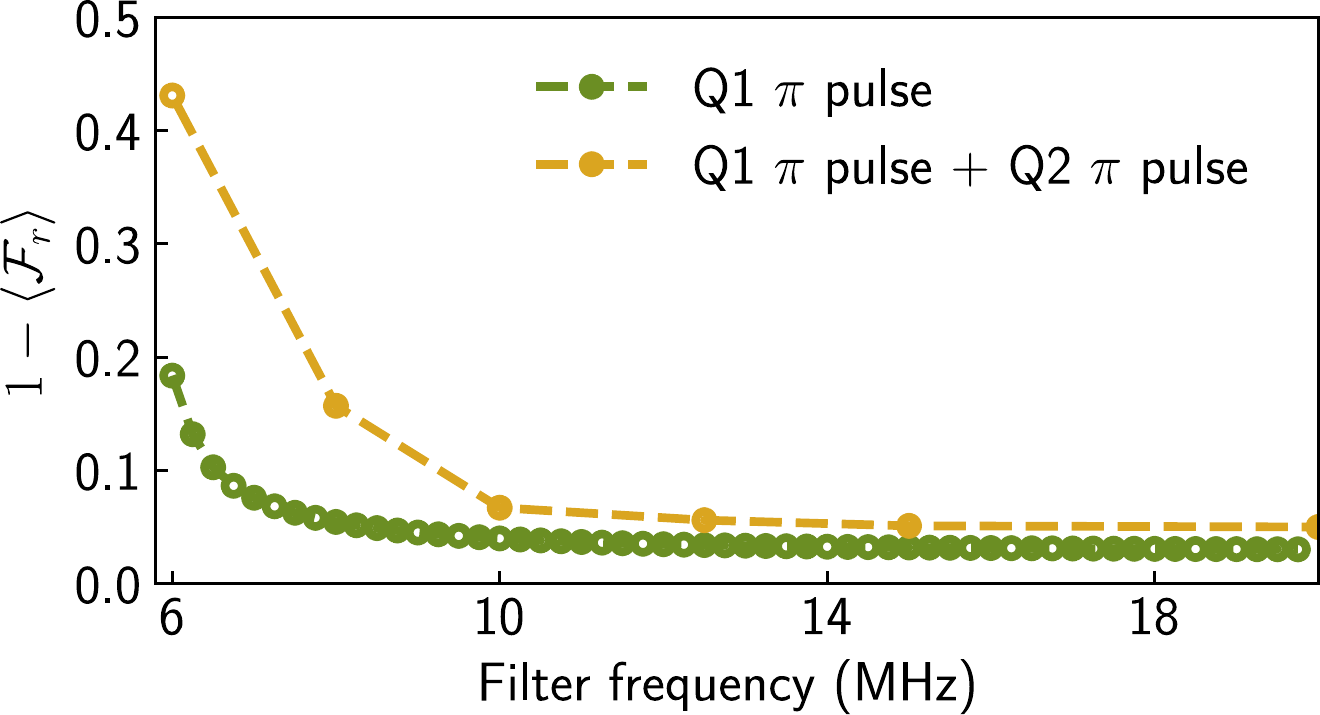}
    \caption{Average reconstruction fidelity vs. critical frequency of low-pass Butterworth filter applied to the data. Dashed line is provided as a guide for the eye. The reconstruction fidelity, averaged over initial states, $\langle\mathcal{F}_r\rangle$ falls when the time series voltage data is filtered too strongly and distorts the dynamics.}
    \label{fig:fig_6}
\end{figure}
The Hamiltonian reconstruction algorithm requires as input the full density matrix of each initial state used, $\{\rho(0)\}$, the qubits' $z(t)$ coordinates during $H_t(t)$, and finally an estimate of the rates of measurement-induced dephasing $\{\Gamma_m\}$. In this section, we detail the steps taken to calculate these $z(t)$ from the resonator field.

The output of the experiment is an ensemble of voltage traces measured with $1\ns$ resolution, which are averaged to estimate the ensemble averaged resonator field. We showed in App.~\ref{supp:qubit_state_recovery} that this amplitude is proportional to the qubit $z$ coordinate retarded by a delay $\tau = 2/\kappa$. However, the data must first be filtered, deconvolved, and rescaled. Filtering is necessary as weak measurements leave residual high frequency noise due to finite averaging, and it is particularly necessary for weak measurement of both qubits, as we use a single feedline to simultaneously probe both qubits' resonators. To achieve good readout isolation, we have designed the resonator frequencies in this device to be 102\,MHz apart and low-pass filter each qubit's data to suppress dynamics of the other qubit as well as other high frequency shot noise. We downsample to $2\ns$ time resolution and use a third-order low-pass Butterworth filter with a 50\,MHz critical frequency to smooth the data. Then, to determine the overall scaling to convert voltages to $z(t)$, we also prepare the qubits along the $z$ axis and measure the field without applying the target Hamiltonian. We use the difference of the mean values of these two measurements to rescale the voltages to lie in $[-1,1]$. Values that extend beyond these bounds are clipped. This produces the $z(t)$ input for the Hamiltonian reconstruction algorithm. The reconstructed amplitudes $\{\Omega_{IX}(t),\Omega_{IY}(t)\dots \Omega_{ZY}(t)\}$ are also filtered using a 5th order low-pass Butterworth filter with critical frequency 50\,MHz, as the update methods described in Sec.~\ref{supp:reconstruction_app} amplify high frequency noise. Naturally, filtering the data more aggressively smooths the inputs but can distort the Hamiltonian. We demonstrate this effect in Fig.~\ref{fig:fig_6} for single qubit Hamiltonian reconstruction applied to a $\pi$ pulse and two qubit Hamiltonian reconstruction applied to simultaneous $\pi$ pulses. When the critical frequency is too low, the voltage data and consequently the input to the algorithm $z(t)$ are distorted. The algorithm then poorly reconstructs the Hamiltonian and fails to predict the final state of the system accurately. It is therefore necessary to keep the critical frequency high and instead decrease shot noise through more averaging, stronger measurement, or more efficient measurement (e.g. through quantum-limited amplifiers).

\section{$XY$ interactions through parametric modulation}\label{supp:coupler}
In this Appendix, we review how parametric modulation of the flux-tunable coupler generates an entangling exchange interaction between the two qubits and derive its particular form. Our starting point for the derivation is the effective two-qubit Hamiltonian in the limit where the coupler frequency far exceeds the qubit frequencies, i.e.
\begin{align}
    \Delta_i(\Phi_\mathrm{ext}) &= \omega_q^{(i)} - \omega_c(\Phi_\mathrm{ext}) \\
    \Sigma_i(\Phi_\mathrm{ext}) &= \omega_q^{(i)} + \omega_c(\Phi_\mathrm{ext}) \\    
    \bigg|\frac{g_{ic}}{\Sigma_i(\Phi_\mathrm{ext})}\bigg|&<\bigg|\frac{g_{ic}}{\Delta_i(\Phi_\mathrm{ext})}\bigg| \ll 1
\end{align}
where $\omega_c(\Phi_\mathrm{ext})$ is the coupler frequency, which tunes with the flux $\Phi_\mathrm{ext}$, and $g_{ic}$ is the direct capacitive couplings between the $i$-th transmon and the coupler. The coupler is also assumed to remain in its ground state. The effective Hamiltonian $H_\mathrm{eff}$ is then
\begin{align}
    \label{eq:effective_2q_Hamiltonian}
    H_\mathrm{eff} &= \tilde{H}_\mathrm{q} + \tilde{H}_\mathrm{qq}\\
    \tilde{H}_\mathrm{q} /\hbar&= \frac{1}{2}\sum_{i=1}^{2} \tilde{\omega}_q^{(i)}(\Phi_\mathrm{ext})\sigma_z^{(i)} \\
    \tilde{H}_{\mathrm{qq}}/\hbar &= \left(J(\Phi_\mathrm{ext}) + g_{12}\right)\sigma_x^{(1)}\sigma_x^{(2)}\\
    \tilde{\omega}_q^{(i)}(\Phi_\mathrm{ext}) &=\omega_q^{(i)} +  g_\mathrm{ic}^2\left(\frac{1}{\Delta_i(\Phi_\mathrm{ext})} - \frac{1}{\Sigma_i(\Phi_\mathrm{ext})}\right) \\    
    J(\Phi_\mathrm{ext}) &= \frac{g_\mathrm{1c} g_\mathrm{2c}}{2}\left(\frac{1}{\Delta(\Phi_\mathrm{ext})} - \frac{1}{\Sigma(\Phi_\mathrm{ext})}\right)
\end{align}
where $\tilde{\omega}_q^{(i)}(\Phi_\mathrm{ext})$ is the qubit frequency, which now tunes with the coupler flux $\Phi_\mathrm{ext}$, $g_{12}$ is the direct capacitive coupling between the transmons, $J(\Phi_\mathrm{ext})$ is the indirect coupling mediated by the coupler, and $1/ \Delta(\Phi_\mathrm{ext}) =1/ \Delta_1(\Phi_\mathrm{ext})+ 1/\Delta_2(\Phi_\mathrm{ext})$ ($\Sigma(\Phi_\mathrm{ext})$ is defined similarly). For a derivation of $H_\mathrm{eff}$ from circuit quantization, see \cite{Yan2018, Sete2021}. 

Suppose we modulate the coupler flux $\Phi_\mathrm{ext}$ as
\begin{equation}
    \label{eq:flux_modulation}
    \Phiext(t) = \PhiDC + \varepsilon\cos(\omega_\Phi t + \phi)
\end{equation}
where $\PhiDC$ is the static bias, $\varepsilon$ is the modulation amplitude, and $\omega_\Phi$ is the modulation frequency. Expanding the fluctuations in the qubit frequencies around $\Phiext=\PhiDC$ in powers of $\varepsilon/\Phi_0$:
\begin{widetext}
\begin{align}
    \tilde{\omega}_q^{(i)}(t) &= \tilde{\omega}_q^{(i)}(\PhiDC) + \sum_{n=1}^\infty \frac{1}{n!} \frac{\partial^n  \tilde{\omega}_q^{(i)}}{\partial \Phiext^n}(\varepsilon \cos(\omega_\Phi t + \phi))^n \\
    &= \tilde{\omega}_q^{(i)}(\PhiDC) + \sum_{n=1}^\infty \frac{1}{n!} \frac{\partial^n  \tilde{\omega}_q^{(i)}}{\partial \Phiext^n}\left(\frac{\varepsilon}{2}\right)^n \sum_{k=0}^n {n\choose k}\cos((n-2k)(\omega_\Phi t + \phi)) \\
    &\approx \tilde{\omega}_q^{(i)}(\PhiDC) + \delta_\mathrm{AC} + \frac{\partial  \tilde{\omega}_q^{(i)}}{\partial \Phiext}\varepsilon \cos(\omega_\Phi t + \phi) + \frac{\partial^2  \tilde{\omega}_q^{(i)}}{\partial \Phiext^2} \frac{\varepsilon^2}{4} \cos(2(\omega_\Phi t + \phi))\\
     \delta\omega^{(i)}_\mathrm{AC} &= \frac{\partial^2  \tilde{\omega}_q^{(i)}}{\partial \Phiext^2}\frac{\varepsilon^2}{4}  \\
     \tilde{H}_\mathrm{q}(t)/\hbar &= \frac{1}{2}\sum_{i=1}^{2} \tilde{\omega}_q^{(i)}(t)\sigma_z^{(i)} 
\end{align}
\end{widetext}
where $\delta\omega^{(i)}_\mathrm{AC}$ is an AC shift of the qubit frequency and all derivatives are evaluated at $\Phiext=\PhiDC$. By similarly expanding $J(\Phiext)$, we obtain a modulated coupling:
\begin{widetext}
\begin{align}
    \tilde{H}_{\mathrm{qq}}(t)/\hbar &\approx \left(J_0+ \frac{\partial  J}{\partial \Phiext}\varepsilon \cos(\omega_\Phi t + \phi) + \frac{\partial^2  J}{\partial \Phiext^2} \frac{\varepsilon^2}{4} \cos(2(\omega_\Phi t + \phi))\right)\sigma_x^{(1)}\sigma_x^{(2)}\\
    J_0 &= J(\PhiDC) + g_{12} + \delta_\mathrm{AC} \\
     \delta J_\mathrm{AC} &= \frac{\partial^2  \tilde{\omega}_q^{(i)}}{\partial \Phiext^2}\frac{\varepsilon^2}{4} 
\end{align}
\end{widetext}
To obtain an exchange interaction, we enter a frame rotating at $ \tilde{\omega}_q^{(i)}(\PhiDC) + \delta\omega^{(i)}_\mathrm{AC}$ and define 
\begin{align}
    \Delta_{12} &= \tilde{\omega}_q^{(1)}(\PhiDC) - \tilde{\omega}_q^{(2)}(\PhiDC) + \delta\omega^{(1)}_\mathrm{AC}-\delta\omega^{(2)}_\mathrm{AC}\\
    \Sigma_{12} &= \tilde{\omega}_q^{(1)}(\PhiDC) + \tilde{\omega}_q^{(2)}(\PhiDC) + \delta\omega^{(1)}_\mathrm{AC}+\delta\omega^{(2)}_\mathrm{AC}    
\end{align}
Then, dropping fast rotating terms, $\tilde{H}_{\mathrm{qq}}(t)$ transforms to
\begin{widetext}
\begin{equation}
    \label{eq:time_dep_exchange}
  \tilde{H}_{\mathrm{qq}}(t)/\hbar \approx \left(\frac{\partial  J}{\partial \Phiext}\varepsilon \cos(\omega_\Phi t + \phi) + \frac{\partial^2  J}{\partial \Phiext^2} \frac{\varepsilon^2}{4} \cos(2(\omega_\Phi t + \phi))\right)(\sigma_+^{(1)}\sigma_+^{(2)}e^{-i\Sigma_{12} t} + \sigma_+^{(1)} \sigma_-^{(2)}e^{-i\Delta_{12} t}+ \text{h.c.})
\end{equation}
\end{widetext}
Eq.~\eqref{eq:time_dep_exchange} suggests that to obtain a static exchange term in this frame, the coupler should be modulated  at $\omega_\Phi = \Delta_{12}$ or $\omega_\Phi = \Delta_{12}/2$. In our experiment, we operate at $\PhiDC=0$ at which point the first term is zero. This is advantageous as coherence away from a sweet spot is decreased by susceptibility to flux noise. Instead, we modulate the coupler at $\omega_\Phi = \Delta_{12}/2$ and set $\phi = \beta/2$ to obtain

\begin{equation}
    \label{eq:modulated_exchange}
  \tilde{H}_{\mathrm{qq}}/\hbar \approx \frac{\partial^2  J}{\partial \Phiext^2} \frac{\varepsilon^2}{8}(\sigma_+^{(1)} \sigma_-^{(2)}e^{i\beta}+ \sigma_-^{(1)} \sigma_+^{(2)}e^{-i\beta})
\end{equation}
By tuning $\beta=0$ and $\beta=\pi/2$, we obtain the two-qubit entangling iSWAP Hamiltonians used in this work.
\begin{align}
    \label{eq:iSWAP_Hamiltonians}
  \tilde{H}_{\mathrm{qq}}/\hbar&= \frac{\partial^2  J}{\partial \Phiext^2} \frac{\varepsilon^2}{16}(\sigma_x^{(1)} \sigma_x^{(2)}+ \sigma_y^{(1)} \sigma_y^{(2)})\\
    \tilde{H}_{\mathrm{qq}}/\hbar&= \frac{\partial^2  J}{\partial \Phiext^2} \frac{\varepsilon^2}{16}(\sigma_x^{(1)} \sigma_y^{(2)}- \sigma_y^{(1)} \sigma_x^{(2)})
\end{align}
From these equations we obtain the constraint that if the pulse is perfect and the reconstruction is accurate, we expect $\Omega_{XX}(t)=\Omega_{YY}(t)$ and $\Omega_{XY}(t) = -\Omega_{YX}(t)$. To go beyond this constraint, one needs to drive the double-excitation terms in Eq.~\eqref{eq:time_dep_exchange}, e.g. by modulating the flux at $\omega_\Phi = \Sigma/2$, activating the so-called bSWAP interaction. We also note that the frame in which the exchange interaction is static differs from the single qubit frames due to $\delta\omega_\text{AC}^{(i)}$, for which we compensate by applying virtual $Z$ phase gates after the coupler pulse. 

\section{Extended Data for Two Qubit Hamiltonian Reconstruction}\label{supp:extended_2_qubit_data}
In this Appendix, we provide data for the extended Hamiltonian reconstruction of two qubits when the Hamiltonian is generated by single qubit pulses and when it is generated by a pulse on the coupler flux line.

\begin{figure*}
    \centering
    \includegraphics[width=0.9\linewidth]{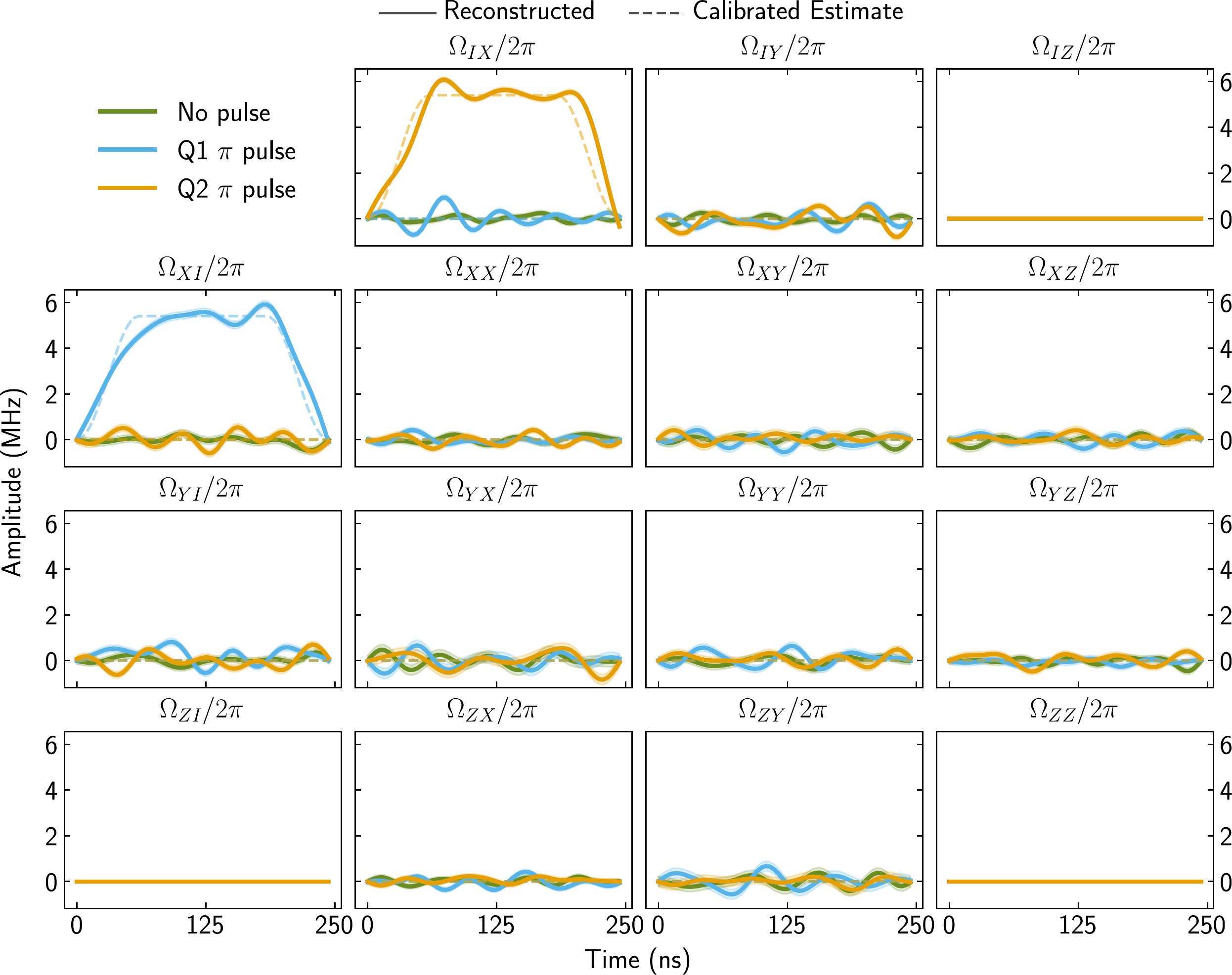}
    \caption{Full set of reconstructed terms for two single-qubit $\pi$ pulses around the $x$ axis.  Solid(dashed) lines represent reconstructed(estimated) amplitudes, with time resolution $\Delta t = 4\ns$. The shaded band indicates an uncertainty, calculated from the ``No pulse" data. $\Omega_{IZ}(t), \Omega_{ZI}(t), \text{ and } \Omega_{ZZ}(t)$ are not solved for, due to the constraint of the first-order update. No significant crosstalk is observed from this reconstruction.}
    \label{fig:fig_7}
\end{figure*}

In Fig.~\ref{fig:fig_7}, we use the two-qubit Hamiltonian reconstruction to measure whether crosstalk occurs in this device when applying a $\pi$ pulse on one of the two qubits at a time. We reconstruct 12 of the 15 possible terms in a two-qubit Hamiltonian (the remaining terms being $\Omega_{IZ}(t)$, $\Omega_{ZI}(t)$, and $\Omega_{ZZ}(t)$) and find no appreciable crosstalk. We also show the result for when no pulse is applied on either qubit. We find some residual fluctuations which may be an artifact of filtering, so we use the RMS value of these curves to estimate an uncertainty for reconstructions of nontrivial pulses, shown as the shaded band extending above and below each curve. 

\begin{figure*}[t!]
    \centering
    \includegraphics[width=0.9\linewidth]{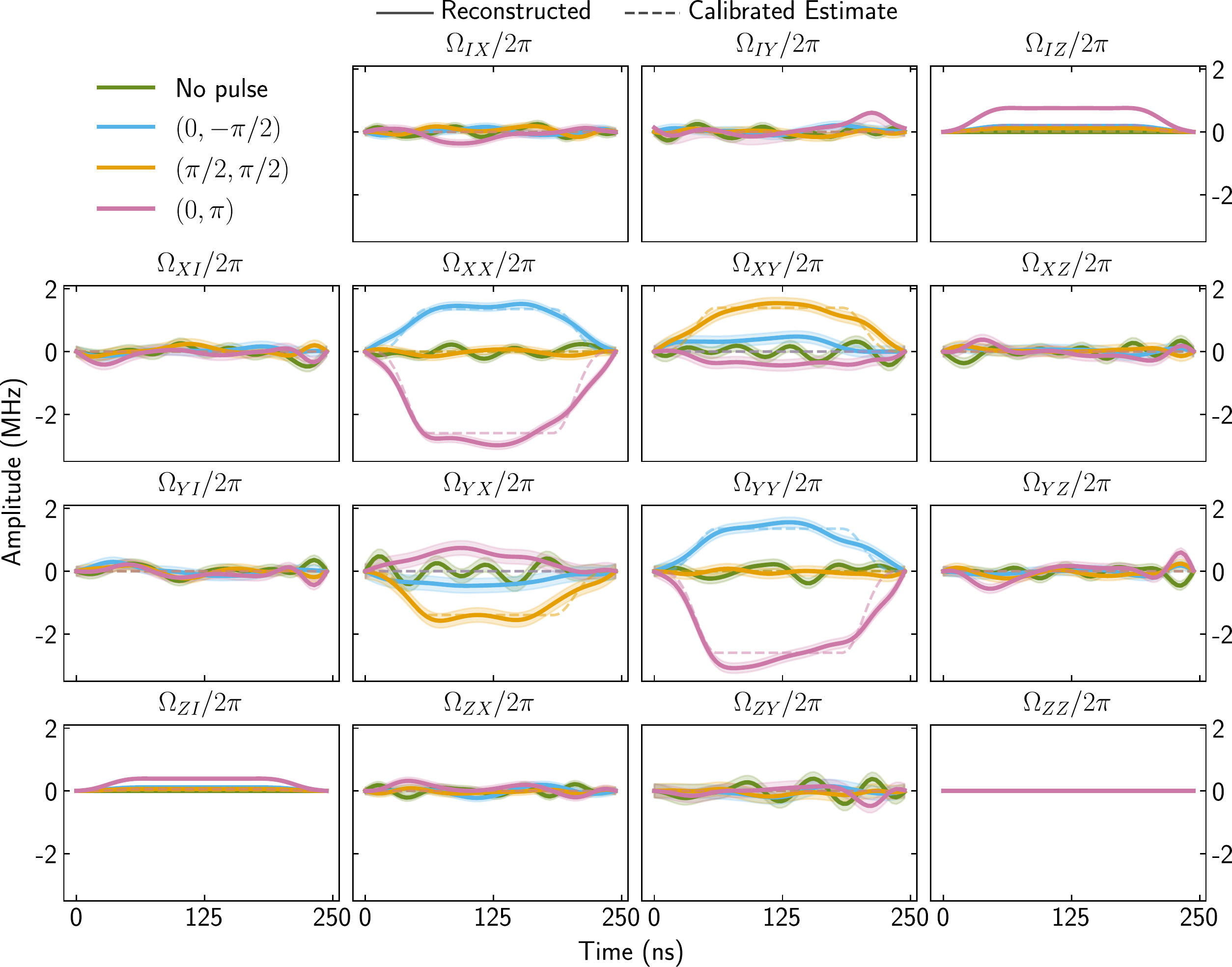}
    \caption{Full set of reconstructed terms for the entangling two-qubit Hamiltonians $XY(\beta,\theta)$ explored in this work. Solid(dashed) lines represent reconstructed(estimated) amplitudes, with time resolution $\Delta t = 4\ns$. $\Omega_{IZ}(t), \Omega_{ZI}(t),\text{ and }\Omega_{ZZ}(t)$ are not solved for, due to the constraint of the first-order update, but $\Omega_{IZ}(t)$ and $\Omega_{ZI}(t)$ are preconditioned as described in the main text. Negligible amplitude appears in the Pauli terms outside the center four, so those are presented in the main text.}
    \label{fig:fig_8}
\end{figure*}

In Fig.~\ref{fig:fig_8}, we show the complete reconstruction data for the entangling two-qubit Hamiltonians generated by modulating the coupler flux, after preconditioning $\Omega_{IZ}(t)$ and $\Omega_{ZI}(t)$ as described in the main text. We find no significant amplitudes outside the main expected four terms.

\bibliography{main}

\begin{thebibliography}{76}%
\makeatletter
\providecommand \@ifxundefined [1]{%
 \@ifx{#1\undefined}
}%
\providecommand \@ifnum [1]{%
 \ifnum #1\expandafter \@firstoftwo
 \else \expandafter \@secondoftwo
 \fi
}%
\providecommand \@ifx [1]{%
 \ifx #1\expandafter \@firstoftwo
 \else \expandafter \@secondoftwo
 \fi
}%
\providecommand \natexlab [1]{#1}%
\providecommand \enquote  [1]{``#1''}%
\providecommand \bibnamefont  [1]{#1}%
\providecommand \bibfnamefont [1]{#1}%
\providecommand \citenamefont [1]{#1}%
\providecommand \href@noop [0]{\@secondoftwo}%
\providecommand \href [0]{\begingroup \@sanitize@url \@href}%
\providecommand \@href[1]{\@@startlink{#1}\@@href}%
\providecommand \@@href[1]{\endgroup#1\@@endlink}%
\providecommand \@sanitize@url [0]{\catcode `\\12\catcode `\$12\catcode
  `\&12\catcode `\#12\catcode `\^12\catcode `\_12\catcode `\%12\relax}%
\providecommand \@@startlink[1]{}%
\providecommand \@@endlink[0]{}%
\providecommand \url  [0]{\begingroup\@sanitize@url \@url }%
\providecommand \@url [1]{\endgroup\@href {#1}{\urlprefix }}%
\providecommand \urlprefix  [0]{URL }%
\providecommand \Eprint [0]{\href }%
\providecommand \doibase [0]{http://dx.doi.org/}%
\providecommand \selectlanguage [0]{\@gobble}%
\providecommand \bibinfo  [0]{\@secondoftwo}%
\providecommand \bibfield  [0]{\@secondoftwo}%
\providecommand \translation [1]{[#1]}%
\providecommand \BibitemOpen [0]{}%
\providecommand \bibitemStop [0]{}%
\providecommand \bibitemNoStop [0]{.\EOS\space}%
\providecommand \EOS [0]{\spacefactor3000\relax}%
\providecommand \BibitemShut  [1]{\csname bibitem#1\endcsname}%
\let\auto@bib@innerbib\@empty
\bibitem [{\citenamefont {Chuang}\ and\ \citenamefont
  {Nielsen}(1997)}]{Chuang1997}%
  \BibitemOpen
  \bibfield  {author} {\bibinfo {author} {\bibfnamefont {Isaac~L.}\
  \bibnamefont {Chuang}}\ and\ \bibinfo {author} {\bibfnamefont {M.~A.}\
  \bibnamefont {Nielsen}},\ }\bibfield  {title} {\enquote {\bibinfo {title}
  {Prescription for experimental determination of the dynamics of a quantum
  black box},}\ }\href {\doibase 10.1080/09500349708231894} {\bibfield
  {journal} {\bibinfo  {journal} {Journal of Modern Optics}\ }\textbf {\bibinfo
  {volume} {44}},\ \bibinfo {pages} {2455--2467} (\bibinfo {year}
  {1997})}\BibitemShut {NoStop}%
\bibitem [{\citenamefont {Merkel}\ \emph {et~al.}(2013)\citenamefont {Merkel},
  \citenamefont {Gambetta}, \citenamefont {Smolin}, \citenamefont {Poletto},
  \citenamefont {C\'orcoles}, \citenamefont {Johnson}, \citenamefont {Ryan},\
  and\ \citenamefont {Steffen}}]{Merkel2013}%
  \BibitemOpen
  \bibfield  {author} {\bibinfo {author} {\bibfnamefont {Seth~T.}\ \bibnamefont
  {Merkel}}, \bibinfo {author} {\bibfnamefont {Jay~M.}\ \bibnamefont
  {Gambetta}}, \bibinfo {author} {\bibfnamefont {John~A.}\ \bibnamefont
  {Smolin}}, \bibinfo {author} {\bibfnamefont {Stefano}\ \bibnamefont
  {Poletto}}, \bibinfo {author} {\bibfnamefont {Antonio~D.}\ \bibnamefont
  {C\'orcoles}}, \bibinfo {author} {\bibfnamefont {Blake~R.}\ \bibnamefont
  {Johnson}}, \bibinfo {author} {\bibfnamefont {Colm~A.}\ \bibnamefont {Ryan}},
  \ and\ \bibinfo {author} {\bibfnamefont {Matthias}\ \bibnamefont {Steffen}},\
  }\bibfield  {title} {\enquote {\bibinfo {title} {Self-consistent quantum
  process tomography},}\ }\href {\doibase 10.1103/PhysRevA.87.062119}
  {\bibfield  {journal} {\bibinfo  {journal} {Phys. Rev. A}\ }\textbf {\bibinfo
  {volume} {87}},\ \bibinfo {pages} {062119} (\bibinfo {year}
  {2013})}\BibitemShut {NoStop}%
\bibitem [{\citenamefont {McKay}\ \emph {et~al.}(2019)\citenamefont {McKay},
  \citenamefont {Sheldon}, \citenamefont {Smolin}, \citenamefont {Chow},\ and\
  \citenamefont {Gambetta}}]{McKay2019}%
  \BibitemOpen
  \bibfield  {author} {\bibinfo {author} {\bibfnamefont {David~C.}\
  \bibnamefont {McKay}}, \bibinfo {author} {\bibfnamefont {Sarah}\ \bibnamefont
  {Sheldon}}, \bibinfo {author} {\bibfnamefont {John~A.}\ \bibnamefont
  {Smolin}}, \bibinfo {author} {\bibfnamefont {Jerry~M.}\ \bibnamefont {Chow}},
  \ and\ \bibinfo {author} {\bibfnamefont {Jay~M.}\ \bibnamefont {Gambetta}},\
  }\bibfield  {title} {\enquote {\bibinfo {title} {Three-qubit randomized
  benchmarking},}\ }\href {\doibase 10.1103/PhysRevLett.122.200502} {\bibfield
  {journal} {\bibinfo  {journal} {Phys. Rev. Lett.}\ }\textbf {\bibinfo
  {volume} {122}},\ \bibinfo {pages} {200502} (\bibinfo {year}
  {2019})}\BibitemShut {NoStop}%
\bibitem [{\citenamefont {Blume-Kohout}\ \emph {et~al.}(2013)\citenamefont
  {Blume-Kohout}, \citenamefont {Gamble}, \citenamefont {Nielsen},
  \citenamefont {Mizrahi}, \citenamefont {Sterk},\ and\ \citenamefont
  {Maunz}}]{Blume-Kohout2013}%
  \BibitemOpen
  \bibfield  {author} {\bibinfo {author} {\bibfnamefont {Robin}\ \bibnamefont
  {Blume-Kohout}}, \bibinfo {author} {\bibfnamefont {John~King}\ \bibnamefont
  {Gamble}}, \bibinfo {author} {\bibfnamefont {Erik}\ \bibnamefont {Nielsen}},
  \bibinfo {author} {\bibfnamefont {Jonathan}\ \bibnamefont {Mizrahi}},
  \bibinfo {author} {\bibfnamefont {Jonathan~D.}\ \bibnamefont {Sterk}}, \ and\
  \bibinfo {author} {\bibfnamefont {Peter}\ \bibnamefont {Maunz}},\ }\href
  {\doibase 10.48550/ARXIV.1310.4492} {\enquote {\bibinfo {title} {Robust,
  self-consistent, closed-form tomography of quantum logic gates on a trapped
  ion qubit},}\ } (\bibinfo {year} {2013})\BibitemShut {NoStop}%
\bibitem [{\citenamefont {Greenbaum}(2015)}]{Greenbaum2015}%
  \BibitemOpen
  \bibfield  {author} {\bibinfo {author} {\bibfnamefont {Daniel}\ \bibnamefont
  {Greenbaum}},\ }\href {\doibase 10.48550/ARXIV.1509.02921} {\enquote
  {\bibinfo {title} {Introduction to quantum gate set tomography},}\ }
  (\bibinfo {year} {2015})\BibitemShut {NoStop}%
\bibitem [{\citenamefont {Blume-Kohout}\ \emph {et~al.}(2017)\citenamefont
  {Blume-Kohout}, \citenamefont {Gamble}, \citenamefont {Nielsen},
  \citenamefont {Rudinger}, \citenamefont {Mizrahi}, \citenamefont {Fortier},\
  and\ \citenamefont {Maunz}}]{Blume-Kohout2017}%
  \BibitemOpen
  \bibfield  {author} {\bibinfo {author} {\bibfnamefont {Robin}\ \bibnamefont
  {Blume-Kohout}}, \bibinfo {author} {\bibfnamefont {John~King}\ \bibnamefont
  {Gamble}}, \bibinfo {author} {\bibfnamefont {Erik}\ \bibnamefont {Nielsen}},
  \bibinfo {author} {\bibfnamefont {Kenneth}\ \bibnamefont {Rudinger}},
  \bibinfo {author} {\bibfnamefont {Jonathan}\ \bibnamefont {Mizrahi}},
  \bibinfo {author} {\bibfnamefont {Kevin}\ \bibnamefont {Fortier}}, \ and\
  \bibinfo {author} {\bibfnamefont {Peter}\ \bibnamefont {Maunz}},\ }\bibfield
  {title} {\enquote {\bibinfo {title} {Demonstration of qubit operations below
  a rigorous fault tolerance threshold with gate set tomography},}\ }\href
  {\doibase 10.1038/ncomms14485} {\bibfield  {journal} {\bibinfo  {journal}
  {Nature Communications}\ }\textbf {\bibinfo {volume} {8}},\ \bibinfo {pages}
  {14485} (\bibinfo {year} {2017})}\BibitemShut {NoStop}%
\bibitem [{\citenamefont {Rudinger}\ \emph {et~al.}(2021)\citenamefont
  {Rudinger}, \citenamefont {Hogle}, \citenamefont {Naik}, \citenamefont
  {Hashim}, \citenamefont {Lobser}, \citenamefont {Santiago}, \citenamefont
  {Grace}, \citenamefont {Nielsen}, \citenamefont {Proctor}, \citenamefont
  {Seritan}, \citenamefont {Clark}, \citenamefont {Blume-Kohout}, \citenamefont
  {Siddiqi},\ and\ \citenamefont {Young}}]{Rudinger2021}%
  \BibitemOpen
  \bibfield  {author} {\bibinfo {author} {\bibfnamefont {Kenneth}\ \bibnamefont
  {Rudinger}}, \bibinfo {author} {\bibfnamefont {Craig~W.}\ \bibnamefont
  {Hogle}}, \bibinfo {author} {\bibfnamefont {Ravi~K.}\ \bibnamefont {Naik}},
  \bibinfo {author} {\bibfnamefont {Akel}\ \bibnamefont {Hashim}}, \bibinfo
  {author} {\bibfnamefont {Daniel}\ \bibnamefont {Lobser}}, \bibinfo {author}
  {\bibfnamefont {David~I.}\ \bibnamefont {Santiago}}, \bibinfo {author}
  {\bibfnamefont {Matthew~D.}\ \bibnamefont {Grace}}, \bibinfo {author}
  {\bibfnamefont {Erik}\ \bibnamefont {Nielsen}}, \bibinfo {author}
  {\bibfnamefont {Timothy}\ \bibnamefont {Proctor}}, \bibinfo {author}
  {\bibfnamefont {Stefan}\ \bibnamefont {Seritan}}, \bibinfo {author}
  {\bibfnamefont {Susan~M.}\ \bibnamefont {Clark}}, \bibinfo {author}
  {\bibfnamefont {Robin}\ \bibnamefont {Blume-Kohout}}, \bibinfo {author}
  {\bibfnamefont {Irfan}\ \bibnamefont {Siddiqi}}, \ and\ \bibinfo {author}
  {\bibfnamefont {Kevin~C.}\ \bibnamefont {Young}},\ }\bibfield  {title}
  {\enquote {\bibinfo {title} {Experimental characterization of crosstalk
  errors with simultaneous gate set tomography},}\ }\href {\doibase
  10.1103/PRXQuantum.2.040338} {\bibfield  {journal} {\bibinfo  {journal} {PRX
  Quantum}\ }\textbf {\bibinfo {volume} {2}},\ \bibinfo {pages} {040338}
  (\bibinfo {year} {2021})}\BibitemShut {NoStop}%
\bibitem [{\citenamefont {Emerson}\ \emph {et~al.}(2005)\citenamefont
  {Emerson}, \citenamefont {Alicki},\ and\ \citenamefont
  {{\.{Z}}yczkowski}}]{Emerson2005}%
  \BibitemOpen
  \bibfield  {author} {\bibinfo {author} {\bibfnamefont {Joseph}\ \bibnamefont
  {Emerson}}, \bibinfo {author} {\bibfnamefont {Robert}\ \bibnamefont
  {Alicki}}, \ and\ \bibinfo {author} {\bibfnamefont {Karol}\ \bibnamefont
  {{\.{Z}}yczkowski}},\ }\bibfield  {title} {\enquote {\bibinfo {title}
  {Scalable noise estimation with random unitary operators},}\ }\href {\doibase
  10.1088/1464-4266/7/10/021} {\bibfield  {journal} {\bibinfo  {journal}
  {Journal of Optics B: Quantum and Semiclassical Optics}\ }\textbf {\bibinfo
  {volume} {7}},\ \bibinfo {pages} {S347--S352} (\bibinfo {year}
  {2005})}\BibitemShut {NoStop}%
\bibitem [{\citenamefont {Knill}\ \emph {et~al.}(2008)\citenamefont {Knill},
  \citenamefont {Leibfried}, \citenamefont {Reichle}, \citenamefont {Britton},
  \citenamefont {Blakestad}, \citenamefont {Jost}, \citenamefont {Langer},
  \citenamefont {Ozeri}, \citenamefont {Seidelin},\ and\ \citenamefont
  {Wineland}}]{Knill2008}%
  \BibitemOpen
  \bibfield  {author} {\bibinfo {author} {\bibfnamefont {E.}~\bibnamefont
  {Knill}}, \bibinfo {author} {\bibfnamefont {D.}~\bibnamefont {Leibfried}},
  \bibinfo {author} {\bibfnamefont {R.}~\bibnamefont {Reichle}}, \bibinfo
  {author} {\bibfnamefont {J.}~\bibnamefont {Britton}}, \bibinfo {author}
  {\bibfnamefont {R.~B.}\ \bibnamefont {Blakestad}}, \bibinfo {author}
  {\bibfnamefont {J.~D.}\ \bibnamefont {Jost}}, \bibinfo {author}
  {\bibfnamefont {C.}~\bibnamefont {Langer}}, \bibinfo {author} {\bibfnamefont
  {R.}~\bibnamefont {Ozeri}}, \bibinfo {author} {\bibfnamefont
  {S.}~\bibnamefont {Seidelin}}, \ and\ \bibinfo {author} {\bibfnamefont
  {D.~J.}\ \bibnamefont {Wineland}},\ }\bibfield  {title} {\enquote {\bibinfo
  {title} {Randomized benchmarking of quantum gates},}\ }\href {\doibase
  10.1103/PhysRevA.77.012307} {\bibfield  {journal} {\bibinfo  {journal} {Phys.
  Rev. A}\ }\textbf {\bibinfo {volume} {77}},\ \bibinfo {pages} {012307}
  (\bibinfo {year} {2008})}\BibitemShut {NoStop}%
\bibitem [{\citenamefont {Dankert}\ \emph {et~al.}(2009)\citenamefont
  {Dankert}, \citenamefont {Cleve}, \citenamefont {Emerson},\ and\
  \citenamefont {Livine}}]{Dankert2009}%
  \BibitemOpen
  \bibfield  {author} {\bibinfo {author} {\bibfnamefont {Christoph}\
  \bibnamefont {Dankert}}, \bibinfo {author} {\bibfnamefont {Richard}\
  \bibnamefont {Cleve}}, \bibinfo {author} {\bibfnamefont {Joseph}\
  \bibnamefont {Emerson}}, \ and\ \bibinfo {author} {\bibfnamefont {Etera}\
  \bibnamefont {Livine}},\ }\bibfield  {title} {\enquote {\bibinfo {title}
  {Exact and approximate unitary 2-designs and their application to fidelity
  estimation},}\ }\href {\doibase 10.1103/PhysRevA.80.012304} {\bibfield
  {journal} {\bibinfo  {journal} {Phys. Rev. A}\ }\textbf {\bibinfo {volume}
  {80}},\ \bibinfo {pages} {012304} (\bibinfo {year} {2009})}\BibitemShut
  {NoStop}%
\bibitem [{\citenamefont {Magesan}\ \emph {et~al.}(2011)\citenamefont
  {Magesan}, \citenamefont {Gambetta},\ and\ \citenamefont
  {Emerson}}]{Magesan2011}%
  \BibitemOpen
  \bibfield  {author} {\bibinfo {author} {\bibfnamefont {Easwar}\ \bibnamefont
  {Magesan}}, \bibinfo {author} {\bibfnamefont {J.~M.}\ \bibnamefont
  {Gambetta}}, \ and\ \bibinfo {author} {\bibfnamefont {Joseph}\ \bibnamefont
  {Emerson}},\ }\bibfield  {title} {\enquote {\bibinfo {title} {Scalable and
  robust randomized benchmarking of quantum processes},}\ }\href {\doibase
  10.1103/PhysRevLett.106.180504} {\bibfield  {journal} {\bibinfo  {journal}
  {Phys. Rev. Lett.}\ }\textbf {\bibinfo {volume} {106}},\ \bibinfo {pages}
  {180504} (\bibinfo {year} {2011})}\BibitemShut {NoStop}%
\bibitem [{\citenamefont {Kim}\ \emph {et~al.}(2014)\citenamefont {Kim},
  \citenamefont {Shi}, \citenamefont {Simmons}, \citenamefont {Ward},
  \citenamefont {Prance}, \citenamefont {Koh}, \citenamefont {Gamble},
  \citenamefont {Savage}, \citenamefont {Lagally}, \citenamefont {Friesen},
  \citenamefont {Coppersmith},\ and\ \citenamefont {Eriksson}}]{Kim2014}%
  \BibitemOpen
  \bibfield  {author} {\bibinfo {author} {\bibfnamefont {Dohun}\ \bibnamefont
  {Kim}}, \bibinfo {author} {\bibfnamefont {Zhan}\ \bibnamefont {Shi}},
  \bibinfo {author} {\bibfnamefont {C.~B.}\ \bibnamefont {Simmons}}, \bibinfo
  {author} {\bibfnamefont {D.~R.}\ \bibnamefont {Ward}}, \bibinfo {author}
  {\bibfnamefont {J.~R.}\ \bibnamefont {Prance}}, \bibinfo {author}
  {\bibfnamefont {Teck~Seng}\ \bibnamefont {Koh}}, \bibinfo {author}
  {\bibfnamefont {John~King}\ \bibnamefont {Gamble}}, \bibinfo {author}
  {\bibfnamefont {D.~E.}\ \bibnamefont {Savage}}, \bibinfo {author}
  {\bibfnamefont {M.~G.}\ \bibnamefont {Lagally}}, \bibinfo {author}
  {\bibfnamefont {Mark}\ \bibnamefont {Friesen}}, \bibinfo {author}
  {\bibfnamefont {S.~N.}\ \bibnamefont {Coppersmith}}, \ and\ \bibinfo {author}
  {\bibfnamefont {Mark~A.}\ \bibnamefont {Eriksson}},\ }\bibfield  {title}
  {\enquote {\bibinfo {title} {Quantum control and process tomography of a
  semiconductor quantum dot hybrid qubit},}\ }\href {\doibase
  10.1038/nature13407} {\bibfield  {journal} {\bibinfo  {journal} {Nature}\
  }\textbf {\bibinfo {volume} {511}},\ \bibinfo {pages} {70--74} (\bibinfo
  {year} {2014})}\BibitemShut {NoStop}%
\bibitem [{\citenamefont {Madzik}\ \emph {et~al.}(2022)\citenamefont {Madzik},
  \citenamefont {Asaad}, \citenamefont {Youssry}, \citenamefont {Joecker},
  \citenamefont {Rudinger}, \citenamefont {Nielsen}, \citenamefont {Young},
  \citenamefont {Proctor}, \citenamefont {Baczewski}, \citenamefont {Laucht},
  \citenamefont {Schmitt}, \citenamefont {Hudson}, \citenamefont {Itoh},
  \citenamefont {Jakob}, \citenamefont {Johnson}, \citenamefont {Jamieson},
  \citenamefont {Dzurak}, \citenamefont {Ferrie}, \citenamefont
  {Blume-Kohout},\ and\ \citenamefont {Morello}}]{Madzik2022}%
  \BibitemOpen
  \bibfield  {author} {\bibinfo {author} {\bibfnamefont {Mateusz~T.}\
  \bibnamefont {Madzik}}, \bibinfo {author} {\bibfnamefont {Serwan}\
  \bibnamefont {Asaad}}, \bibinfo {author} {\bibfnamefont {Akram}\ \bibnamefont
  {Youssry}}, \bibinfo {author} {\bibfnamefont {Benjamin}\ \bibnamefont
  {Joecker}}, \bibinfo {author} {\bibfnamefont {Kenneth~M.}\ \bibnamefont
  {Rudinger}}, \bibinfo {author} {\bibfnamefont {Erik}\ \bibnamefont
  {Nielsen}}, \bibinfo {author} {\bibfnamefont {Kevin~C.}\ \bibnamefont
  {Young}}, \bibinfo {author} {\bibfnamefont {Timothy~J.}\ \bibnamefont
  {Proctor}}, \bibinfo {author} {\bibfnamefont {Andrew~D.}\ \bibnamefont
  {Baczewski}}, \bibinfo {author} {\bibfnamefont {Arne}\ \bibnamefont
  {Laucht}}, \bibinfo {author} {\bibfnamefont {Vivien}\ \bibnamefont
  {Schmitt}}, \bibinfo {author} {\bibfnamefont {Fay~E.}\ \bibnamefont
  {Hudson}}, \bibinfo {author} {\bibfnamefont {Kohei~M.}\ \bibnamefont {Itoh}},
  \bibinfo {author} {\bibfnamefont {Alexander~M.}\ \bibnamefont {Jakob}},
  \bibinfo {author} {\bibfnamefont {Brett~C.}\ \bibnamefont {Johnson}},
  \bibinfo {author} {\bibfnamefont {David~N.}\ \bibnamefont {Jamieson}},
  \bibinfo {author} {\bibfnamefont {Andrew~S.}\ \bibnamefont {Dzurak}},
  \bibinfo {author} {\bibfnamefont {Christopher}\ \bibnamefont {Ferrie}},
  \bibinfo {author} {\bibfnamefont {Robin}\ \bibnamefont {Blume-Kohout}}, \
  and\ \bibinfo {author} {\bibfnamefont {Andrea}\ \bibnamefont {Morello}},\
  }\bibfield  {title} {\enquote {\bibinfo {title} {Precision tomography of a
  three-qubit donor quantum processor in silicon},}\ }\href {\doibase
  10.1038/s41586-021-04292-7} {\bibfield  {journal} {\bibinfo  {journal}
  {Nature}\ }\textbf {\bibinfo {volume} {601}},\ \bibinfo {pages} {348--353}
  (\bibinfo {year} {2022})}\BibitemShut {NoStop}%
\bibitem [{\citenamefont {Xue}\ \emph {et~al.}(2022)\citenamefont {Xue},
  \citenamefont {Russ}, \citenamefont {Samkharadze}, \citenamefont {Undseth},
  \citenamefont {Sammak}, \citenamefont {Scappucci},\ and\ \citenamefont
  {Vandersypen}}]{Xue2022}%
  \BibitemOpen
  \bibfield  {author} {\bibinfo {author} {\bibfnamefont {Xiao}\ \bibnamefont
  {Xue}}, \bibinfo {author} {\bibfnamefont {Maximilian}\ \bibnamefont {Russ}},
  \bibinfo {author} {\bibfnamefont {Nodar}\ \bibnamefont {Samkharadze}},
  \bibinfo {author} {\bibfnamefont {Brennan}\ \bibnamefont {Undseth}}, \bibinfo
  {author} {\bibfnamefont {Amir}\ \bibnamefont {Sammak}}, \bibinfo {author}
  {\bibfnamefont {Giordano}\ \bibnamefont {Scappucci}}, \ and\ \bibinfo
  {author} {\bibfnamefont {Lieven M.~K.}\ \bibnamefont {Vandersypen}},\
  }\bibfield  {title} {\enquote {\bibinfo {title} {Quantum logic with spin
  qubits crossing the surface code threshold},}\ }\href {\doibase
  10.1038/s41586-021-04273-w} {\bibfield  {journal} {\bibinfo  {journal}
  {Nature}\ }\textbf {\bibinfo {volume} {601}},\ \bibinfo {pages} {343--347}
  (\bibinfo {year} {2022})}\BibitemShut {NoStop}%
\bibitem [{\citenamefont {Yan}\ \emph {et~al.}(2016)\citenamefont {Yan},
  \citenamefont {Kamal}, \citenamefont {Birenbaum}, \citenamefont {Sears},
  \citenamefont {Hover}, \citenamefont {Gudmundsen}, \citenamefont {Rosenberg},
  \citenamefont {Samach}, \citenamefont {Weber}, \citenamefont {Yoder},
  \citenamefont {Orlando}, \citenamefont {Clarke}, \citenamefont {Kerman},\
  and\ \citenamefont {Oliver}}]{Yan2016}%
  \BibitemOpen
  \bibfield  {author} {\bibinfo {author} {\bibfnamefont {S.}~\bibnamefont
  {Yan}, \bibfnamefont {F.and~Gustavsson}}, \bibinfo {author} {\bibfnamefont
  {A.}~\bibnamefont {Kamal}}, \bibinfo {author} {\bibfnamefont
  {J.}~\bibnamefont {Birenbaum}}, \bibinfo {author} {\bibfnamefont {A.~P.}\
  \bibnamefont {Sears}}, \bibinfo {author} {\bibfnamefont {D.}~\bibnamefont
  {Hover}}, \bibinfo {author} {\bibfnamefont {T.~J.}\ \bibnamefont
  {Gudmundsen}}, \bibinfo {author} {\bibfnamefont {D.}~\bibnamefont
  {Rosenberg}}, \bibinfo {author} {\bibfnamefont {G.}~\bibnamefont {Samach}},
  \bibinfo {author} {\bibfnamefont {S.}~\bibnamefont {Weber}}, \bibinfo
  {author} {\bibfnamefont {J.~L.}\ \bibnamefont {Yoder}}, \bibinfo {author}
  {\bibfnamefont {T.~P.}\ \bibnamefont {Orlando}}, \bibinfo {author}
  {\bibfnamefont {J.}~\bibnamefont {Clarke}}, \bibinfo {author} {\bibfnamefont
  {A.~J.}\ \bibnamefont {Kerman}}, \ and\ \bibinfo {author} {\bibfnamefont
  {W.~D.}\ \bibnamefont {Oliver}},\ }\bibfield  {title} {\enquote {\bibinfo
  {title} {The flux qubit revisited to enhance coherence and
  reproducibility},}\ }\href {\doibase 10.1038/ncomms12964} {\bibfield
  {journal} {\bibinfo  {journal} {Nat. Commun.}\ }\textbf {\bibinfo {volume}
  {7}},\ \bibinfo {pages} {12964} (\bibinfo {year} {2016})}\BibitemShut
  {NoStop}%
\bibitem [{\citenamefont {Willsch}\ \emph {et~al.}(2017)\citenamefont
  {Willsch}, \citenamefont {Nocon}, \citenamefont {Jin}, \citenamefont
  {De~Raedt},\ and\ \citenamefont {Michielsen}}]{Willsch2017}%
  \BibitemOpen
  \bibfield  {author} {\bibinfo {author} {\bibfnamefont {D.}~\bibnamefont
  {Willsch}}, \bibinfo {author} {\bibfnamefont {M.}~\bibnamefont {Nocon}},
  \bibinfo {author} {\bibfnamefont {F.}~\bibnamefont {Jin}}, \bibinfo {author}
  {\bibfnamefont {H.}~\bibnamefont {De~Raedt}}, \ and\ \bibinfo {author}
  {\bibfnamefont {K.}~\bibnamefont {Michielsen}},\ }\bibfield  {title}
  {\enquote {\bibinfo {title} {Gate-error analysis in simulations of quantum
  computers with transmon qubits},}\ }\href {\doibase
  10.1103/PhysRevA.96.062302} {\bibfield  {journal} {\bibinfo  {journal} {Phys.
  Rev. A}\ }\textbf {\bibinfo {volume} {96}},\ \bibinfo {pages} {062302}
  (\bibinfo {year} {2017})}\BibitemShut {NoStop}%
\bibitem [{\citenamefont {Klimov}\ \emph {et~al.}(2018)\citenamefont {Klimov},
  \citenamefont {Kelly}, \citenamefont {Chen}, \citenamefont {Neeley},
  \citenamefont {Megrant}, \citenamefont {Burkett}, \citenamefont {Barends},
  \citenamefont {Arya}, \citenamefont {Chiaro}, \citenamefont {Chen},
  \citenamefont {Dunsworth}, \citenamefont {Fowler}, \citenamefont {Foxen},
  \citenamefont {Gidney}, \citenamefont {Giustina}, \citenamefont {Graff},
  \citenamefont {Huang}, \citenamefont {Jeffrey}, \citenamefont {Lucero},
  \citenamefont {Mutus}, \citenamefont {Naaman}, \citenamefont {Neill},
  \citenamefont {Quintana}, \citenamefont {Roushan}, \citenamefont {Sank},
  \citenamefont {Vainsencher}, \citenamefont {Wenner}, \citenamefont {White},
  \citenamefont {Boixo}, \citenamefont {Babbush}, \citenamefont {Smelyanskiy},
  \citenamefont {Neven},\ and\ \citenamefont {Martinis}}]{Klimov2018}%
  \BibitemOpen
  \bibfield  {author} {\bibinfo {author} {\bibfnamefont {P.~V.}\ \bibnamefont
  {Klimov}}, \bibinfo {author} {\bibfnamefont {J.}~\bibnamefont {Kelly}},
  \bibinfo {author} {\bibfnamefont {Z.}~\bibnamefont {Chen}}, \bibinfo {author}
  {\bibfnamefont {M.}~\bibnamefont {Neeley}}, \bibinfo {author} {\bibfnamefont
  {A.}~\bibnamefont {Megrant}}, \bibinfo {author} {\bibfnamefont
  {B.}~\bibnamefont {Burkett}}, \bibinfo {author} {\bibfnamefont
  {R.}~\bibnamefont {Barends}}, \bibinfo {author} {\bibfnamefont
  {K.}~\bibnamefont {Arya}}, \bibinfo {author} {\bibfnamefont {B.}~\bibnamefont
  {Chiaro}}, \bibinfo {author} {\bibfnamefont {Yu}~\bibnamefont {Chen}},
  \bibinfo {author} {\bibfnamefont {A.}~\bibnamefont {Dunsworth}}, \bibinfo
  {author} {\bibfnamefont {A.}~\bibnamefont {Fowler}}, \bibinfo {author}
  {\bibfnamefont {B.}~\bibnamefont {Foxen}}, \bibinfo {author} {\bibfnamefont
  {C.}~\bibnamefont {Gidney}}, \bibinfo {author} {\bibfnamefont
  {M.}~\bibnamefont {Giustina}}, \bibinfo {author} {\bibfnamefont
  {R.}~\bibnamefont {Graff}}, \bibinfo {author} {\bibfnamefont
  {T.}~\bibnamefont {Huang}}, \bibinfo {author} {\bibfnamefont
  {E.}~\bibnamefont {Jeffrey}}, \bibinfo {author} {\bibfnamefont {Erik}\
  \bibnamefont {Lucero}}, \bibinfo {author} {\bibfnamefont {J.~Y.}\
  \bibnamefont {Mutus}}, \bibinfo {author} {\bibfnamefont {O.}~\bibnamefont
  {Naaman}}, \bibinfo {author} {\bibfnamefont {C.}~\bibnamefont {Neill}},
  \bibinfo {author} {\bibfnamefont {C.}~\bibnamefont {Quintana}}, \bibinfo
  {author} {\bibfnamefont {P.}~\bibnamefont {Roushan}}, \bibinfo {author}
  {\bibfnamefont {Daniel}\ \bibnamefont {Sank}}, \bibinfo {author}
  {\bibfnamefont {A.}~\bibnamefont {Vainsencher}}, \bibinfo {author}
  {\bibfnamefont {J.}~\bibnamefont {Wenner}}, \bibinfo {author} {\bibfnamefont
  {T.~C.}\ \bibnamefont {White}}, \bibinfo {author} {\bibfnamefont
  {S.}~\bibnamefont {Boixo}}, \bibinfo {author} {\bibfnamefont
  {R.}~\bibnamefont {Babbush}}, \bibinfo {author} {\bibfnamefont {V.~N.}\
  \bibnamefont {Smelyanskiy}}, \bibinfo {author} {\bibfnamefont
  {H.}~\bibnamefont {Neven}}, \ and\ \bibinfo {author} {\bibfnamefont
  {John~M.}\ \bibnamefont {Martinis}},\ }\bibfield  {title} {\enquote {\bibinfo
  {title} {Fluctuations of energy-relaxation times in superconducting
  qubits},}\ }\href {\doibase 10.1103/PhysRevLett.121.090502} {\bibfield
  {journal} {\bibinfo  {journal} {Phys. Rev. Lett.}\ }\textbf {\bibinfo
  {volume} {121}},\ \bibinfo {pages} {090502} (\bibinfo {year}
  {2018})}\BibitemShut {NoStop}%
\bibitem [{\citenamefont {Arute}\ \emph {et~al.}(2019)\citenamefont {Arute},
  \citenamefont {Arya},\ and\ \citenamefont {Babbush}}]{Arute2019}%
  \BibitemOpen
  \bibfield  {author} {\bibinfo {author} {\bibfnamefont {F.}~\bibnamefont
  {Arute}}, \bibinfo {author} {\bibfnamefont {K.}~\bibnamefont {Arya}}, \ and\
  \bibinfo {author} {\bibfnamefont {et.~al.}\ \bibnamefont {Babbush}},\
  }\bibfield  {title} {\enquote {\bibinfo {title} {Quantum supremacy using a
  programmable superconducting processor},}\ }\href {\doibase
  10.1038/s41586-019-1666-5} {\bibfield  {journal} {\bibinfo  {journal}
  {Nature}\ }\textbf {\bibinfo {volume} {574}},\ \bibinfo {pages} {505--510}
  (\bibinfo {year} {2019})}\BibitemShut {NoStop}%
\bibitem [{\citenamefont {Lao}\ \emph {et~al.}(2022)\citenamefont {Lao},
  \citenamefont {Korotkov}, \citenamefont {Jiang}, \citenamefont
  {Mruczkiewicz}, \citenamefont {O{\textquotesingle}Brien},\ and\ \citenamefont
  {Browne}}]{Lao2022}%
  \BibitemOpen
  \bibfield  {author} {\bibinfo {author} {\bibfnamefont {Lingling}\
  \bibnamefont {Lao}}, \bibinfo {author} {\bibfnamefont {Alexander}\
  \bibnamefont {Korotkov}}, \bibinfo {author} {\bibfnamefont {Zhang}\
  \bibnamefont {Jiang}}, \bibinfo {author} {\bibfnamefont {Wojciech}\
  \bibnamefont {Mruczkiewicz}}, \bibinfo {author} {\bibfnamefont {Thomas~E}\
  \bibnamefont {O{\textquotesingle}Brien}}, \ and\ \bibinfo {author}
  {\bibfnamefont {Dan~E}\ \bibnamefont {Browne}},\ }\bibfield  {title}
  {\enquote {\bibinfo {title} {Software mitigation of coherent two-qubit gate
  errors},}\ }\href {\doibase 10.1088/2058-9565/ac57f1} {\bibfield  {journal}
  {\bibinfo  {journal} {Quantum Science and Technology}\ }\textbf {\bibinfo
  {volume} {7}},\ \bibinfo {pages} {025021} (\bibinfo {year}
  {2022})}\BibitemShut {NoStop}%
\bibitem [{\citenamefont {Khaneja}\ \emph {et~al.}(2005)\citenamefont
  {Khaneja}, \citenamefont {Reiss}, \citenamefont {Kehlet}, \citenamefont
  {Schulte-Herbr\"uggen},\ and\ \citenamefont {Glaser}}]{Khaneja2005}%
  \BibitemOpen
  \bibfield  {author} {\bibinfo {author} {\bibfnamefont {Navin}\ \bibnamefont
  {Khaneja}}, \bibinfo {author} {\bibfnamefont {Timo}\ \bibnamefont {Reiss}},
  \bibinfo {author} {\bibfnamefont {Cindie}\ \bibnamefont {Kehlet}}, \bibinfo
  {author} {\bibfnamefont {Thomas}\ \bibnamefont {Schulte-Herbr\"uggen}}, \
  and\ \bibinfo {author} {\bibfnamefont {Steffen~J.}\ \bibnamefont {Glaser}},\
  }\bibfield  {title} {\enquote {\bibinfo {title} {Optimal control of coupled
  spin dynamics: design of nmr pulse sequences by gradient ascent
  algorithms},}\ }\href {\doibase https://doi.org/10.1016/j.jmr.2004.11.004}
  {\bibfield  {journal} {\bibinfo  {journal} {Journal of Magnetic Resonance}\
  }\textbf {\bibinfo {volume} {172}},\ \bibinfo {pages} {296--305} (\bibinfo
  {year} {2005})}\BibitemShut {NoStop}%
\bibitem [{\citenamefont {Caneva}\ \emph {et~al.}(2011)\citenamefont {Caneva},
  \citenamefont {Calarco},\ and\ \citenamefont {Montangero}}]{Caneva2011}%
  \BibitemOpen
  \bibfield  {author} {\bibinfo {author} {\bibfnamefont {Tommaso}\ \bibnamefont
  {Caneva}}, \bibinfo {author} {\bibfnamefont {Tommaso}\ \bibnamefont
  {Calarco}}, \ and\ \bibinfo {author} {\bibfnamefont {Simone}\ \bibnamefont
  {Montangero}},\ }\bibfield  {title} {\enquote {\bibinfo {title} {Chopped
  random-basis quantum optimization},}\ }\href {\doibase
  10.1103/PhysRevA.84.022326} {\bibfield  {journal} {\bibinfo  {journal} {Phys.
  Rev. A}\ }\textbf {\bibinfo {volume} {84}},\ \bibinfo {pages} {022326}
  (\bibinfo {year} {2011})}\BibitemShut {NoStop}%
\bibitem [{\citenamefont {Doria}\ \emph {et~al.}(2011)\citenamefont {Doria},
  \citenamefont {Calarco},\ and\ \citenamefont {Montangero}}]{Doria2011}%
  \BibitemOpen
  \bibfield  {author} {\bibinfo {author} {\bibfnamefont {Patrick}\ \bibnamefont
  {Doria}}, \bibinfo {author} {\bibfnamefont {Tommaso}\ \bibnamefont
  {Calarco}}, \ and\ \bibinfo {author} {\bibfnamefont {Simone}\ \bibnamefont
  {Montangero}},\ }\bibfield  {title} {\enquote {\bibinfo {title} {Optimal
  control technique for many-body quantum dynamics},}\ }\href {\doibase
  10.1103/PhysRevLett.106.190501} {\bibfield  {journal} {\bibinfo  {journal}
  {Phys. Rev. Lett.}\ }\textbf {\bibinfo {volume} {106}},\ \bibinfo {pages}
  {190501} (\bibinfo {year} {2011})}\BibitemShut {NoStop}%
\bibitem [{\citenamefont {M\"uller}\ \emph {et~al.}(2011)\citenamefont
  {M\"uller}, \citenamefont {Reich}, \citenamefont {Murphy}, \citenamefont
  {Yuan}, \citenamefont {Vala}, \citenamefont {Whaley}, \citenamefont
  {Calarco},\ and\ \citenamefont {Koch}}]{Muller2011}%
  \BibitemOpen
  \bibfield  {author} {\bibinfo {author} {\bibfnamefont {M.~M.}\ \bibnamefont
  {M\"uller}}, \bibinfo {author} {\bibfnamefont {D.~M.}\ \bibnamefont {Reich}},
  \bibinfo {author} {\bibfnamefont {M.}~\bibnamefont {Murphy}}, \bibinfo
  {author} {\bibfnamefont {H.}~\bibnamefont {Yuan}}, \bibinfo {author}
  {\bibfnamefont {J.}~\bibnamefont {Vala}}, \bibinfo {author} {\bibfnamefont
  {K.~B.}\ \bibnamefont {Whaley}}, \bibinfo {author} {\bibfnamefont
  {T.}~\bibnamefont {Calarco}}, \ and\ \bibinfo {author} {\bibfnamefont
  {C.~P.}\ \bibnamefont {Koch}},\ }\bibfield  {title} {\enquote {\bibinfo
  {title} {Optimizing entangling quantum gates for physical systems},}\ }\href
  {\doibase 10.1103/PhysRevA.84.042315} {\bibfield  {journal} {\bibinfo
  {journal} {Phys. Rev. A}\ }\textbf {\bibinfo {volume} {84}},\ \bibinfo
  {pages} {042315} (\bibinfo {year} {2011})}\BibitemShut {NoStop}%
\bibitem [{\citenamefont {Greiner}\ \emph {et~al.}(2002)\citenamefont
  {Greiner}, \citenamefont {Mandel}, \citenamefont {Esslinger}, \citenamefont
  {H{\"a}nsch},\ and\ \citenamefont {Bloch}}]{Greiner2002}%
  \BibitemOpen
  \bibfield  {author} {\bibinfo {author} {\bibfnamefont {Markus}\ \bibnamefont
  {Greiner}}, \bibinfo {author} {\bibfnamefont {Olaf}\ \bibnamefont {Mandel}},
  \bibinfo {author} {\bibfnamefont {Tilman}\ \bibnamefont {Esslinger}},
  \bibinfo {author} {\bibfnamefont {Theodor~W.}\ \bibnamefont {H{\"a}nsch}}, \
  and\ \bibinfo {author} {\bibfnamefont {Immanuel}\ \bibnamefont {Bloch}},\
  }\bibfield  {title} {\enquote {\bibinfo {title} {Quantum phase transition
  from a superfluid to a mott insulator in a gas of ultracold atoms},}\ }\href
  {\doibase 10.1038/415039a} {\bibfield  {journal} {\bibinfo  {journal}
  {Nature}\ }\textbf {\bibinfo {volume} {415}},\ \bibinfo {pages} {39--44}
  (\bibinfo {year} {2002})}\BibitemShut {NoStop}%
\bibitem [{\citenamefont {yoon Choi}\ \emph {et~al.}(2016)\citenamefont {yoon
  Choi}, \citenamefont {Hild}, \citenamefont {Zeiher}, \citenamefont {Schauß},
  \citenamefont {Rubio-Abadal}, \citenamefont {Yefsah}, \citenamefont
  {Khemani}, \citenamefont {Huse}, \citenamefont {Bloch},\ and\ \citenamefont
  {Gross}}]{Choi2016}%
  \BibitemOpen
  \bibfield  {author} {\bibinfo {author} {\bibfnamefont {Jae}\ \bibnamefont
  {yoon Choi}}, \bibinfo {author} {\bibfnamefont {Sebastian}\ \bibnamefont
  {Hild}}, \bibinfo {author} {\bibfnamefont {Johannes}\ \bibnamefont {Zeiher}},
  \bibinfo {author} {\bibfnamefont {Peter}\ \bibnamefont {Schauß}}, \bibinfo
  {author} {\bibfnamefont {Antonio}\ \bibnamefont {Rubio-Abadal}}, \bibinfo
  {author} {\bibfnamefont {Tarik}\ \bibnamefont {Yefsah}}, \bibinfo {author}
  {\bibfnamefont {Vedika}\ \bibnamefont {Khemani}}, \bibinfo {author}
  {\bibfnamefont {David~A.}\ \bibnamefont {Huse}}, \bibinfo {author}
  {\bibfnamefont {Immanuel}\ \bibnamefont {Bloch}}, \ and\ \bibinfo {author}
  {\bibfnamefont {Christian}\ \bibnamefont {Gross}},\ }\bibfield  {title}
  {\enquote {\bibinfo {title} {Exploring the many-body localization transition
  in two dimensions},}\ }\href {\doibase 10.1126/science.aaf8834} {\bibfield
  {journal} {\bibinfo  {journal} {Science}\ }\textbf {\bibinfo {volume}
  {352}},\ \bibinfo {pages} {1547--1552} (\bibinfo {year} {2016})}\BibitemShut
  {NoStop}%
\bibitem [{\citenamefont {Bluvstein}\ \emph {et~al.}(2021)\citenamefont
  {Bluvstein}, \citenamefont {Omran}, \citenamefont {Levine}, \citenamefont
  {Keesling}, \citenamefont {Semeghini}, \citenamefont {Ebadi}, \citenamefont
  {Wang}, \citenamefont {Michailidis}, \citenamefont {Maskara}, \citenamefont
  {Ho}, \citenamefont {Choi}, \citenamefont {Serbyn}, \citenamefont {Greiner},
  \citenamefont {Vuletić},\ and\ \citenamefont {Lukin}}]{Bluvstein2021}%
  \BibitemOpen
  \bibfield  {author} {\bibinfo {author} {\bibfnamefont {D.}~\bibnamefont
  {Bluvstein}}, \bibinfo {author} {\bibfnamefont {A.}~\bibnamefont {Omran}},
  \bibinfo {author} {\bibfnamefont {H.}~\bibnamefont {Levine}}, \bibinfo
  {author} {\bibfnamefont {A.}~\bibnamefont {Keesling}}, \bibinfo {author}
  {\bibfnamefont {G.}~\bibnamefont {Semeghini}}, \bibinfo {author}
  {\bibfnamefont {S.}~\bibnamefont {Ebadi}}, \bibinfo {author} {\bibfnamefont
  {T.~T.}\ \bibnamefont {Wang}}, \bibinfo {author} {\bibfnamefont {A.~A.}\
  \bibnamefont {Michailidis}}, \bibinfo {author} {\bibfnamefont
  {N.}~\bibnamefont {Maskara}}, \bibinfo {author} {\bibfnamefont {W.~W.}\
  \bibnamefont {Ho}}, \bibinfo {author} {\bibfnamefont {S.}~\bibnamefont
  {Choi}}, \bibinfo {author} {\bibfnamefont {M.}~\bibnamefont {Serbyn}},
  \bibinfo {author} {\bibfnamefont {M.}~\bibnamefont {Greiner}}, \bibinfo
  {author} {\bibfnamefont {V.}~\bibnamefont {Vuletić}}, \ and\ \bibinfo
  {author} {\bibfnamefont {M.~D.}\ \bibnamefont {Lukin}},\ }\bibfield  {title}
  {\enquote {\bibinfo {title} {Controlling quantum many-body dynamics in driven
  rydberg atom arrays},}\ }\href {\doibase 10.1126/science.abg2530} {\bibfield
  {journal} {\bibinfo  {journal} {Science}\ }\textbf {\bibinfo {volume}
  {371}},\ \bibinfo {pages} {1355--1359} (\bibinfo {year} {2021})}\BibitemShut
  {NoStop}%
\bibitem [{\citenamefont {Semeghini}\ \emph {et~al.}(2021)\citenamefont
  {Semeghini}, \citenamefont {Levine}, \citenamefont {Keesling}, \citenamefont
  {Ebadi}, \citenamefont {Wang}, \citenamefont {Bluvstein}, \citenamefont
  {Verresen}, \citenamefont {Pichler}, \citenamefont {Kalinowski},
  \citenamefont {Samajdar}, \citenamefont {Omran}, \citenamefont {Sachdev},
  \citenamefont {Vishwanath}, \citenamefont {Greiner}, \citenamefont
  {Vuletić},\ and\ \citenamefont {Lukin}}]{Semeghini2021}%
  \BibitemOpen
  \bibfield  {author} {\bibinfo {author} {\bibfnamefont {G.}~\bibnamefont
  {Semeghini}}, \bibinfo {author} {\bibfnamefont {H.}~\bibnamefont {Levine}},
  \bibinfo {author} {\bibfnamefont {A.}~\bibnamefont {Keesling}}, \bibinfo
  {author} {\bibfnamefont {S.}~\bibnamefont {Ebadi}}, \bibinfo {author}
  {\bibfnamefont {T.~T.}\ \bibnamefont {Wang}}, \bibinfo {author}
  {\bibfnamefont {D.}~\bibnamefont {Bluvstein}}, \bibinfo {author}
  {\bibfnamefont {R.}~\bibnamefont {Verresen}}, \bibinfo {author}
  {\bibfnamefont {H.}~\bibnamefont {Pichler}}, \bibinfo {author} {\bibfnamefont
  {M.}~\bibnamefont {Kalinowski}}, \bibinfo {author} {\bibfnamefont
  {R.}~\bibnamefont {Samajdar}}, \bibinfo {author} {\bibfnamefont
  {A.}~\bibnamefont {Omran}}, \bibinfo {author} {\bibfnamefont
  {S.}~\bibnamefont {Sachdev}}, \bibinfo {author} {\bibfnamefont
  {A.}~\bibnamefont {Vishwanath}}, \bibinfo {author} {\bibfnamefont
  {M.}~\bibnamefont {Greiner}}, \bibinfo {author} {\bibfnamefont
  {V.}~\bibnamefont {Vuletić}}, \ and\ \bibinfo {author} {\bibfnamefont
  {M.~D.}\ \bibnamefont {Lukin}},\ }\bibfield  {title} {\enquote {\bibinfo
  {title} {Probing topological spin liquids on a programmable quantum
  simulator},}\ }\href {\doibase 10.1126/science.abi8794} {\bibfield  {journal}
  {\bibinfo  {journal} {Science}\ }\textbf {\bibinfo {volume} {374}},\ \bibinfo
  {pages} {1242--1247} (\bibinfo {year} {2021})}\BibitemShut {NoStop}%
\bibitem [{\citenamefont {Scholl}\ \emph {et~al.}(2021)\citenamefont {Scholl},
  \citenamefont {Schuler}, \citenamefont {Williams}, \citenamefont
  {Eberharter}, \citenamefont {Barredo}, \citenamefont {Schymik}, \citenamefont
  {Lienhard}, \citenamefont {Henry}, \citenamefont {Lang}, \citenamefont
  {Lahaye}, \citenamefont {L{\"a}uchli},\ and\ \citenamefont
  {Browaeys}}]{Scholl2021}%
  \BibitemOpen
  \bibfield  {author} {\bibinfo {author} {\bibfnamefont {Pascal}\ \bibnamefont
  {Scholl}}, \bibinfo {author} {\bibfnamefont {Michael}\ \bibnamefont
  {Schuler}}, \bibinfo {author} {\bibfnamefont {Hannah~J.}\ \bibnamefont
  {Williams}}, \bibinfo {author} {\bibfnamefont {Alexander~A.}\ \bibnamefont
  {Eberharter}}, \bibinfo {author} {\bibfnamefont {Daniel}\ \bibnamefont
  {Barredo}}, \bibinfo {author} {\bibfnamefont {Kai-Niklas}\ \bibnamefont
  {Schymik}}, \bibinfo {author} {\bibfnamefont {Vincent}\ \bibnamefont
  {Lienhard}}, \bibinfo {author} {\bibfnamefont {Louis-Paul}\ \bibnamefont
  {Henry}}, \bibinfo {author} {\bibfnamefont {Thomas~C.}\ \bibnamefont {Lang}},
  \bibinfo {author} {\bibfnamefont {Thierry}\ \bibnamefont {Lahaye}}, \bibinfo
  {author} {\bibfnamefont {Andreas~M.}\ \bibnamefont {L{\"a}uchli}}, \ and\
  \bibinfo {author} {\bibfnamefont {Antoine}\ \bibnamefont {Browaeys}},\
  }\bibfield  {title} {\enquote {\bibinfo {title} {Quantum simulation of 2d
  antiferromagnets with hundreds of rydberg atoms},}\ }\href {\doibase
  10.1038/s41586-021-03585-1} {\bibfield  {journal} {\bibinfo  {journal}
  {Nature}\ }\textbf {\bibinfo {volume} {595}},\ \bibinfo {pages} {233--238}
  (\bibinfo {year} {2021})}\BibitemShut {NoStop}%
\bibitem [{\citenamefont {Altman}\ \emph {et~al.}(2021)\citenamefont {Altman},
  \citenamefont {Brown}, \citenamefont {Carleo}, \citenamefont {Carr},
  \citenamefont {Demler}, \citenamefont {Chin}, \citenamefont {DeMarco},
  \citenamefont {Economou}, \citenamefont {Eriksson}, \citenamefont {Fu},
  \citenamefont {Greiner}, \citenamefont {Hazzard}, \citenamefont {Hulet},
  \citenamefont {Koll\'ar}, \citenamefont {Lev}, \citenamefont {Lukin},
  \citenamefont {Ma}, \citenamefont {Mi}, \citenamefont {Misra}, \citenamefont
  {Monroe}, \citenamefont {Murch}, \citenamefont {Nazario}, \citenamefont {Ni},
  \citenamefont {Potter}, \citenamefont {Roushan}, \citenamefont {Saffman},
  \citenamefont {Schleier-Smith}, \citenamefont {Siddiqi}, \citenamefont
  {Simmonds}, \citenamefont {Singh}, \citenamefont {Spielman}, \citenamefont
  {Temme}, \citenamefont {Weiss}, \citenamefont {Vu\ifmmode \check{c}\else
  \v{c}\fi{}kovi\ifmmode~\acute{c}\else \'{c}\fi{}}, \citenamefont
  {Vuleti\ifmmode~\acute{c}\else \'{c}\fi{}}, \citenamefont {Ye},\ and\
  \citenamefont {Zwierlein}}]{Altman2021}%
  \BibitemOpen
  \bibfield  {author} {\bibinfo {author} {\bibfnamefont {Ehud}\ \bibnamefont
  {Altman}}, \bibinfo {author} {\bibfnamefont {Kenneth~R.}\ \bibnamefont
  {Brown}}, \bibinfo {author} {\bibfnamefont {Giuseppe}\ \bibnamefont
  {Carleo}}, \bibinfo {author} {\bibfnamefont {Lincoln~D.}\ \bibnamefont
  {Carr}}, \bibinfo {author} {\bibfnamefont {Eugene}\ \bibnamefont {Demler}},
  \bibinfo {author} {\bibfnamefont {Cheng}\ \bibnamefont {Chin}}, \bibinfo
  {author} {\bibfnamefont {Brian}\ \bibnamefont {DeMarco}}, \bibinfo {author}
  {\bibfnamefont {Sophia~E.}\ \bibnamefont {Economou}}, \bibinfo {author}
  {\bibfnamefont {Mark~A.}\ \bibnamefont {Eriksson}}, \bibinfo {author}
  {\bibfnamefont {Kai-Mei~C.}\ \bibnamefont {Fu}}, \bibinfo {author}
  {\bibfnamefont {Markus}\ \bibnamefont {Greiner}}, \bibinfo {author}
  {\bibfnamefont {Kaden~R.A.}\ \bibnamefont {Hazzard}}, \bibinfo {author}
  {\bibfnamefont {Randall~G.}\ \bibnamefont {Hulet}}, \bibinfo {author}
  {\bibfnamefont {Alicia~J.}\ \bibnamefont {Koll\'ar}}, \bibinfo {author}
  {\bibfnamefont {Benjamin~L.}\ \bibnamefont {Lev}}, \bibinfo {author}
  {\bibfnamefont {Mikhail~D.}\ \bibnamefont {Lukin}}, \bibinfo {author}
  {\bibfnamefont {Ruichao}\ \bibnamefont {Ma}}, \bibinfo {author}
  {\bibfnamefont {Xiao}\ \bibnamefont {Mi}}, \bibinfo {author} {\bibfnamefont
  {Shashank}\ \bibnamefont {Misra}}, \bibinfo {author} {\bibfnamefont
  {Christopher}\ \bibnamefont {Monroe}}, \bibinfo {author} {\bibfnamefont
  {Kater}\ \bibnamefont {Murch}}, \bibinfo {author} {\bibfnamefont {Zaira}\
  \bibnamefont {Nazario}}, \bibinfo {author} {\bibfnamefont {Kang-Kuen}\
  \bibnamefont {Ni}}, \bibinfo {author} {\bibfnamefont {Andrew~C.}\
  \bibnamefont {Potter}}, \bibinfo {author} {\bibfnamefont {Pedram}\
  \bibnamefont {Roushan}}, \bibinfo {author} {\bibfnamefont {Mark}\
  \bibnamefont {Saffman}}, \bibinfo {author} {\bibfnamefont {Monika}\
  \bibnamefont {Schleier-Smith}}, \bibinfo {author} {\bibfnamefont {Irfan}\
  \bibnamefont {Siddiqi}}, \bibinfo {author} {\bibfnamefont {Raymond}\
  \bibnamefont {Simmonds}}, \bibinfo {author} {\bibfnamefont {Meenakshi}\
  \bibnamefont {Singh}}, \bibinfo {author} {\bibfnamefont {I.B.}\ \bibnamefont
  {Spielman}}, \bibinfo {author} {\bibfnamefont {Kristan}\ \bibnamefont
  {Temme}}, \bibinfo {author} {\bibfnamefont {David~S.}\ \bibnamefont {Weiss}},
  \bibinfo {author} {\bibfnamefont {Jelena}\ \bibnamefont {Vu\ifmmode
  \check{c}\else \v{c}\fi{}kovi\ifmmode~\acute{c}\else \'{c}\fi{}}}, \bibinfo
  {author} {\bibfnamefont {Vladan}\ \bibnamefont {Vuleti\ifmmode~\acute{c}\else
  \'{c}\fi{}}}, \bibinfo {author} {\bibfnamefont {Jun}\ \bibnamefont {Ye}}, \
  and\ \bibinfo {author} {\bibfnamefont {Martin}\ \bibnamefont {Zwierlein}},\
  }\bibfield  {title} {\enquote {\bibinfo {title} {Quantum simulators:
  Architectures and opportunities},}\ }\href {\doibase
  10.1103/PRXQuantum.2.017003} {\bibfield  {journal} {\bibinfo  {journal} {PRX
  Quantum}\ }\textbf {\bibinfo {volume} {2}},\ \bibinfo {pages} {017003}
  (\bibinfo {year} {2021})}\BibitemShut {NoStop}%
\bibitem [{\citenamefont {Monroe}\ \emph {et~al.}(2021)\citenamefont {Monroe},
  \citenamefont {Campbell}, \citenamefont {Duan}, \citenamefont {Gong},
  \citenamefont {Gorshkov}, \citenamefont {Hess}, \citenamefont {Islam},
  \citenamefont {Kim}, \citenamefont {Linke}, \citenamefont {Pagano},
  \citenamefont {Richerme}, \citenamefont {Senko},\ and\ \citenamefont
  {Yao}}]{Monroe2021}%
  \BibitemOpen
  \bibfield  {author} {\bibinfo {author} {\bibfnamefont {C.}~\bibnamefont
  {Monroe}}, \bibinfo {author} {\bibfnamefont {W.~C.}\ \bibnamefont
  {Campbell}}, \bibinfo {author} {\bibfnamefont {L.-M.}\ \bibnamefont {Duan}},
  \bibinfo {author} {\bibfnamefont {Z.-X.}\ \bibnamefont {Gong}}, \bibinfo
  {author} {\bibfnamefont {A.~V.}\ \bibnamefont {Gorshkov}}, \bibinfo {author}
  {\bibfnamefont {P.~W.}\ \bibnamefont {Hess}}, \bibinfo {author}
  {\bibfnamefont {R.}~\bibnamefont {Islam}}, \bibinfo {author} {\bibfnamefont
  {K.}~\bibnamefont {Kim}}, \bibinfo {author} {\bibfnamefont {N.~M.}\
  \bibnamefont {Linke}}, \bibinfo {author} {\bibfnamefont {G.}~\bibnamefont
  {Pagano}}, \bibinfo {author} {\bibfnamefont {P.}~\bibnamefont {Richerme}},
  \bibinfo {author} {\bibfnamefont {C.}~\bibnamefont {Senko}}, \ and\ \bibinfo
  {author} {\bibfnamefont {N.~Y.}\ \bibnamefont {Yao}},\ }\bibfield  {title}
  {\enquote {\bibinfo {title} {Programmable quantum simulations of spin systems
  with trapped ions},}\ }\href {\doibase 10.1103/RevModPhys.93.025001}
  {\bibfield  {journal} {\bibinfo  {journal} {Rev. Mod. Phys.}\ }\textbf
  {\bibinfo {volume} {93}},\ \bibinfo {pages} {025001} (\bibinfo {year}
  {2021})}\BibitemShut {NoStop}%
\bibitem [{\citenamefont {Bairey}\ \emph {et~al.}(2019)\citenamefont {Bairey},
  \citenamefont {Arad},\ and\ \citenamefont {Lindner}}]{Bairey2019}%
  \BibitemOpen
  \bibfield  {author} {\bibinfo {author} {\bibfnamefont {Eyal}\ \bibnamefont
  {Bairey}}, \bibinfo {author} {\bibfnamefont {Itai}\ \bibnamefont {Arad}}, \
  and\ \bibinfo {author} {\bibfnamefont {Netanel~H.}\ \bibnamefont {Lindner}},\
  }\bibfield  {title} {\enquote {\bibinfo {title} {Learning a local hamiltonian
  from local measurements},}\ }\href {\doibase 10.1103/PhysRevLett.122.020504}
  {\bibfield  {journal} {\bibinfo  {journal} {Phys. Rev. Lett.}\ }\textbf
  {\bibinfo {volume} {122}},\ \bibinfo {pages} {020504} (\bibinfo {year}
  {2019})}\BibitemShut {NoStop}%
\bibitem [{\citenamefont {Qi}\ and\ \citenamefont {Ranard}(2019)}]{Qi2019}%
  \BibitemOpen
  \bibfield  {author} {\bibinfo {author} {\bibfnamefont {Xiao-Liang}\
  \bibnamefont {Qi}}\ and\ \bibinfo {author} {\bibfnamefont {Daniel}\
  \bibnamefont {Ranard}},\ }\bibfield  {title} {\enquote {\bibinfo {title}
  {Determining a local {H}amiltonian from a single eigenstate},}\ }\href
  {\doibase 10.22331/q-2019-07-08-159} {\bibfield  {journal} {\bibinfo
  {journal} {{Quantum}}\ }\textbf {\bibinfo {volume} {3}},\ \bibinfo {pages}
  {159} (\bibinfo {year} {2019})}\BibitemShut {NoStop}%
\bibitem [{\citenamefont {Li}\ \emph {et~al.}(2020)\citenamefont {Li},
  \citenamefont {Zou},\ and\ \citenamefont {Hsieh}}]{Li2020}%
  \BibitemOpen
  \bibfield  {author} {\bibinfo {author} {\bibfnamefont {Zhi}\ \bibnamefont
  {Li}}, \bibinfo {author} {\bibfnamefont {Liujun}\ \bibnamefont {Zou}}, \ and\
  \bibinfo {author} {\bibfnamefont {Timothy~H.}\ \bibnamefont {Hsieh}},\
  }\bibfield  {title} {\enquote {\bibinfo {title} {Hamiltonian tomography via
  quantum quench},}\ }\href {\doibase 10.1103/PhysRevLett.124.160502}
  {\bibfield  {journal} {\bibinfo  {journal} {Phys. Rev. Lett.}\ }\textbf
  {\bibinfo {volume} {124}},\ \bibinfo {pages} {160502} (\bibinfo {year}
  {2020})}\BibitemShut {NoStop}%
\bibitem [{\citenamefont {Diósi}(1988)}]{Diosi88}%
  \BibitemOpen
  \bibfield  {author} {\bibinfo {author} {\bibfnamefont {L.}~\bibnamefont
  {Diósi}},\ }\bibfield  {title} {\enquote {\bibinfo {title} {Continuous
  quantum measurement and itô formalism},}\ }\href {\doibase
  https://doi.org/10.1016/0375-9601(88)90309-X} {\bibfield  {journal} {\bibinfo
   {journal} {Physics Letters A}\ }\textbf {\bibinfo {volume} {129}},\ \bibinfo
  {pages} {419--423} (\bibinfo {year} {1988})}\BibitemShut {NoStop}%
\bibitem [{\citenamefont {Wiseman}\ and\ \citenamefont
  {Milburn}(1993)}]{Wiseman1993}%
  \BibitemOpen
  \bibfield  {author} {\bibinfo {author} {\bibfnamefont {H.~M.}\ \bibnamefont
  {Wiseman}}\ and\ \bibinfo {author} {\bibfnamefont {G.~J.}\ \bibnamefont
  {Milburn}},\ }\bibfield  {title} {\enquote {\bibinfo {title} {Quantum theory
  of field-quadrature measurements},}\ }\href {\doibase
  10.1103/PhysRevA.47.642} {\bibfield  {journal} {\bibinfo  {journal} {Phys.
  Rev. A}\ }\textbf {\bibinfo {volume} {47}},\ \bibinfo {pages} {642--662}
  (\bibinfo {year} {1993})}\BibitemShut {NoStop}%
\bibitem [{\citenamefont {Jacobs}\ and\ \citenamefont
  {Steck}(2006)}]{Jacobs2006}%
  \BibitemOpen
  \bibfield  {author} {\bibinfo {author} {\bibfnamefont {Kurt}\ \bibnamefont
  {Jacobs}}\ and\ \bibinfo {author} {\bibfnamefont {Daniel~A.}\ \bibnamefont
  {Steck}},\ }\bibfield  {title} {\enquote {\bibinfo {title} {A straightforward
  introduction to continuous quantum measurement},}\ }\href {\doibase
  10.1080/00107510601101934} {\bibfield  {journal} {\bibinfo  {journal}
  {Contemporary Physics}\ }\textbf {\bibinfo {volume} {47}},\ \bibinfo {pages}
  {279--303} (\bibinfo {year} {2006})}\BibitemShut {NoStop}%
\bibitem [{\citenamefont {Gambetta}\ \emph {et~al.}(2008)\citenamefont
  {Gambetta}, \citenamefont {Blais}, \citenamefont {Boissonneault},
  \citenamefont {Houck}, \citenamefont {Schuster},\ and\ \citenamefont
  {Girvin}}]{Gambetta2008}%
  \BibitemOpen
  \bibfield  {author} {\bibinfo {author} {\bibfnamefont {Jay}\ \bibnamefont
  {Gambetta}}, \bibinfo {author} {\bibfnamefont {Alexandre}\ \bibnamefont
  {Blais}}, \bibinfo {author} {\bibfnamefont {M.}~\bibnamefont
  {Boissonneault}}, \bibinfo {author} {\bibfnamefont {A.~A.}\ \bibnamefont
  {Houck}}, \bibinfo {author} {\bibfnamefont {D.~I.}\ \bibnamefont {Schuster}},
  \ and\ \bibinfo {author} {\bibfnamefont {S.~M.}\ \bibnamefont {Girvin}},\
  }\bibfield  {title} {\enquote {\bibinfo {title} {Quantum trajectory approach
  to circuit qed: Quantum jumps and the zeno effect},}\ }\href {\doibase
  10.1103/PhysRevA.77.012112} {\bibfield  {journal} {\bibinfo  {journal} {Phys.
  Rev. A}\ }\textbf {\bibinfo {volume} {77}},\ \bibinfo {pages} {012112}
  (\bibinfo {year} {2008})}\BibitemShut {NoStop}%
\bibitem [{\citenamefont {Murch}\ \emph {et~al.}(2013)\citenamefont {Murch},
  \citenamefont {Weber}, \citenamefont {Macklin},\ and\ \citenamefont
  {Siddiqi}}]{Murch2013}%
  \BibitemOpen
  \bibfield  {author} {\bibinfo {author} {\bibfnamefont {K.~W.}\ \bibnamefont
  {Murch}}, \bibinfo {author} {\bibfnamefont {S.~J.}\ \bibnamefont {Weber}},
  \bibinfo {author} {\bibfnamefont {C.}~\bibnamefont {Macklin}}, \ and\
  \bibinfo {author} {\bibfnamefont {I.}~\bibnamefont {Siddiqi}},\ }\bibfield
  {title} {\enquote {\bibinfo {title} {Observing single quantum trajectories of
  a superconducting quantum bit},}\ }\href {\doibase 10.1038/nature12539}
  {\bibfield  {journal} {\bibinfo  {journal} {Nature}\ }\textbf {\bibinfo
  {volume} {502}},\ \bibinfo {pages} {211--214} (\bibinfo {year}
  {2013})}\BibitemShut {NoStop}%
\bibitem [{\citenamefont {Jacobs}(2014)}]{Jacobs2014}%
  \BibitemOpen
  \bibfield  {author} {\bibinfo {author} {\bibfnamefont {Kurt}\ \bibnamefont
  {Jacobs}},\ }\href@noop {} {\emph {\bibinfo {title} {Quantum Measurement
  Theory and Its Applications}}}\ (\bibinfo  {publisher} {Cambridge University
  Press},\ \bibinfo {address} {Cambridge},\ \bibinfo {year} {2014})\BibitemShut
  {NoStop}%
\bibitem [{\citenamefont {Korotkov}(2014)}]{Korotkov2014}%
  \BibitemOpen
  \bibfield  {author} {\bibinfo {author} {\bibfnamefont {A.~N.}\ \bibnamefont
  {Korotkov}},\ }\bibfield  {title} {\enquote {\bibinfo {title} {Quantum
  bayesian approach to circuit qed measurement},}\ }in\ \href {\doibase
  10.1093/acprof:oso/9780199681181.003.0017} {\emph {\bibinfo {booktitle}
  {{Quantum Machines: Measurement and Control of Engineered Quantum Systems:
  Lecture Notes of the Les Houches Summer School: Volume 96, July 2011}}}}\
  (\bibinfo  {publisher} {Oxford University Press},\ \bibinfo {year}
  {2014})\BibitemShut {NoStop}%
\bibitem [{\citenamefont {Korotkov}(2016)}]{Korotkov2016}%
  \BibitemOpen
  \bibfield  {author} {\bibinfo {author} {\bibfnamefont {Alexander~N.}\
  \bibnamefont {Korotkov}},\ }\bibfield  {title} {\enquote {\bibinfo {title}
  {Quantum bayesian approach to circuit qed measurement with moderate
  bandwidth},}\ }\href {\doibase 10.1103/PhysRevA.94.042326} {\bibfield
  {journal} {\bibinfo  {journal} {Phys. Rev. A}\ }\textbf {\bibinfo {volume}
  {94}},\ \bibinfo {pages} {042326} (\bibinfo {year} {2016})}\BibitemShut
  {NoStop}%
\bibitem [{\citenamefont {McKay}\ \emph {et~al.}(2016)\citenamefont {McKay},
  \citenamefont {Filipp}, \citenamefont {Mezzacapo}, \citenamefont {Magesan},
  \citenamefont {Chow},\ and\ \citenamefont {Gambetta}}]{McKay2016}%
  \BibitemOpen
  \bibfield  {author} {\bibinfo {author} {\bibfnamefont {David~C.}\
  \bibnamefont {McKay}}, \bibinfo {author} {\bibfnamefont {Stefan}\
  \bibnamefont {Filipp}}, \bibinfo {author} {\bibfnamefont {Antonio}\
  \bibnamefont {Mezzacapo}}, \bibinfo {author} {\bibfnamefont {Easwar}\
  \bibnamefont {Magesan}}, \bibinfo {author} {\bibfnamefont {Jerry~M.}\
  \bibnamefont {Chow}}, \ and\ \bibinfo {author} {\bibfnamefont {Jay~M.}\
  \bibnamefont {Gambetta}},\ }\bibfield  {title} {\enquote {\bibinfo {title}
  {{Universal Gate for Fixed-Frequency Qubits via a Tunable Bus}},}\ }\href
  {\doibase 10.1103/PhysRevApplied.6.064007} {\bibfield  {journal} {\bibinfo
  {journal} {Physical Review Applied}\ }\textbf {\bibinfo {volume} {6}},\
  \bibinfo {pages} {1--10} (\bibinfo {year} {2016})}\BibitemShut {NoStop}%
\bibitem [{\citenamefont {Lu}\ \emph {et~al.}(2017)\citenamefont {Lu},
  \citenamefont {Chakram}, \citenamefont {Leung}, \citenamefont {Earnest},
  \citenamefont {Naik}, \citenamefont {Huang}, \citenamefont {Groszkowski},
  \citenamefont {Kapit}, \citenamefont {Koch},\ and\ \citenamefont
  {Schuster}}]{Lu2017}%
  \BibitemOpen
  \bibfield  {author} {\bibinfo {author} {\bibfnamefont {Yao}\ \bibnamefont
  {Lu}}, \bibinfo {author} {\bibfnamefont {S.}~\bibnamefont {Chakram}},
  \bibinfo {author} {\bibfnamefont {N.}~\bibnamefont {Leung}}, \bibinfo
  {author} {\bibfnamefont {N.}~\bibnamefont {Earnest}}, \bibinfo {author}
  {\bibfnamefont {R.~K.}\ \bibnamefont {Naik}}, \bibinfo {author}
  {\bibfnamefont {Ziwen}\ \bibnamefont {Huang}}, \bibinfo {author}
  {\bibfnamefont {Peter}\ \bibnamefont {Groszkowski}}, \bibinfo {author}
  {\bibfnamefont {Eliot}\ \bibnamefont {Kapit}}, \bibinfo {author}
  {\bibfnamefont {Jens}\ \bibnamefont {Koch}}, \ and\ \bibinfo {author}
  {\bibfnamefont {David~I.}\ \bibnamefont {Schuster}},\ }\bibfield  {title}
  {\enquote {\bibinfo {title} {Universal stabilization of a parametrically
  coupled qubit},}\ }\href {\doibase 10.1103/PhysRevLett.119.150502} {\bibfield
   {journal} {\bibinfo  {journal} {Phys. Rev. Lett.}\ }\textbf {\bibinfo
  {volume} {119}},\ \bibinfo {pages} {150502} (\bibinfo {year}
  {2017})}\BibitemShut {NoStop}%
\bibitem [{\citenamefont {Yan}\ \emph {et~al.}(2018)\citenamefont {Yan},
  \citenamefont {Krantz}, \citenamefont {Sung}, \citenamefont {Kjaergaard},
  \citenamefont {Campbell}, \citenamefont {Orlando}, \citenamefont
  {Gustavsson},\ and\ \citenamefont {Oliver}}]{Yan2018}%
  \BibitemOpen
  \bibfield  {author} {\bibinfo {author} {\bibfnamefont {Fei}\ \bibnamefont
  {Yan}}, \bibinfo {author} {\bibfnamefont {Philip}\ \bibnamefont {Krantz}},
  \bibinfo {author} {\bibfnamefont {Youngkyu}\ \bibnamefont {Sung}}, \bibinfo
  {author} {\bibfnamefont {Morten}\ \bibnamefont {Kjaergaard}}, \bibinfo
  {author} {\bibfnamefont {Daniel~L.}\ \bibnamefont {Campbell}}, \bibinfo
  {author} {\bibfnamefont {Terry~P.}\ \bibnamefont {Orlando}}, \bibinfo
  {author} {\bibfnamefont {Simon}\ \bibnamefont {Gustavsson}}, \ and\ \bibinfo
  {author} {\bibfnamefont {William~D.}\ \bibnamefont {Oliver}},\ }\bibfield
  {title} {\enquote {\bibinfo {title} {Tunable coupling scheme for implementing
  high-fidelity two-qubit gates},}\ }\href {\doibase
  10.1103/PhysRevApplied.10.054062} {\bibfield  {journal} {\bibinfo  {journal}
  {Phys. Rev. Applied}\ }\textbf {\bibinfo {volume} {10}},\ \bibinfo {pages}
  {054062} (\bibinfo {year} {2018})}\BibitemShut {NoStop}%
\bibitem [{\citenamefont {Abrams}\ \emph {et~al.}(2020)\citenamefont {Abrams},
  \citenamefont {Didier}, \citenamefont {Johnson}, \citenamefont {Silva},\ and\
  \citenamefont {Ryan}}]{Abrams2020}%
  \BibitemOpen
  \bibfield  {author} {\bibinfo {author} {\bibfnamefont {Deanna~M.}\
  \bibnamefont {Abrams}}, \bibinfo {author} {\bibfnamefont {Nicolas}\
  \bibnamefont {Didier}}, \bibinfo {author} {\bibfnamefont {Blake~R.}\
  \bibnamefont {Johnson}}, \bibinfo {author} {\bibfnamefont {Marcus P.~da}\
  \bibnamefont {Silva}}, \ and\ \bibinfo {author} {\bibfnamefont {Colm~A.}\
  \bibnamefont {Ryan}},\ }\bibfield  {title} {\enquote {\bibinfo {title}
  {Implementation of xy entangling gates with a single calibrated pulse},}\
  }\href {\doibase 10.1038/s41928-020-00498-1} {\bibfield  {journal} {\bibinfo
  {journal} {Nature Electronics}\ }\textbf {\bibinfo {volume} {3}},\ \bibinfo
  {pages} {744--750} (\bibinfo {year} {2020})}\BibitemShut {NoStop}%
\bibitem [{\citenamefont {Hong}\ \emph {et~al.}(2020)\citenamefont {Hong},
  \citenamefont {Papageorge}, \citenamefont {Sivarajah}, \citenamefont
  {Crossman}, \citenamefont {Didier}, \citenamefont {Polloreno}, \citenamefont
  {Sete}, \citenamefont {Turkowski}, \citenamefont {da~Silva},\ and\
  \citenamefont {Johnson}}]{Hong2020}%
  \BibitemOpen
  \bibfield  {author} {\bibinfo {author} {\bibfnamefont {Sabrina~S.}\
  \bibnamefont {Hong}}, \bibinfo {author} {\bibfnamefont {Alexander~T.}\
  \bibnamefont {Papageorge}}, \bibinfo {author} {\bibfnamefont {Prasahnt}\
  \bibnamefont {Sivarajah}}, \bibinfo {author} {\bibfnamefont {Genya}\
  \bibnamefont {Crossman}}, \bibinfo {author} {\bibfnamefont {Nicolas}\
  \bibnamefont {Didier}}, \bibinfo {author} {\bibfnamefont {Anthony~M.}\
  \bibnamefont {Polloreno}}, \bibinfo {author} {\bibfnamefont {Eyob~A.}\
  \bibnamefont {Sete}}, \bibinfo {author} {\bibfnamefont {Stefan~W.}\
  \bibnamefont {Turkowski}}, \bibinfo {author} {\bibfnamefont {Marcus~P.}\
  \bibnamefont {da~Silva}}, \ and\ \bibinfo {author} {\bibfnamefont {Blake~R.}\
  \bibnamefont {Johnson}},\ }\bibfield  {title} {\enquote {\bibinfo {title}
  {Demonstration of a parametrically activated entangling gate protected from
  flux noise},}\ }\href {\doibase 10.1103/PhysRevA.101.012302} {\bibfield
  {journal} {\bibinfo  {journal} {Phys. Rev. A}\ }\textbf {\bibinfo {volume}
  {101}},\ \bibinfo {pages} {012302} (\bibinfo {year} {2020})}\BibitemShut
  {NoStop}%
\bibitem [{\citenamefont {Foxen}\ \emph {et~al.}(2020)\citenamefont {Foxen},
  \citenamefont {Neill}, \citenamefont {Dunsworth}, \citenamefont {Roushan},
  \citenamefont {Chiaro}, \citenamefont {Megrant}, \citenamefont {Kelly},
  \citenamefont {Chen}, \citenamefont {Satzinger}, \citenamefont {Barends},
  \citenamefont {Arute}, \citenamefont {Arya}, \citenamefont {Babbush},
  \citenamefont {Bacon}, \citenamefont {Bardin}, \citenamefont {Boixo},
  \citenamefont {Buell}, \citenamefont {Burkett}, \citenamefont {Chen},
  \citenamefont {Collins}, \citenamefont {Farhi}, \citenamefont {Fowler},
  \citenamefont {Gidney}, \citenamefont {Giustina}, \citenamefont {Graff},
  \citenamefont {Harrigan}, \citenamefont {Huang}, \citenamefont {Isakov},
  \citenamefont {Jeffrey}, \citenamefont {Jiang}, \citenamefont {Kafri},
  \citenamefont {Kechedzhi}, \citenamefont {Klimov}, \citenamefont {Korotkov},
  \citenamefont {Kostritsa}, \citenamefont {Landhuis}, \citenamefont {Lucero},
  \citenamefont {McClean}, \citenamefont {McEwen}, \citenamefont {Mi},
  \citenamefont {Mohseni}, \citenamefont {Mutus}, \citenamefont {Naaman},
  \citenamefont {Neeley}, \citenamefont {Niu}, \citenamefont {Petukhov},
  \citenamefont {Quintana}, \citenamefont {Rubin}, \citenamefont {Sank},
  \citenamefont {Smelyanskiy}, \citenamefont {Vainsencher}, \citenamefont
  {White}, \citenamefont {Yao}, \citenamefont {Yeh}, \citenamefont {Zalcman},
  \citenamefont {Neven},\ and\ \citenamefont {Martinis}}]{Foxen2020}%
  \BibitemOpen
  \bibfield  {author} {\bibinfo {author} {\bibfnamefont {B.}~\bibnamefont
  {Foxen}}, \bibinfo {author} {\bibfnamefont {C.}~\bibnamefont {Neill}},
  \bibinfo {author} {\bibfnamefont {A.}~\bibnamefont {Dunsworth}}, \bibinfo
  {author} {\bibfnamefont {P.}~\bibnamefont {Roushan}}, \bibinfo {author}
  {\bibfnamefont {B.}~\bibnamefont {Chiaro}}, \bibinfo {author} {\bibfnamefont
  {A.}~\bibnamefont {Megrant}}, \bibinfo {author} {\bibfnamefont
  {J.}~\bibnamefont {Kelly}}, \bibinfo {author} {\bibfnamefont {Zijun}\
  \bibnamefont {Chen}}, \bibinfo {author} {\bibfnamefont {K.}~\bibnamefont
  {Satzinger}}, \bibinfo {author} {\bibfnamefont {R.}~\bibnamefont {Barends}},
  \bibinfo {author} {\bibfnamefont {F.}~\bibnamefont {Arute}}, \bibinfo
  {author} {\bibfnamefont {K.}~\bibnamefont {Arya}}, \bibinfo {author}
  {\bibfnamefont {R.}~\bibnamefont {Babbush}}, \bibinfo {author} {\bibfnamefont
  {D.}~\bibnamefont {Bacon}}, \bibinfo {author} {\bibfnamefont {J.~C.}\
  \bibnamefont {Bardin}}, \bibinfo {author} {\bibfnamefont {S.}~\bibnamefont
  {Boixo}}, \bibinfo {author} {\bibfnamefont {D.}~\bibnamefont {Buell}},
  \bibinfo {author} {\bibfnamefont {B.}~\bibnamefont {Burkett}}, \bibinfo
  {author} {\bibfnamefont {Yu}~\bibnamefont {Chen}}, \bibinfo {author}
  {\bibfnamefont {R.}~\bibnamefont {Collins}}, \bibinfo {author} {\bibfnamefont
  {E.}~\bibnamefont {Farhi}}, \bibinfo {author} {\bibfnamefont
  {A.}~\bibnamefont {Fowler}}, \bibinfo {author} {\bibfnamefont
  {C.}~\bibnamefont {Gidney}}, \bibinfo {author} {\bibfnamefont
  {M.}~\bibnamefont {Giustina}}, \bibinfo {author} {\bibfnamefont
  {R.}~\bibnamefont {Graff}}, \bibinfo {author} {\bibfnamefont
  {M.}~\bibnamefont {Harrigan}}, \bibinfo {author} {\bibfnamefont
  {T.}~\bibnamefont {Huang}}, \bibinfo {author} {\bibfnamefont {S.~V.}\
  \bibnamefont {Isakov}}, \bibinfo {author} {\bibfnamefont {E.}~\bibnamefont
  {Jeffrey}}, \bibinfo {author} {\bibfnamefont {Z.}~\bibnamefont {Jiang}},
  \bibinfo {author} {\bibfnamefont {D.}~\bibnamefont {Kafri}}, \bibinfo
  {author} {\bibfnamefont {K.}~\bibnamefont {Kechedzhi}}, \bibinfo {author}
  {\bibfnamefont {P.}~\bibnamefont {Klimov}}, \bibinfo {author} {\bibfnamefont
  {A.}~\bibnamefont {Korotkov}}, \bibinfo {author} {\bibfnamefont
  {F.}~\bibnamefont {Kostritsa}}, \bibinfo {author} {\bibfnamefont
  {D.}~\bibnamefont {Landhuis}}, \bibinfo {author} {\bibfnamefont
  {E.}~\bibnamefont {Lucero}}, \bibinfo {author} {\bibfnamefont
  {J.}~\bibnamefont {McClean}}, \bibinfo {author} {\bibfnamefont
  {M.}~\bibnamefont {McEwen}}, \bibinfo {author} {\bibfnamefont
  {X.}~\bibnamefont {Mi}}, \bibinfo {author} {\bibfnamefont {M.}~\bibnamefont
  {Mohseni}}, \bibinfo {author} {\bibfnamefont {J.~Y.}\ \bibnamefont {Mutus}},
  \bibinfo {author} {\bibfnamefont {O.}~\bibnamefont {Naaman}}, \bibinfo
  {author} {\bibfnamefont {M.}~\bibnamefont {Neeley}}, \bibinfo {author}
  {\bibfnamefont {M.}~\bibnamefont {Niu}}, \bibinfo {author} {\bibfnamefont
  {A.}~\bibnamefont {Petukhov}}, \bibinfo {author} {\bibfnamefont
  {C.}~\bibnamefont {Quintana}}, \bibinfo {author} {\bibfnamefont
  {N.}~\bibnamefont {Rubin}}, \bibinfo {author} {\bibfnamefont
  {D.}~\bibnamefont {Sank}}, \bibinfo {author} {\bibfnamefont {V.}~\bibnamefont
  {Smelyanskiy}}, \bibinfo {author} {\bibfnamefont {A.}~\bibnamefont
  {Vainsencher}}, \bibinfo {author} {\bibfnamefont {T.~C.}\ \bibnamefont
  {White}}, \bibinfo {author} {\bibfnamefont {Z.}~\bibnamefont {Yao}}, \bibinfo
  {author} {\bibfnamefont {P.}~\bibnamefont {Yeh}}, \bibinfo {author}
  {\bibfnamefont {A.}~\bibnamefont {Zalcman}}, \bibinfo {author} {\bibfnamefont
  {H.}~\bibnamefont {Neven}}, \ and\ \bibinfo {author} {\bibfnamefont {J.~M.}\
  \bibnamefont {Martinis}} (\bibinfo {collaboration} {Google AI Quantum}),\
  }\bibfield  {title} {\enquote {\bibinfo {title} {Demonstrating a continuous
  set of two-qubit gates for near-term quantum algorithms},}\ }\href {\doibase
  10.1103/PhysRevLett.125.120504} {\bibfield  {journal} {\bibinfo  {journal}
  {Phys. Rev. Lett.}\ }\textbf {\bibinfo {volume} {125}},\ \bibinfo {pages}
  {120504} (\bibinfo {year} {2020})}\BibitemShut {NoStop}%
\bibitem [{\citenamefont {Sung}\ \emph {et~al.}(2021)\citenamefont {Sung},
  \citenamefont {Ding}, \citenamefont {Braum\"uller}, \citenamefont
  {Veps\"al\"ainen}, \citenamefont {Kannan}, \citenamefont {Kjaergaard},
  \citenamefont {Greene}, \citenamefont {Samach}, \citenamefont {McNally},
  \citenamefont {Kim}, \citenamefont {Melville}, \citenamefont {Niedzielski},
  \citenamefont {Schwartz}, \citenamefont {Yoder}, \citenamefont {Orlando},
  \citenamefont {Gustavsson},\ and\ \citenamefont {Oliver}}]{Sung2021}%
  \BibitemOpen
  \bibfield  {author} {\bibinfo {author} {\bibfnamefont {Youngkyu}\
  \bibnamefont {Sung}}, \bibinfo {author} {\bibfnamefont {Leon}\ \bibnamefont
  {Ding}}, \bibinfo {author} {\bibfnamefont {Jochen}\ \bibnamefont
  {Braum\"uller}}, \bibinfo {author} {\bibfnamefont {Antti}\ \bibnamefont
  {Veps\"al\"ainen}}, \bibinfo {author} {\bibfnamefont {Bharath}\ \bibnamefont
  {Kannan}}, \bibinfo {author} {\bibfnamefont {Morten}\ \bibnamefont
  {Kjaergaard}}, \bibinfo {author} {\bibfnamefont {Ami}\ \bibnamefont
  {Greene}}, \bibinfo {author} {\bibfnamefont {Gabriel~O.}\ \bibnamefont
  {Samach}}, \bibinfo {author} {\bibfnamefont {Chris}\ \bibnamefont {McNally}},
  \bibinfo {author} {\bibfnamefont {David}\ \bibnamefont {Kim}}, \bibinfo
  {author} {\bibfnamefont {Alexander}\ \bibnamefont {Melville}}, \bibinfo
  {author} {\bibfnamefont {Bethany~M.}\ \bibnamefont {Niedzielski}}, \bibinfo
  {author} {\bibfnamefont {Mollie~E.}\ \bibnamefont {Schwartz}}, \bibinfo
  {author} {\bibfnamefont {Jonilyn~L.}\ \bibnamefont {Yoder}}, \bibinfo
  {author} {\bibfnamefont {Terry~P.}\ \bibnamefont {Orlando}}, \bibinfo
  {author} {\bibfnamefont {Simon}\ \bibnamefont {Gustavsson}}, \ and\ \bibinfo
  {author} {\bibfnamefont {William~D.}\ \bibnamefont {Oliver}},\ }\bibfield
  {title} {\enquote {\bibinfo {title} {Realization of high-fidelity cz and
  $zz$-free iswap gates with a tunable coupler},}\ }\href {\doibase
  10.1103/PhysRevX.11.021058} {\bibfield  {journal} {\bibinfo  {journal} {Phys.
  Rev. X}\ }\textbf {\bibinfo {volume} {11}},\ \bibinfo {pages} {021058}
  (\bibinfo {year} {2021})}\BibitemShut {NoStop}%
\bibitem [{\citenamefont {Sete}\ \emph {et~al.}(2021)\citenamefont {Sete},
  \citenamefont {Didier}, \citenamefont {Chen}, \citenamefont {Kulshreshtha},
  \citenamefont {Manenti},\ and\ \citenamefont {Poletto}}]{Sete2021}%
  \BibitemOpen
  \bibfield  {author} {\bibinfo {author} {\bibfnamefont {Eyob~A.}\ \bibnamefont
  {Sete}}, \bibinfo {author} {\bibfnamefont {Nicolas}\ \bibnamefont {Didier}},
  \bibinfo {author} {\bibfnamefont {Angela~Q.}\ \bibnamefont {Chen}}, \bibinfo
  {author} {\bibfnamefont {Shobhan}\ \bibnamefont {Kulshreshtha}}, \bibinfo
  {author} {\bibfnamefont {Riccardo}\ \bibnamefont {Manenti}}, \ and\ \bibinfo
  {author} {\bibfnamefont {Stefano}\ \bibnamefont {Poletto}},\ }\bibfield
  {title} {\enquote {\bibinfo {title} {Parametric-resonance entangling gates
  with a tunable coupler},}\ }\href {\doibase 10.1103/PhysRevApplied.16.024050}
  {\bibfield  {journal} {\bibinfo  {journal} {Phys. Rev. Applied}\ }\textbf
  {\bibinfo {volume} {16}},\ \bibinfo {pages} {024050} (\bibinfo {year}
  {2021})}\BibitemShut {NoStop}%
\bibitem [{\citenamefont {Krastanov}\ \emph {et~al.}(2019)\citenamefont
  {Krastanov}, \citenamefont {Zhou}, \citenamefont {Flammia},\ and\
  \citenamefont {Jiang}}]{Krastanov2019}%
  \BibitemOpen
  \bibfield  {author} {\bibinfo {author} {\bibfnamefont {Stefan}\ \bibnamefont
  {Krastanov}}, \bibinfo {author} {\bibfnamefont {Sisi}\ \bibnamefont {Zhou}},
  \bibinfo {author} {\bibfnamefont {Steven~T}\ \bibnamefont {Flammia}}, \ and\
  \bibinfo {author} {\bibfnamefont {Liang}\ \bibnamefont {Jiang}},\ }\bibfield
  {title} {\enquote {\bibinfo {title} {Stochastic estimation of dynamical
  variables},}\ }\href {\doibase 10.1088/2058-9565/ab18d5} {\bibfield
  {journal} {\bibinfo  {journal} {Quantum Science and Technology}\ }\textbf
  {\bibinfo {volume} {4}},\ \bibinfo {pages} {035003} (\bibinfo {year}
  {2019})}\BibitemShut {NoStop}%
\bibitem [{\citenamefont {Barchielli}\ and\ \citenamefont
  {Gregoratti}(2009)}]{Barchielli2009}%
  \BibitemOpen
  \bibfield  {author} {\bibinfo {author} {\bibfnamefont {Alberto}\ \bibnamefont
  {Barchielli}}\ and\ \bibinfo {author} {\bibfnamefont {Matteo}\ \bibnamefont
  {Gregoratti}},\ }\href@noop {} {\emph {\bibinfo {title} {Quantum Trajectories
  and Measurements in Continuous Time: The Diffusive Case}}},\ \bibinfo
  {series} {Lecture Notes in Physics}, Vol.\ \bibinfo {volume} {782}\ (\bibinfo
   {publisher} {Springer},\ \bibinfo {address} {Berlin},\ \bibinfo {year}
  {2009})\BibitemShut {NoStop}%
\bibitem [{\citenamefont {Vijay}\ \emph {et~al.}(2011)\citenamefont {Vijay},
  \citenamefont {Slichter},\ and\ \citenamefont {Siddiqi}}]{Vijay2011}%
  \BibitemOpen
  \bibfield  {author} {\bibinfo {author} {\bibfnamefont {R.}~\bibnamefont
  {Vijay}}, \bibinfo {author} {\bibfnamefont {D.~H.}\ \bibnamefont {Slichter}},
  \ and\ \bibinfo {author} {\bibfnamefont {I.}~\bibnamefont {Siddiqi}},\
  }\bibfield  {title} {\enquote {\bibinfo {title} {Observation of quantum jumps
  in a superconducting artificial atom},}\ }\href {\doibase
  10.1103/PhysRevLett.106.110502} {\bibfield  {journal} {\bibinfo  {journal}
  {Phys. Rev. Lett.}\ }\textbf {\bibinfo {volume} {106}},\ \bibinfo {pages}
  {110502} (\bibinfo {year} {2011})}\BibitemShut {NoStop}%
\bibitem [{\citenamefont {Vijay}\ \emph {et~al.}(2012)\citenamefont {Vijay},
  \citenamefont {Macklin}, \citenamefont {Slichter}, \citenamefont {Weber},
  \citenamefont {Murch}, \citenamefont {Naik}, \citenamefont {Korotkov},\ and\
  \citenamefont {Siddiqi}}]{Vijay2012}%
  \BibitemOpen
  \bibfield  {author} {\bibinfo {author} {\bibfnamefont {R.}~\bibnamefont
  {Vijay}}, \bibinfo {author} {\bibfnamefont {C.}~\bibnamefont {Macklin}},
  \bibinfo {author} {\bibfnamefont {D.~H.}\ \bibnamefont {Slichter}}, \bibinfo
  {author} {\bibfnamefont {S.~J.}\ \bibnamefont {Weber}}, \bibinfo {author}
  {\bibfnamefont {K.~W.}\ \bibnamefont {Murch}}, \bibinfo {author}
  {\bibfnamefont {R.}~\bibnamefont {Naik}}, \bibinfo {author} {\bibfnamefont
  {A.~N.}\ \bibnamefont {Korotkov}}, \ and\ \bibinfo {author} {\bibfnamefont
  {I.}~\bibnamefont {Siddiqi}},\ }\bibfield  {title} {\enquote {\bibinfo
  {title} {Stabilizing rabi oscillations in a superconducting qubit using
  quantum feedback},}\ }\href {\doibase 10.1038/nature11505} {\bibfield
  {journal} {\bibinfo  {journal} {Nature}\ }\textbf {\bibinfo {volume} {490}},\
  \bibinfo {pages} {77--80} (\bibinfo {year} {2012})}\BibitemShut {NoStop}%
\bibitem [{\citenamefont {Hatridge}\ \emph {et~al.}(2013)\citenamefont
  {Hatridge}, \citenamefont {Shankar}, \citenamefont {Mirrahimi}, \citenamefont
  {Schackert}, \citenamefont {Geerlings}, \citenamefont {Brecht}, \citenamefont
  {Sliwa}, \citenamefont {Abdo}, \citenamefont {Frunzio}, \citenamefont
  {Girvin}, \citenamefont {Schoelkopf},\ and\ \citenamefont
  {Devoret}}]{Hatridge2013}%
  \BibitemOpen
  \bibfield  {author} {\bibinfo {author} {\bibfnamefont {M.}~\bibnamefont
  {Hatridge}}, \bibinfo {author} {\bibfnamefont {S.}~\bibnamefont {Shankar}},
  \bibinfo {author} {\bibfnamefont {M.}~\bibnamefont {Mirrahimi}}, \bibinfo
  {author} {\bibfnamefont {F.}~\bibnamefont {Schackert}}, \bibinfo {author}
  {\bibfnamefont {K.}~\bibnamefont {Geerlings}}, \bibinfo {author}
  {\bibfnamefont {T.}~\bibnamefont {Brecht}}, \bibinfo {author} {\bibfnamefont
  {K.~M.}\ \bibnamefont {Sliwa}}, \bibinfo {author} {\bibfnamefont
  {B.}~\bibnamefont {Abdo}}, \bibinfo {author} {\bibfnamefont {L.}~\bibnamefont
  {Frunzio}}, \bibinfo {author} {\bibfnamefont {S.~M.}\ \bibnamefont {Girvin}},
  \bibinfo {author} {\bibfnamefont {R.~J.}\ \bibnamefont {Schoelkopf}}, \ and\
  \bibinfo {author} {\bibfnamefont {M.~H.}\ \bibnamefont {Devoret}},\
  }\bibfield  {title} {\enquote {\bibinfo {title} {Quantum back-action of an
  individual variable-strength measurement},}\ }\href {\doibase
  10.1126/science.1226897} {\bibfield  {journal} {\bibinfo  {journal}
  {Science}\ }\textbf {\bibinfo {volume} {339}},\ \bibinfo {pages} {178--181}
  (\bibinfo {year} {2013})}\BibitemShut {NoStop}%
\bibitem [{\citenamefont {Roch}\ \emph {et~al.}(2014)\citenamefont {Roch},
  \citenamefont {Schwartz}, \citenamefont {Motzoi}, \citenamefont {Macklin},
  \citenamefont {Vijay}, \citenamefont {Eddins}, \citenamefont {Korotkov},
  \citenamefont {Whaley}, \citenamefont {Sarovar},\ and\ \citenamefont
  {Siddiqi}}]{Roch2014}%
  \BibitemOpen
  \bibfield  {author} {\bibinfo {author} {\bibfnamefont {N.}~\bibnamefont
  {Roch}}, \bibinfo {author} {\bibfnamefont {M.~E.}\ \bibnamefont {Schwartz}},
  \bibinfo {author} {\bibfnamefont {F.}~\bibnamefont {Motzoi}}, \bibinfo
  {author} {\bibfnamefont {C.}~\bibnamefont {Macklin}}, \bibinfo {author}
  {\bibfnamefont {R.}~\bibnamefont {Vijay}}, \bibinfo {author} {\bibfnamefont
  {A.~W.}\ \bibnamefont {Eddins}}, \bibinfo {author} {\bibfnamefont {A.~N.}\
  \bibnamefont {Korotkov}}, \bibinfo {author} {\bibfnamefont {K.~B.}\
  \bibnamefont {Whaley}}, \bibinfo {author} {\bibfnamefont {M.}~\bibnamefont
  {Sarovar}}, \ and\ \bibinfo {author} {\bibfnamefont {I.}~\bibnamefont
  {Siddiqi}},\ }\bibfield  {title} {\enquote {\bibinfo {title} {Observation of
  measurement-induced entanglement and quantum trajectories of remote
  superconducting qubits},}\ }\href {\doibase 10.1103/PhysRevLett.112.170501}
  {\bibfield  {journal} {\bibinfo  {journal} {Phys. Rev. Lett.}\ }\textbf
  {\bibinfo {volume} {112}},\ \bibinfo {pages} {170501} (\bibinfo {year}
  {2014})}\BibitemShut {NoStop}%
\bibitem [{\citenamefont {Sun}\ \emph {et~al.}(2014)\citenamefont {Sun},
  \citenamefont {Petrenko}, \citenamefont {Leghtas}, \citenamefont {Vlastakis},
  \citenamefont {Kirchmair}, \citenamefont {Sliwa}, \citenamefont {Narla},
  \citenamefont {Hatridge}, \citenamefont {Shankar}, \citenamefont {Blumoff},
  \citenamefont {Frunzio}, \citenamefont {Mirrahimi}, \citenamefont {Devoret},\
  and\ \citenamefont {Schoelkopf}}]{Sun2014}%
  \BibitemOpen
  \bibfield  {author} {\bibinfo {author} {\bibfnamefont {L.}~\bibnamefont
  {Sun}}, \bibinfo {author} {\bibfnamefont {A.}~\bibnamefont {Petrenko}},
  \bibinfo {author} {\bibfnamefont {Z.}~\bibnamefont {Leghtas}}, \bibinfo
  {author} {\bibfnamefont {B.}~\bibnamefont {Vlastakis}}, \bibinfo {author}
  {\bibfnamefont {G.}~\bibnamefont {Kirchmair}}, \bibinfo {author}
  {\bibfnamefont {K.~M.}\ \bibnamefont {Sliwa}}, \bibinfo {author}
  {\bibfnamefont {A.}~\bibnamefont {Narla}}, \bibinfo {author} {\bibfnamefont
  {M.}~\bibnamefont {Hatridge}}, \bibinfo {author} {\bibfnamefont
  {S.}~\bibnamefont {Shankar}}, \bibinfo {author} {\bibfnamefont
  {J.}~\bibnamefont {Blumoff}}, \bibinfo {author} {\bibfnamefont
  {L.}~\bibnamefont {Frunzio}}, \bibinfo {author} {\bibfnamefont
  {M.}~\bibnamefont {Mirrahimi}}, \bibinfo {author} {\bibfnamefont {M.~H.}\
  \bibnamefont {Devoret}}, \ and\ \bibinfo {author} {\bibfnamefont {R.~J.}\
  \bibnamefont {Schoelkopf}},\ }\bibfield  {title} {\enquote {\bibinfo {title}
  {Tracking photon jumps with repeated quantum non-demolition parity
  measurements},}\ }\href {\doibase 10.1038/nature13436} {\bibfield  {journal}
  {\bibinfo  {journal} {Nature}\ }\textbf {\bibinfo {volume} {511}},\ \bibinfo
  {pages} {444--448} (\bibinfo {year} {2014})}\BibitemShut {NoStop}%
\bibitem [{\citenamefont {Chantasri}\ \emph {et~al.}(2016)\citenamefont
  {Chantasri}, \citenamefont {Kimchi-Schwartz}, \citenamefont {Roch},
  \citenamefont {Siddiqi},\ and\ \citenamefont {Jordan}}]{Chantasri2016}%
  \BibitemOpen
  \bibfield  {author} {\bibinfo {author} {\bibfnamefont {Areeya}\ \bibnamefont
  {Chantasri}}, \bibinfo {author} {\bibfnamefont {Mollie~E.}\ \bibnamefont
  {Kimchi-Schwartz}}, \bibinfo {author} {\bibfnamefont {Nicolas}\ \bibnamefont
  {Roch}}, \bibinfo {author} {\bibfnamefont {Irfan}\ \bibnamefont {Siddiqi}}, \
  and\ \bibinfo {author} {\bibfnamefont {Andrew~N.}\ \bibnamefont {Jordan}},\
  }\bibfield  {title} {\enquote {\bibinfo {title} {Quantum trajectories and
  their statistics for remotely entangled quantum bits},}\ }\href {\doibase
  10.1103/PhysRevX.6.041052} {\bibfield  {journal} {\bibinfo  {journal} {Phys.
  Rev. X}\ }\textbf {\bibinfo {volume} {6}},\ \bibinfo {pages} {041052}
  (\bibinfo {year} {2016})}\BibitemShut {NoStop}%
\bibitem [{\citenamefont {Hacohen-Gourgy}\ \emph {et~al.}(2016)\citenamefont
  {Hacohen-Gourgy}, \citenamefont {Martin}, \citenamefont {Flurin},
  \citenamefont {Ramasesh}, \citenamefont {Whaley},\ and\ \citenamefont
  {Siddiqi}}]{Hacohen-Gourgy2016}%
  \BibitemOpen
  \bibfield  {author} {\bibinfo {author} {\bibfnamefont {Shay}\ \bibnamefont
  {Hacohen-Gourgy}}, \bibinfo {author} {\bibfnamefont {Leigh~S.}\ \bibnamefont
  {Martin}}, \bibinfo {author} {\bibfnamefont {Emmanuel}\ \bibnamefont
  {Flurin}}, \bibinfo {author} {\bibfnamefont {Vinay~V.}\ \bibnamefont
  {Ramasesh}}, \bibinfo {author} {\bibfnamefont {K.~Birgitta}\ \bibnamefont
  {Whaley}}, \ and\ \bibinfo {author} {\bibfnamefont {Irfan}\ \bibnamefont
  {Siddiqi}},\ }\bibfield  {title} {\enquote {\bibinfo {title} {Quantum
  dynamics of simultaneously measured non-commuting observables},}\ }\href
  {\doibase 10.1038/nature19762} {\bibfield  {journal} {\bibinfo  {journal}
  {Nature}\ }\textbf {\bibinfo {volume} {538}},\ \bibinfo {pages} {491--494}
  (\bibinfo {year} {2016})}\BibitemShut {NoStop}%
\bibitem [{\citenamefont {Campagne-Ibarcq}\ \emph {et~al.}(2016)\citenamefont
  {Campagne-Ibarcq}, \citenamefont {Six}, \citenamefont {Bretheau},
  \citenamefont {Sarlette}, \citenamefont {Mirrahimi}, \citenamefont
  {Rouchon},\ and\ \citenamefont {Huard}}]{Campagne-Ibarcq2016}%
  \BibitemOpen
  \bibfield  {author} {\bibinfo {author} {\bibfnamefont {P.}~\bibnamefont
  {Campagne-Ibarcq}}, \bibinfo {author} {\bibfnamefont {P.}~\bibnamefont
  {Six}}, \bibinfo {author} {\bibfnamefont {L.}~\bibnamefont {Bretheau}},
  \bibinfo {author} {\bibfnamefont {A.}~\bibnamefont {Sarlette}}, \bibinfo
  {author} {\bibfnamefont {M.}~\bibnamefont {Mirrahimi}}, \bibinfo {author}
  {\bibfnamefont {P.}~\bibnamefont {Rouchon}}, \ and\ \bibinfo {author}
  {\bibfnamefont {B.}~\bibnamefont {Huard}},\ }\bibfield  {title} {\enquote
  {\bibinfo {title} {Observing quantum state diffusion by heterodyne detection
  of fluorescence},}\ }\href {\doibase 10.1103/PhysRevX.6.011002} {\bibfield
  {journal} {\bibinfo  {journal} {Phys. Rev. X}\ }\textbf {\bibinfo {volume}
  {6}},\ \bibinfo {pages} {011002} (\bibinfo {year} {2016})}\BibitemShut
  {NoStop}%
\bibitem [{\citenamefont {Vool}\ \emph {et~al.}(2016)\citenamefont {Vool},
  \citenamefont {Shankar}, \citenamefont {Mundhada}, \citenamefont {Ofek},
  \citenamefont {Narla}, \citenamefont {Sliwa}, \citenamefont {Zalys-Geller},
  \citenamefont {Liu}, \citenamefont {Frunzio}, \citenamefont {Schoelkopf},
  \citenamefont {Girvin},\ and\ \citenamefont {Devoret}}]{Vool2016}%
  \BibitemOpen
  \bibfield  {author} {\bibinfo {author} {\bibfnamefont {U.}~\bibnamefont
  {Vool}}, \bibinfo {author} {\bibfnamefont {S.}~\bibnamefont {Shankar}},
  \bibinfo {author} {\bibfnamefont {S.~O.}\ \bibnamefont {Mundhada}}, \bibinfo
  {author} {\bibfnamefont {N.}~\bibnamefont {Ofek}}, \bibinfo {author}
  {\bibfnamefont {A.}~\bibnamefont {Narla}}, \bibinfo {author} {\bibfnamefont
  {K.}~\bibnamefont {Sliwa}}, \bibinfo {author} {\bibfnamefont
  {E.}~\bibnamefont {Zalys-Geller}}, \bibinfo {author} {\bibfnamefont
  {Y.}~\bibnamefont {Liu}}, \bibinfo {author} {\bibfnamefont {L.}~\bibnamefont
  {Frunzio}}, \bibinfo {author} {\bibfnamefont {R.~J.}\ \bibnamefont
  {Schoelkopf}}, \bibinfo {author} {\bibfnamefont {S.~M.}\ \bibnamefont
  {Girvin}}, \ and\ \bibinfo {author} {\bibfnamefont {M.~H.}\ \bibnamefont
  {Devoret}},\ }\bibfield  {title} {\enquote {\bibinfo {title} {Continuous
  quantum nondemolition measurement of the transverse component of a qubit},}\
  }\href {\doibase 10.1103/PhysRevLett.117.133601} {\bibfield  {journal}
  {\bibinfo  {journal} {Phys. Rev. Lett.}\ }\textbf {\bibinfo {volume} {117}},\
  \bibinfo {pages} {133601} (\bibinfo {year} {2016})}\BibitemShut {NoStop}%
\bibitem [{\citenamefont {Weber}\ \emph {et~al.}(2016)\citenamefont {Weber},
  \citenamefont {Murch}, \citenamefont {Kimchi-Schwartz}, \citenamefont
  {Roch},\ and\ \citenamefont {Siddiqi}}]{Weber2016}%
  \BibitemOpen
  \bibfield  {author} {\bibinfo {author} {\bibfnamefont {Steven~J.}\
  \bibnamefont {Weber}}, \bibinfo {author} {\bibfnamefont {Kater~W.}\
  \bibnamefont {Murch}}, \bibinfo {author} {\bibfnamefont {Mollie~E.}\
  \bibnamefont {Kimchi-Schwartz}}, \bibinfo {author} {\bibfnamefont {Nicolas}\
  \bibnamefont {Roch}}, \ and\ \bibinfo {author} {\bibfnamefont {Irfan}\
  \bibnamefont {Siddiqi}},\ }\bibfield  {title} {\enquote {\bibinfo {title}
  {Quantum trajectories of superconducting qubits},}\ }\href {\doibase
  https://doi.org/10.1016/j.crhy.2016.07.007} {\bibfield  {journal} {\bibinfo
  {journal} {Comptes Rendus Physique}\ }\textbf {\bibinfo {volume} {17}},\
  \bibinfo {pages} {766--777} (\bibinfo {year} {2016})},\ \bibinfo {note}
  {quantum microwaves / Micro-ondes quantiques}\BibitemShut {NoStop}%
\bibitem [{\citenamefont {Ficheux}\ \emph {et~al.}(2018)\citenamefont
  {Ficheux}, \citenamefont {Jezouin}, \citenamefont {Leghtas},\ and\
  \citenamefont {Huard}}]{Ficheux2018}%
  \BibitemOpen
  \bibfield  {author} {\bibinfo {author} {\bibfnamefont {Q.}~\bibnamefont
  {Ficheux}}, \bibinfo {author} {\bibfnamefont {S.}~\bibnamefont {Jezouin}},
  \bibinfo {author} {\bibfnamefont {Z.}~\bibnamefont {Leghtas}}, \ and\
  \bibinfo {author} {\bibfnamefont {B.}~\bibnamefont {Huard}},\ }\bibfield
  {title} {\enquote {\bibinfo {title} {Dynamics of a qubit while simultaneously
  monitoring its relaxation and dephasing},}\ }\href {\doibase
  10.1038/s41467-018-04372-9} {\bibfield  {journal} {\bibinfo  {journal}
  {Nature Communications}\ }\textbf {\bibinfo {volume} {9}},\ \bibinfo {pages}
  {1926} (\bibinfo {year} {2018})}\BibitemShut {NoStop}%
\bibitem [{\citenamefont {Steinmetz}\ \emph {et~al.}(2022)\citenamefont
  {Steinmetz}, \citenamefont {Das}, \citenamefont {Siddiqi},\ and\
  \citenamefont {Jordan}}]{Steinmetz2022}%
  \BibitemOpen
  \bibfield  {author} {\bibinfo {author} {\bibfnamefont {John}\ \bibnamefont
  {Steinmetz}}, \bibinfo {author} {\bibfnamefont {Debmalya}\ \bibnamefont
  {Das}}, \bibinfo {author} {\bibfnamefont {Irfan}\ \bibnamefont {Siddiqi}}, \
  and\ \bibinfo {author} {\bibfnamefont {Andrew~N.}\ \bibnamefont {Jordan}},\
  }\bibfield  {title} {\enquote {\bibinfo {title} {Continuous measurement of a
  qudit using dispersively coupled radiation},}\ }\href {\doibase
  10.1103/PhysRevA.105.052229} {\bibfield  {journal} {\bibinfo  {journal}
  {Phys. Rev. A}\ }\textbf {\bibinfo {volume} {105}},\ \bibinfo {pages}
  {052229} (\bibinfo {year} {2022})}\BibitemShut {NoStop}%
\bibitem [{\citenamefont {Koch}\ \emph {et~al.}(2007)\citenamefont {Koch},
  \citenamefont {Yu}, \citenamefont {Gambetta}, \citenamefont {Houck},
  \citenamefont {Schuster}, \citenamefont {Majer}, \citenamefont {Blais},
  \citenamefont {Devoret}, \citenamefont {Girvin},\ and\ \citenamefont
  {Schoelkopf}}]{Koch2007}%
  \BibitemOpen
  \bibfield  {author} {\bibinfo {author} {\bibfnamefont {Jens}\ \bibnamefont
  {Koch}}, \bibinfo {author} {\bibfnamefont {Terri~M.}\ \bibnamefont {Yu}},
  \bibinfo {author} {\bibfnamefont {Jay}\ \bibnamefont {Gambetta}}, \bibinfo
  {author} {\bibfnamefont {A.~A.}\ \bibnamefont {Houck}}, \bibinfo {author}
  {\bibfnamefont {D.~I.}\ \bibnamefont {Schuster}}, \bibinfo {author}
  {\bibfnamefont {J.}~\bibnamefont {Majer}}, \bibinfo {author} {\bibfnamefont
  {Alexandre}\ \bibnamefont {Blais}}, \bibinfo {author} {\bibfnamefont {M.~H.}\
  \bibnamefont {Devoret}}, \bibinfo {author} {\bibfnamefont {S.~M.}\
  \bibnamefont {Girvin}}, \ and\ \bibinfo {author} {\bibfnamefont {R.~J.}\
  \bibnamefont {Schoelkopf}},\ }\bibfield  {title} {\enquote {\bibinfo {title}
  {Charge-insensitive qubit design derived from the cooper pair box},}\ }\href
  {\doibase 10.1103/PhysRevA.76.042319} {\bibfield  {journal} {\bibinfo
  {journal} {Phys. Rev. A}\ }\textbf {\bibinfo {volume} {76}},\ \bibinfo
  {pages} {042319} (\bibinfo {year} {2007})}\BibitemShut {NoStop}%
\bibitem [{\citenamefont {Jeffrey}\ \emph {et~al.}(2014)\citenamefont
  {Jeffrey}, \citenamefont {Sank}, \citenamefont {Mutus}, \citenamefont
  {White}, \citenamefont {Kelly}, \citenamefont {Barends}, \citenamefont
  {Chen}, \citenamefont {Chen}, \citenamefont {Chiaro}, \citenamefont
  {Dunsworth}, \citenamefont {Megrant}, \citenamefont {O'Malley}, \citenamefont
  {Neill}, \citenamefont {Roushan}, \citenamefont {Vainsencher}, \citenamefont
  {Wenner}, \citenamefont {Cleland},\ and\ \citenamefont
  {Martinis}}]{Jeffrey2014}%
  \BibitemOpen
  \bibfield  {author} {\bibinfo {author} {\bibfnamefont {Evan}\ \bibnamefont
  {Jeffrey}}, \bibinfo {author} {\bibfnamefont {Daniel}\ \bibnamefont {Sank}},
  \bibinfo {author} {\bibfnamefont {J.~Y.}\ \bibnamefont {Mutus}}, \bibinfo
  {author} {\bibfnamefont {T.~C.}\ \bibnamefont {White}}, \bibinfo {author}
  {\bibfnamefont {J.}~\bibnamefont {Kelly}}, \bibinfo {author} {\bibfnamefont
  {R.}~\bibnamefont {Barends}}, \bibinfo {author} {\bibfnamefont
  {Y.}~\bibnamefont {Chen}}, \bibinfo {author} {\bibfnamefont {Z.}~\bibnamefont
  {Chen}}, \bibinfo {author} {\bibfnamefont {B.}~\bibnamefont {Chiaro}},
  \bibinfo {author} {\bibfnamefont {A.}~\bibnamefont {Dunsworth}}, \bibinfo
  {author} {\bibfnamefont {A.}~\bibnamefont {Megrant}}, \bibinfo {author}
  {\bibfnamefont {P.~J.~J.}\ \bibnamefont {O'Malley}}, \bibinfo {author}
  {\bibfnamefont {C.}~\bibnamefont {Neill}}, \bibinfo {author} {\bibfnamefont
  {P.}~\bibnamefont {Roushan}}, \bibinfo {author} {\bibfnamefont
  {A.}~\bibnamefont {Vainsencher}}, \bibinfo {author} {\bibfnamefont
  {J.}~\bibnamefont {Wenner}}, \bibinfo {author} {\bibfnamefont {A.~N.}\
  \bibnamefont {Cleland}}, \ and\ \bibinfo {author} {\bibfnamefont {John~M.}\
  \bibnamefont {Martinis}},\ }\bibfield  {title} {\enquote {\bibinfo {title}
  {Fast accurate state measurement with superconducting qubits},}\ }\href
  {\doibase 10.1103/PhysRevLett.112.190504} {\bibfield  {journal} {\bibinfo
  {journal} {Phys. Rev. Lett.}\ }\textbf {\bibinfo {volume} {112}},\ \bibinfo
  {pages} {190504} (\bibinfo {year} {2014})}\BibitemShut {NoStop}%
\bibitem [{\citenamefont {Sete}\ \emph {et~al.}(2015)\citenamefont {Sete},
  \citenamefont {Martinis},\ and\ \citenamefont {Korotkov}}]{Sete2015}%
  \BibitemOpen
  \bibfield  {author} {\bibinfo {author} {\bibfnamefont {Eyob~A.}\ \bibnamefont
  {Sete}}, \bibinfo {author} {\bibfnamefont {John~M.}\ \bibnamefont
  {Martinis}}, \ and\ \bibinfo {author} {\bibfnamefont {Alexander~N.}\
  \bibnamefont {Korotkov}},\ }\bibfield  {title} {\enquote {\bibinfo {title}
  {Quantum theory of a bandpass purcell filter for qubit readout},}\ }\href
  {\doibase 10.1103/PhysRevA.92.012325} {\bibfield  {journal} {\bibinfo
  {journal} {Phys. Rev. A}\ }\textbf {\bibinfo {volume} {92}},\ \bibinfo
  {pages} {012325} (\bibinfo {year} {2015})}\BibitemShut {NoStop}%
\bibitem [{\citenamefont {Bronn}\ \emph {et~al.}(2015)\citenamefont {Bronn},
  \citenamefont {Liu}, \citenamefont {Hertzberg}, \citenamefont {Córcoles},
  \citenamefont {Houck}, \citenamefont {Gambetta},\ and\ \citenamefont
  {Chow}}]{Bronn2015}%
  \BibitemOpen
  \bibfield  {author} {\bibinfo {author} {\bibfnamefont {Nicholas~T.}\
  \bibnamefont {Bronn}}, \bibinfo {author} {\bibfnamefont {Yanbing}\
  \bibnamefont {Liu}}, \bibinfo {author} {\bibfnamefont {Jared~B.}\
  \bibnamefont {Hertzberg}}, \bibinfo {author} {\bibfnamefont {Antonio~D.}\
  \bibnamefont {Córcoles}}, \bibinfo {author} {\bibfnamefont {Andrew~A.}\
  \bibnamefont {Houck}}, \bibinfo {author} {\bibfnamefont {Jay~M.}\
  \bibnamefont {Gambetta}}, \ and\ \bibinfo {author} {\bibfnamefont {Jerry~M.}\
  \bibnamefont {Chow}},\ }\bibfield  {title} {\enquote {\bibinfo {title}
  {Broadband filters for abatement of spontaneous emission in circuit quantum
  electrodynamics},}\ }\href {\doibase 10.1063/1.4934867} {\bibfield  {journal}
  {\bibinfo  {journal} {Applied Physics Letters}\ }\textbf {\bibinfo {volume}
  {107}},\ \bibinfo {pages} {172601} (\bibinfo {year} {2015})}\BibitemShut
  {NoStop}%
\bibitem [{\citenamefont {Macklin}\ \emph {et~al.}(2015)\citenamefont
  {Macklin}, \citenamefont {O’Brien}, \citenamefont {Hover}, \citenamefont
  {Schwartz}, \citenamefont {Bolkhovsky}, \citenamefont {Zhang}, \citenamefont
  {Oliver},\ and\ \citenamefont {Siddiqi}}]{Macklin2015}%
  \BibitemOpen
  \bibfield  {author} {\bibinfo {author} {\bibfnamefont {C.}~\bibnamefont
  {Macklin}}, \bibinfo {author} {\bibfnamefont {K.}~\bibnamefont {O’Brien}},
  \bibinfo {author} {\bibfnamefont {D.}~\bibnamefont {Hover}}, \bibinfo
  {author} {\bibfnamefont {M.~E.}\ \bibnamefont {Schwartz}}, \bibinfo {author}
  {\bibfnamefont {V.}~\bibnamefont {Bolkhovsky}}, \bibinfo {author}
  {\bibfnamefont {X.}~\bibnamefont {Zhang}}, \bibinfo {author} {\bibfnamefont
  {W.~D.}\ \bibnamefont {Oliver}}, \ and\ \bibinfo {author} {\bibfnamefont
  {I.}~\bibnamefont {Siddiqi}},\ }\bibfield  {title} {\enquote {\bibinfo
  {title} {A near quantum-limited josephson traveling-wave parametric
  amplifier},}\ }\href {\doibase 10.1126/science.aaa8525} {\bibfield  {journal}
  {\bibinfo  {journal} {Science}\ }\textbf {\bibinfo {volume} {350}},\ \bibinfo
  {pages} {307--310} (\bibinfo {year} {2015})}\BibitemShut {NoStop}%
\bibitem [{\citenamefont {Koolstra}\ \emph {et~al.}(2022)\citenamefont
  {Koolstra}, \citenamefont {Stevenson}, \citenamefont {Barzili}, \citenamefont
  {Burns}, \citenamefont {Siva}, \citenamefont {Greenfield}, \citenamefont
  {Livingston}, \citenamefont {Hashim}, \citenamefont {Naik}, \citenamefont
  {Kreikebaum}, \citenamefont {O'Brien}, \citenamefont {Santiago},
  \citenamefont {Dressel},\ and\ \citenamefont {Siddiqi}}]{Koolstra2022}%
  \BibitemOpen
  \bibfield  {author} {\bibinfo {author} {\bibfnamefont {G.}~\bibnamefont
  {Koolstra}}, \bibinfo {author} {\bibfnamefont {N.}~\bibnamefont {Stevenson}},
  \bibinfo {author} {\bibfnamefont {S.}~\bibnamefont {Barzili}}, \bibinfo
  {author} {\bibfnamefont {L.}~\bibnamefont {Burns}}, \bibinfo {author}
  {\bibfnamefont {K.}~\bibnamefont {Siva}}, \bibinfo {author} {\bibfnamefont
  {S.}~\bibnamefont {Greenfield}}, \bibinfo {author} {\bibfnamefont
  {W.}~\bibnamefont {Livingston}}, \bibinfo {author} {\bibfnamefont
  {A.}~\bibnamefont {Hashim}}, \bibinfo {author} {\bibfnamefont {R.~K.}\
  \bibnamefont {Naik}}, \bibinfo {author} {\bibfnamefont {J.~M.}\ \bibnamefont
  {Kreikebaum}}, \bibinfo {author} {\bibfnamefont {K.~P.}\ \bibnamefont
  {O'Brien}}, \bibinfo {author} {\bibfnamefont {D.~I.}\ \bibnamefont
  {Santiago}}, \bibinfo {author} {\bibfnamefont {J.}~\bibnamefont {Dressel}}, \
  and\ \bibinfo {author} {\bibfnamefont {I.}~\bibnamefont {Siddiqi}},\
  }\bibfield  {title} {\enquote {\bibinfo {title} {Monitoring fast
  superconducting qubit dynamics using a neural network},}\ }\href {\doibase
  10.1103/PhysRevX.12.031017} {\bibfield  {journal} {\bibinfo  {journal} {Phys.
  Rev. X}\ }\textbf {\bibinfo {volume} {12}},\ \bibinfo {pages} {031017}
  (\bibinfo {year} {2022})}\BibitemShut {NoStop}%
\bibitem [{Note1()}]{Note1}%
  \BibitemOpen
  \bibinfo {note} {The Moore-Penrose left pseudoinverse $A^+$ of an $m\times n$
  matrix $A$ with rank $n$ is given by $A^+ = (A^T A)^{-1} A^T$ such that $A^+
  A = I$.}\BibitemShut {Stop}%
\bibitem [{\citenamefont {Johansson}\ \emph {et~al.}(2012)\citenamefont
  {Johansson}, \citenamefont {Nation},\ and\ \citenamefont
  {Nori}}]{Johansson2012}%
  \BibitemOpen
  \bibfield  {author} {\bibinfo {author} {\bibfnamefont {J.R.}\ \bibnamefont
  {Johansson}}, \bibinfo {author} {\bibfnamefont {P.D.}\ \bibnamefont
  {Nation}}, \ and\ \bibinfo {author} {\bibfnamefont {Franco}\ \bibnamefont
  {Nori}},\ }\bibfield  {title} {\enquote {\bibinfo {title} {Qutip: An
  open-source python framework for the dynamics of open quantum systems},}\
  }\href {\doibase https://doi.org/10.1016/j.cpc.2012.02.021} {\bibfield
  {journal} {\bibinfo  {journal} {Computer Physics Communications}\ }\textbf
  {\bibinfo {volume} {183}},\ \bibinfo {pages} {1760--1772} (\bibinfo {year}
  {2012})}\BibitemShut {NoStop}%
\bibitem [{\citenamefont {Johansson}\ \emph {et~al.}(2013)\citenamefont
  {Johansson}, \citenamefont {Nation},\ and\ \citenamefont
  {Nori}}]{Johansson2013}%
  \BibitemOpen
  \bibfield  {author} {\bibinfo {author} {\bibfnamefont {J.R.}\ \bibnamefont
  {Johansson}}, \bibinfo {author} {\bibfnamefont {P.D.}\ \bibnamefont
  {Nation}}, \ and\ \bibinfo {author} {\bibfnamefont {Franco}\ \bibnamefont
  {Nori}},\ }\bibfield  {title} {\enquote {\bibinfo {title} {Qutip 2: A python
  framework for the dynamics of open quantum systems},}\ }\href {\doibase
  https://doi.org/10.1016/j.cpc.2012.11.019} {\bibfield  {journal} {\bibinfo
  {journal} {Computer Physics Communications}\ }\textbf {\bibinfo {volume}
  {184}},\ \bibinfo {pages} {1234--1240} (\bibinfo {year} {2013})}\BibitemShut
  {NoStop}%
\bibitem [{\citenamefont {Szombati}\ \emph {et~al.}(2020)\citenamefont
  {Szombati}, \citenamefont {Gomez~Frieiro}, \citenamefont {M\"uller},
  \citenamefont {Jones}, \citenamefont {Jerger},\ and\ \citenamefont
  {Fedorov}}]{Szombati2020}%
  \BibitemOpen
  \bibfield  {author} {\bibinfo {author} {\bibfnamefont {Daniel}\ \bibnamefont
  {Szombati}}, \bibinfo {author} {\bibfnamefont {Alejandro}\ \bibnamefont
  {Gomez~Frieiro}}, \bibinfo {author} {\bibfnamefont {Clemens}\ \bibnamefont
  {M\"uller}}, \bibinfo {author} {\bibfnamefont {Tyler}\ \bibnamefont {Jones}},
  \bibinfo {author} {\bibfnamefont {Markus}\ \bibnamefont {Jerger}}, \ and\
  \bibinfo {author} {\bibfnamefont {Arkady}\ \bibnamefont {Fedorov}},\
  }\bibfield  {title} {\enquote {\bibinfo {title} {Quantum rifling: Protecting
  a qubit from measurement back action},}\ }\href {\doibase
  10.1103/PhysRevLett.124.070401} {\bibfield  {journal} {\bibinfo  {journal}
  {Phys. Rev. Lett.}\ }\textbf {\bibinfo {volume} {124}},\ \bibinfo {pages}
  {070401} (\bibinfo {year} {2020})}\BibitemShut {NoStop}%
\bibitem [{\citenamefont {Takita}\ \emph {et~al.}(2017)\citenamefont {Takita},
  \citenamefont {Cross}, \citenamefont {C\'orcoles}, \citenamefont {Chow},\
  and\ \citenamefont {Gambetta}}]{Takita2017}%
  \BibitemOpen
  \bibfield  {author} {\bibinfo {author} {\bibfnamefont {Maika}\ \bibnamefont
  {Takita}}, \bibinfo {author} {\bibfnamefont {Andrew~W.}\ \bibnamefont
  {Cross}}, \bibinfo {author} {\bibfnamefont {A.~D.}\ \bibnamefont
  {C\'orcoles}}, \bibinfo {author} {\bibfnamefont {Jerry~M.}\ \bibnamefont
  {Chow}}, \ and\ \bibinfo {author} {\bibfnamefont {Jay~M.}\ \bibnamefont
  {Gambetta}},\ }\bibfield  {title} {\enquote {\bibinfo {title} {Experimental
  demonstration of fault-tolerant state preparation with superconducting
  qubits},}\ }\href {\doibase 10.1103/PhysRevLett.119.180501} {\bibfield
  {journal} {\bibinfo  {journal} {Phys. Rev. Lett.}\ }\textbf {\bibinfo
  {volume} {119}},\ \bibinfo {pages} {180501} (\bibinfo {year}
  {2017})}\BibitemShut {NoStop}%
\bibitem [{\citenamefont {McKay}\ \emph {et~al.}(2017)\citenamefont {McKay},
  \citenamefont {Wood}, \citenamefont {Sheldon}, \citenamefont {Chow},\ and\
  \citenamefont {Gambetta}}]{McKay2017}%
  \BibitemOpen
  \bibfield  {author} {\bibinfo {author} {\bibfnamefont {David~C.}\
  \bibnamefont {McKay}}, \bibinfo {author} {\bibfnamefont {Christopher~J.}\
  \bibnamefont {Wood}}, \bibinfo {author} {\bibfnamefont {Sarah}\ \bibnamefont
  {Sheldon}}, \bibinfo {author} {\bibfnamefont {Jerry~M.}\ \bibnamefont
  {Chow}}, \ and\ \bibinfo {author} {\bibfnamefont {Jay~M.}\ \bibnamefont
  {Gambetta}},\ }\bibfield  {title} {\enquote {\bibinfo {title} {Efficient $z$
  gates for quantum computing},}\ }\href {\doibase 10.1103/PhysRevA.96.022330}
  {\bibfield  {journal} {\bibinfo  {journal} {Phys. Rev. A}\ }\textbf {\bibinfo
  {volume} {96}},\ \bibinfo {pages} {022330} (\bibinfo {year}
  {2017})}\BibitemShut {NoStop}%
\bibitem [{\citenamefont {Magnard}\ \emph {et~al.}(2018)\citenamefont
  {Magnard}, \citenamefont {Kurpiers}, \citenamefont {Royer}, \citenamefont
  {Walter}, \citenamefont {Besse}, \citenamefont {Gasparinetti}, \citenamefont
  {Pechal}, \citenamefont {Heinsoo}, \citenamefont {Storz}, \citenamefont
  {Blais},\ and\ \citenamefont {Wallraff}}]{Magnard2018}%
  \BibitemOpen
  \bibfield  {author} {\bibinfo {author} {\bibfnamefont {P.}~\bibnamefont
  {Magnard}}, \bibinfo {author} {\bibfnamefont {P.}~\bibnamefont {Kurpiers}},
  \bibinfo {author} {\bibfnamefont {B.}~\bibnamefont {Royer}}, \bibinfo
  {author} {\bibfnamefont {T.}~\bibnamefont {Walter}}, \bibinfo {author}
  {\bibfnamefont {J.-C.}\ \bibnamefont {Besse}}, \bibinfo {author}
  {\bibfnamefont {S.}~\bibnamefont {Gasparinetti}}, \bibinfo {author}
  {\bibfnamefont {M.}~\bibnamefont {Pechal}}, \bibinfo {author} {\bibfnamefont
  {J.}~\bibnamefont {Heinsoo}}, \bibinfo {author} {\bibfnamefont
  {S.}~\bibnamefont {Storz}}, \bibinfo {author} {\bibfnamefont
  {A.}~\bibnamefont {Blais}}, \ and\ \bibinfo {author} {\bibfnamefont
  {A.}~\bibnamefont {Wallraff}},\ }\bibfield  {title} {\enquote {\bibinfo
  {title} {Fast and unconditional all-microwave reset of a superconducting
  qubit},}\ }\href {\doibase 10.1103/PhysRevLett.121.060502} {\bibfield
  {journal} {\bibinfo  {journal} {Phys. Rev. Lett.}\ }\textbf {\bibinfo
  {volume} {121}},\ \bibinfo {pages} {060502} (\bibinfo {year}
  {2018})}\BibitemShut {NoStop}%
\end{thebibliography}%
\end{document}